\documentclass[fleqn]{LMCS}

\def\doi{8 (1:04) 2012}
\lmcsheading%
{\doi}
{1--61}
{}
{}
{Mar.~\phantom09, 2011}
{Feb.~16, 2012}
{}

\usepackage{enumerate}
\usepackage{hyperref}

\usepackage{tech2,zed2,cspsymb,psfrag,graphicx,infrule2}
\usepackage[all]{xy}

\theoremstyle{definition}
\newtheorem{remark}[thm]{Remark}
\newtheorem{example}[thm]{Example}

\newcommand{\newnot}[1]{}
\newcommand{\Spec}{Spec}
\newcommand{\shiftedhat}[1]{\hspace{0.10em}\hat{\hspace{-0.10em}{#1}}}
\newcommand{\True}{\mathit{True}}
\newcommand{\False}{\mathit{False}}
\newcommand{\Norm}{\textbf{Seq\-Norm}}
\newcommand{\construct}{c\S_1x_1\mathrm{:}X_1\ldots\S_kx_k\mathrm{:}X_k}
\newcommand{\nont}{{\mbox{\scriptsize \it{non-t}}}}
\newcommand{\Value}{\mathit{Value}}
\newcommand{\Var}{\mathit{Var}}
\newcommand{\FV}{\mathit{FV}}
\newcommand{\Visible}{\mathit{Visible}}
\newcommand{\symtrans}[1]{\trans[#1][s]}
\newcommand{\strace}[1]{\trace{#1}}
\newcommand{\SymTrans}[1]{\mapstotrans[#1][s]}
\newcommand{\stsem}[1]{\it SymbolicTraces(#1)}
\newcommand{\nontauequiv}{\equiv_{non\mbox{-}\tau}}
\newcommand{\nontequiv}{\equiv_{non\mbox{-}t}}
\newcommand{\envrename}[1]{{[} #1 {]}}
\newcommand{\eval}[2]{\leftsemb #1 \rightsemb_{#2}}
\newcommand{\tsem}[1]{\traces(#1)}
\newcommand{\fsem}[1]{\failures(#1)}
\newcommand{\generates}[1]{\mathrel{\it{generates}_{#1}}}
\newcommand{\symtransstar}[1]{\starit{\trans[#1]}[s]}
\newcommand{\Thresh}{Thresh}
\newcommand{\In}{\mathrel{\mathrm{in}}}
\newcommand{\verproblem}{\hspace{-0.8em}\raisebox{2ex}{\tiny ?}\hspace{0.4em}}

\def\from{\mbox{:}}
\def\hatT{\shiftedhat{T}}

\def\para#1{\paragraph{\textbf{#1}}}
\def\subpara#1{\subparagraph{\textit{#1}}}

\def\Rule#1{\medskip\begin{trivlist}\item[]\textit{#1.}}
\def\endRule{\end{trivlist}\medskip}
\def\Seq{\textbf{Seq}}
\newcounter{enumctr}

\def\romanenumerate{
  \begin{list}{\rm(\roman{enumctr})}{\usecounter{enumctr}
  \setlength{\leftmargin}{3em} \setlength{\labelwidth}{4em}
}}
\def\endromanenumerate{\end{list}}

\begin{document}

\title[A type reduction theory for systems with replicated components]{A type
  reduction theory for systems with replicated components} 

\author[T.~Mazur]{Tomasz Mazur}	

\author[G.~Lowe]{Gavin Lowe}	
\address{Department of Computer Science, University of Oxford, Wolfson
  Building, Parks Road, 
  Oxford, OX1 3QD, United Kingdom.}	
\email{tomasz.mazur@gmail.com, gavin.lowe@cs.ox.ac.uk}  



\begin{abstract}
\noindent The Parameterised Model Checking Problem asks whether an
implementation $Impl(t)$ satisfies a specification $\Spec(t)$ for all
instantiations of parameter~$t$. In general, $t$~can determine numerous
entities: the number of processes used in a network, the type of data, the
capacities of buffers, etc.  The main theme of this paper is automation of
uniform verification of a subclass of PMCP with the parameter of the first
kind, i.e.~the number of processes in the network.  We use CSP as our
formalism.

We present a type reduction theory, which, for a given verification problem,
establishes a function $\phi$ that maps all (sufficiently large)
instantiations~$T$ of the parameter to some fixed type $\shiftedhat{T}$ and
allows us to deduce that if $\Spec(\shiftedhat{T})$ is refined by
$\phi(Impl(T))$, then (subject to certain assumptions) $\Spec(T)$ is refined
by $Impl(T)$. The theory can be used in practice by combining it with a
suitable abstraction method that produces a $t$-independent process $Abstr$
that is refined by $\phi(Impl(T))$ for all sufficiently large $T$. Then, by
testing (with a model checker) if the abstract model $Abstr$ refines
$\Spec(\shiftedhat{T})$, we can deduce a positive answer to the original
uniform verification problem.

The type reduction theory relies on symbolic representation of process
behaviour. We develop a symbolic operational semantics for CSP processes that
satisfy certain normality requirements, and we provide a set of translation
rules that allow us to concretise symbolic transition graphs.  Based on this,
we prove results that allow us to infer behaviours of a process instantiated
with uncollapsed types from known behaviours of the same process instantiated
with a reduced type.  

One of the main advantages of our symbolic operational semantics and the type
reduction theory is their generality, which makes them applicable in a wide
range of settings. 
\end{abstract}

\maketitle

\section{Introduction}

Until recently the primary method of correctness verification was
\emph{testing}, which, given an input, checks the produced output against the
expected outcome. This approach suffers from two main problems. Firstly, it is
almost always impossible to test every possible input and execution
path. Secondly, testing works only for completed implementations. This makes
it particularly unsuitable for verification of safety-critical systems; it is
highly unlikely that someone would ever want to perform testing to verify that
a nuclear power plant never blows up, for example. 

In contrast to the above, \emph{formal verification} methods concentrate on
\emph{proving} the correctness of a given system. One approach to formal
verification is \emph{model checking}. Given a model $Impl$ of an
implementation and a specification $\Spec$ that the model should satisfy,
verification via model checking occurs by exploring (explicitly or
symbolically) all states of $Impl$ and checking if they satisfy $\Spec$. The
greatest advantage of model checking is a large scope for automation, at the
cost of being applicable only to finite-state systems and a few families of
infinite systems. In addition, if the implementation fails to satisfy the
specification, then model checking can produce a counterexample (a behaviour
of the implementation that is not allowed by the specification) that can be
used for debugging purposes. On the other hand, this approach to formal
verification suffers from the \emph{state explosion problem}: the time
complexity of verification algorithms depends on the size of the
implementation, which is typically exponential in the size of its
description. This means that standard model checking algorithms can only work
in cases where the system to be verified is of finite and (relatively) small
size. 


One approach to model checking, highly popularised by Clarke,
Emerson 
and Grumberg~\cite{Clarke:1981,Clarke:1986,Clarke:1992,Clarke:1999}, is based
on temporal logics, where specifications are formulated as expressions in a
linear time logic (e.g.\ LTL~\cite{Pnueli:1977}) or a branching time logic
(e.g.~CTL~\cite{EC80,Pnueli:1981}). Another approach defines a partial order
$\refinedby$ on the set of all expressible systems. The intuitive meaning of
$P \refinedby Q$ (pronounced ``$P$ refined by $Q$'') for systems $P$ and $Q$
is that $Q$ is in some sense ``better'' than $P$, e.g.\ it is more
deterministic, less abstract or contains more implementation details (see
Section~\ref{section:denotationalmodelsintro} for the formal definition). In
this approach $\Spec$ and $Impl$ are modelled using the same formalism and
$Impl$ is said to satisfy $\Spec$ if and only if $\Spec \refinedby Impl$. An
immediate advantage of refinement checking over temporal logic formulae
satisfaction is the fact that what constitutes a specification for a given
implementation in one context can be treated as its abstraction in
another. This is a very useful feature when working with compositional
construction of implementations.

In this paper we use the refinement-based approach to model checking, where
all implementations and specifications are modelled using the CSP process
algebra~\cite{Hoare:1985, Roscoe:1997, Roscoe:2010} (see
Section~\ref{section:introductiontoCSP}) and refinement checks are performed
automatically using the FDR model checker~\cite{FDR:Manual}.

It is often the case that specifications or implementations contain free
variables. These can be parameters that affect the topology of the system
(e.g.\ the number of nodes in a network or the number of users of a system),
the types of data variables (e.g.\ datatypes of database records or memory
contents), performance parameters (e.g.\ bandwidths, response times, clock
speeds), or capacities of buffers or queues used. One is often interested in
the \emph{uniform verification} of a given parameterised pair of a
specification $\Spec$ and an implementation $Impl$, i.e.~in checking whether
$Impl$ satisfies $\Spec$ for \emph{all} instantiations of the
parameters. Given such $\Spec$ and $Impl$, the \emph{Parameterised
  Verification Problem} (\emph{PVP}) asks whether $Impl$ can be uniformly
verified against $\Spec$. The \emph{Parameterised Model Checking Problem}
(\emph{PMCP}) is a subclass of PVP, where we insist on the verification
occurring via model checking. 

In this paper we concentrate on a subclass of PMCP, where specifications and
implementations contains a single parameter $t$, called the
\emph{distinguished type}, which denotes the type of identities of node
processes running concurrently to form a network, possibly within some larger
system.  More precisely, every family of implementations that we consider is
of the form\footnote{The process $\mathop{\Parallel} i \in I \spot [A(i)]
  P(i)$ denotes the parallel composition of the processes $P(i)$ for $i \in
  I$, where $A(i)$ is the alphabet of $P(i)$, and where nodes synchronise on
  all events in common between their alphabets; see
  Section~\ref{section:introductiontoCSP}.}
\begin{eqnarray*}
Impl(t) & = & C_t \left[ \,
  \mathop{\Parallel} i \in t \spot [A(i,t)]\ \mathcal{N}_{i}(t) \, \right],
\end{eqnarray*}
where:
\begin{iteMize}{$\bullet$}
\item
$\mathcal{N}_{i}(t)$ models a single, finite-state node with identity $i$, and
  that can receive, store and send node identities from $t$;

\item
$A(i,t)$ is the set of all visible events that $\mathcal{N}_{i}(t)$ can
  communicate (its alphabet);

\item 
$C_t[\cdot]$ is some CSP context, for example that places the nodes in
  parallel with a controller (possibly parameterised by $t$) and may hide some
  communication.
\end{iteMize}
In fact, the results of this paper apply to more general implementation
processes that the above $Impl(t)$, namely all that are fully symmetric in~$t$
(informally, that renaming the elements of~$t$ under an arbitrary bijection
gives an equivalent process; see Definition~\ref{def:typesym}, below);
however, $Impl(t)$ captures those processes that we are particularly
interested in.

Our overall aim, then, is to verify that for all sufficiently large
instantiations~$T$ of~$t$:
\begin{eqnarray}\label{eqn:goal}
Spec(T) & \refinedby & Impl(T),
\end{eqnarray}
where $Spec(t)$ is a suitable specification process. 

Throughout this paper we assume that every instantiation $T$ of type parameter
$t$ is non-empty and finite. In addition, without loss of generality, we
assume that every instantiation $T$ of~$t$ is an initial segment of the
natural numbers, i.e.\ $T$ is of the form $\set{0 \upto n-1}$ for
some~$n$. Our results and techniques extend to other discrete and finite types
$T$ of size $n$ via simple bijections from $\set{0 \upto n-1}$ to $T$. We
allow processes to contain other parameters in their syntax, but their values
must be known and fixed at the time of writing the process definition, or an
additional technique for handling parameters (e.g.\ data independence
\cite{Lazic:1999, Roscoe:1997}) must be used for complete correctness
analysis.

PMCP is, in general, undecidable~\cite{Apt:1986}, as the Halting
Problem~\cite{Dav58} can be shown to reduce to it. Therefore, we focus
on sound (but incomplete) verification methods.



One general approach is to build a
$t$-independent abstraction process $Abstr$ that captures the behaviours of
all the $Impl(T)$ processes, in a sense that we now explain.  The alphabets of
$Impl(T)$ are (in general) unbounded as a function of~$T$; however, the
alphabet of $Abstr$ needs to be fixed.  Therefore, the construction of $Abstr$
collapses $T$ to some fixed type $\shiftedhat{T} = \set{0 \upto B}$ for
some non-negative integer~$B$, treating all identities in $\set{0 \upto B-1}$
faithfully, but mapping all other identities onto~$B$.  More precisely, for
all sufficiently large instantiations $T$ of type $t$,\, $Abstr$ is such
that\footnote{The process $f(P)$ is a process that acts like~$P$, except every
  event~$a$ is renamed to~$f(a)$; see
  Section~\ref{section:introductiontoCSP}.}
\begin{eqnarray}\label{equation:introabstrrefinement}
	Abstr & \refinedby & \phi(Impl(T))
\end{eqnarray}
holds by construction, where $\phi$ is a $B$-collapsing function:
\begin{defi}\label{definition:collapsing}
A \emph{$B$-collapsing function} is a function $\phi : T \rightarrow \set{0
  \upto B}$ such that
\begin{iteMize}{$\bullet$}
\item $\phi(v) = v$ for $v \in \set{0 \upto B-1}$;
\item $\phi(v) = B$ for $v \in \set{B \upto \#T-1}$.
\end{iteMize}
\end{defi}
\noindent
Having constructed such an $Abstr$, we can use a CSP model checker, such as
FDR, to verify that
\begin{eqnarray*}
\Spec(\shiftedhat{T}) & \refinedby & Abstr.
\end{eqnarray*}
Transitivity of refinement then allows us to deduce that
\begin{eqnarray}\label{equation:introdeduction}
	\Spec(\shiftedhat{T}) & \refinedby & \phi(Impl(T)) 
\end{eqnarray}
for all sufficiently large $T$.  An example of such an abstraction method
(based on counter abstraction techniques~\cite{Lub84,Pnueli:2002,Mazur:2007})
can be found in~\cite{Tomasz-thesis, Mazur:2010}.

The aim of this paper is to bridge the gap between
equations~(\ref{equation:introdeduction}) and~(\ref{eqn:goal}).  We present a
theory that, under suitable assumptions on the specification and
implementation processes, allows us to calculate a suitable value for~$B$ such
that if equation~(\ref{equation:introdeduction}) holds (for the values of
$\phi$ and $\shiftedhat{T}$ corresponding to~$B$), then
equation~(\ref{eqn:goal}) holds for all $T$ such that $\#T > B$ (smaller
values of $T$ can be tested directly).  In particular, the value of~$B$ turns
out to depend only on the syntax of the specification, and is independent of
the implementation. 

Our theory is general, allowing us to combine it with an arbitrary abstraction
method that can produce an abstraction $Abstr$ such that
(\ref{equation:introabstrrefinement}) holds.

The rest of this paper is structured as follows. In
Section~\ref{section:introductiontoCSP} we introduce the syntax of the CSP
process algebra, describe two of its denotational semantics models (traces and
stable failures) and briefly talk about FDR, a model checker for CSP\@.  We
also give an example to illustrate the goals of this paper.  In
Section~\ref{sec:conditions} we define the conditions we will require the
specification process to satisfy, and also the condition of symmetry in~$t$
that we will require the implementation to satisfy.

Proving the main theorems will require us to develop quite a lot of supporting
machinery, in order to relate behaviours of the specification process for
different values of the parameter~$t$.  To this end,
Section~\ref{section:operationalsemantics} is devoted to developing a suitable
operational semantics for CSP\@.  The main part of this section presents a
symbolic operational semantics that allows us to reason about behaviour of
processes without the need for instantiating parameters. We also provide a set
of translation rules for instantiating symbolic transition graphs into
concrete ones, and we prove that this results in an operational semantics
congruent to a fairly standard one.

Being able to reason about process behaviour in a symbolic way is a
prerequisite for our main theory.  We present a number of regularity results
for specifications in Section~\ref{section:regularity}, which show that
specifications exhibit certain clarity in their behaviour.  Our main type
reduction theory is in Section~\ref{section:typereduction}, where we provide
type reduction theorems for the traces and stable failures models. Finally, we
conclude in Section~\ref{section:conclusions}.  In the interests of
readability, we relegate most proofs to appendices.

\section{Introduction to CSP}\label{section:introductiontoCSP}

CSP~\cite{Hoare:1985, Roscoe:1997, Roscoe:2010} is a process algebra used for
modelling and verification of concurrent reactive systems with communication
based on synchronous message passing.

CSP processes interact with each other and the environment within which they
operate by communicating \emph{events}.  Events occur on \emph{channels}; for
example, $c.a.3$ is an event over channel~$c$, passing data~$a$ and~3.  We
assume that each channel has a fixed type (i.e.~can pass a fixed number of
pieces of data, and the type of the data passed in each position is fixed).
The notation $\set{|c|}$ represents the set of events passed over
channel~$c$. 

We let $\Sigma$ be the set of all visible events.  We let $\tau$ denote a
special, internal event (not in~$\Sigma$).  We write $\Sigma^{\tau}$ to mean
$\Sigma \union \set{\tau}$. We also write $\Sigma^{*}$ to mean the set of all
finite sequences of events from~$\Sigma$.

\subsection{Syntax}\label{section:cspsyntax}

In this paper we use the fragment of CSP with the following  syntax.
\[ P ::=
\align
\STOP | \alpha \then P | P \extchoice P | 
  \Extchoice i \in \mathcal{I} \spot P(i) | P \intchoice P | 
  \Intchoice i \in \mathcal{I} \spot P(i) | P \timeout P \\
| \If b \Then P \Else P | b\ \&\ P | P \hide X | 
P \rename{\mathcal{R}} | P \parallel[X][Y] P \\
| \mathop{\Parallel} i \in \mathcal{I} \spot [A(i)]\ P(i) | 
P \parallel[X] P 
| P \interleave P | \mathop{\Interleave} i \in \mathcal{I} \spot P(i) | X
\endalign
\]

The process $\STOP$ is a synonym for deadlock, i.e.~it is the process that
cannot engage in any communication with the environment and cannot perform any
events on its own. 
 
The process $\alpha \then P$ can perform any event that the construct $\alpha$
describes, and then subsequently behaves like~$P$.  The construct~$\alpha$ is
an expression of the form $c\S_1x_1\mathrm{:}X_1\ldots\S_kx_k\mathrm{:}X_k$,
where
\begin{iteMize}{$\bullet$}
\item $c$ is a channel name;

\item $\S_i \in \{\$,?,!\}$ is an input/output symbol\footnote{Standard
  CSP commonly also uses the .\ symbol, but this is only syntactic sugar and
  can always be replaced by one of $\$, ?, !$.};

\item if $\S_i \in \set{\$,?}$, then $x_i$ is an input variable, otherwise it
  is an output value; and

\item if $\S_i \in \set{\$,?}$, then $X_i$ is a type parameter or type of input, otherwise it is $null$.
\end{iteMize} 
The $!$ symbol denotes an output; $?$ denotes an input; $\$$ denotes a
nondeterministic choice (which we sometimes call a nondeterministic input).
The $?$ and $\$$ operators both bind variables to concrete values.
For example, the process $c\$x\mathrm{:}\set{0,1}?y\mathrm{:}\set{2,3}!4 \then
d!(x\mathord+y) \then STOP$ 
nondeterministically chooses a value $v \in \set{0,1}$ and binds the
variable~$x$ to that value; it is then willing to perform any event of the
form $c.v.w.4$ for $w \in \set{2,3}$, and binds the variable $y$ to the
value~$w$; it then performs the event $d.(v\mathord+w)$, and deadlocks. 
For constructs where $\S_i = \mathord !$ for every $i$, we use the more
traditional .\ output symbol instead, e.g.~we write $c.v_1.v_2.v_3$ to mean
$c!v_1!v_2!v_3$. Whenever $X_i$ is $null$, we omit it in practice, e.g.~we
write $c!v$ instead of $c!v\mathrm{:}null$.  The
only way a process can communicate a visible event is via a prefix
construct. 


For two processes~$P$~and~$Q$, the \emph{external} (or \emph{deterministic})
choice $P \extchoice Q$ is a process that offers the environment the choice of
performing any initial event of~$P$ or $Q$; if an initial event of $P$ is
performed, then the choice is resolved to~$P$, and if an initial event of~$Q$
is performed, then the choice is resolved to~$Q$.  We can define a  replicated
version of the operator: $\Extchoice i \in \mathcal{I} \spot P(i)$ is an
external choice between processes $P(i)$ for each $i$ in some finite
indexing set $\mathcal{I}$; we consider this as syntactic sugar for repeated
use of the binary operator.  

$P \intchoice Q$ represents
an \emph{internal} (or \emph{nondeterministic}) choice, where the process
behaves either like~$P$ or like~$Q$, where the choice is made by some
mechanism that we do not model and which cannot be influenced by the
environment.  We define a replicated version: 
\mbox{$\Intchoice i\in\mathcal{I} \spot P(i)$} is an internal choice between
processes $P(i)$ for each $i$ in some finite, non-empty indexing
set~$\mathcal{I}$.  

The \emph{sliding} choice (or \emph{timeout}) $P \timeout Q$ is a process
that behaves like~$P$ for a nondeterministically long period of time, but if
the environment does not engage in any activity with $P$ within this time, it
switches to behaving like $Q$. 

The process $\If b \Then P \Else Q$ is a conditional choice between processes
$P$~and~$Q$. If $b$ evaluates to $\True$, then this process behaves like $P$;
otherwise it behaves like $Q$. The process $b\ \&\ P$ is syntactic sugar for
$\If b \Then P \Else \STOP$, i.e.~$P$ is enabled if and only if guard~$b$ is
true. We say ``a conditional choice on $t$'' to mean a conditional choice
whose boolean condition involves only variables and/or values of type $t$.  

For any set $X \subseteq \Sigma$, $P \hide X$ is a process which behaves like
$P$ except that whenever $P$ would normally communicate an event from set~$X$,
$P \hide X$ performs the internal action,~$\tau$, instead.  

The process $P \rename{\mathcal{R}}$, where $\mathcal{R}$ is a relation
over~$\Sigma$, is a process that behaves like~$P$ except that whenever~$P$
would perform an event~$a$, the renamed process performs an event~$b$ such
that $a~\mathcal{R}~b$ instead.  We sometimes define the renaming relation
using notation similar to substitution: $P \leftsemb ^b / _a \rightsemb$ is a
process that behaves like~$P$ except that whenever~$P$ would normally
perform~$a$, the renamed process performs $b$ instead.  If $\mathcal{R}$ is a
function, we sometimes write the renaming using functional notation,
$\mathcal{R}(P)$.

The notion of parallel composition of processes is key to CSP, allowing one to
model concurrency. The process $P \parallel[X][Y] Q$ is a parallel composition
of $P$ and $Q$, where $P$ is allowed to communicate only members of the set of
visible events $X$,\, $Q$ is allowed to communicate only members of the set of
visible events $Y$, and synchronisation occurs on all common events
(i.e.~those in $X \cap Y$).  We can define its replicated version:
$\Parallel i\in\mathcal{I} \spot [A(i)]\ P(i)$ is the parallel composition of
processes $P(i)$ indexed over a finite, non-empty set $\mathcal{I}$, where
each $P(i)$ is allowed to perform only events from $A(i)$, and synchronises on
event $e \in A(i)$ with each process $P(j)$ such that $e \in A(j)$. 
The process $P \parallel[X] Q$ is the parallel composition of~$P$~and~$Q$ with
handshaken synchronisation on all the members of the set of visible
events~$X$. 
Finally, $P \interleave Q$ is the interleaving of~$P$ and $Q$: the
processes run in parallel, but do not synchronise on any event (note that this
is equivalent to $P \parallel[\set{}] Q$).  We write $\Interleave
i\in\mathcal{I} \spot  P(i)$ for the replicated version. 

Processes are defined by means of equations, such as $P = a \then P$.  We
assume a global environment~$E$, mapping identifiers to process definitions,
capturing these equations.  When a process identifier $X$ is encountered in
syntax, $E$ is used to look up which process definition should be substituted
for~$X$. 

So far we have used the term ``process'' loosely. We now make an important
distinction between \emph{process syntaxes} (also called process definitions)
and \emph{concrete processes}. A process syntax is an open CSP term (i.e.~one
with free variables). On the other hand, every closed CSP term represents a
process. For example, if $Proc(t)$ is a term where $t$ is free, then it is a
process syntax and it represents a family of processes $Proc(T)$, one for each
concrete instantiation $T$.


\subsection{Denotational models and
  refinement}\label{section:denotationalmodelsintro}

A \emph{trace} of a process is a sequence of visible events that it can
perform.  We write $\traces(P)$ for the traces of~$P$.  

Given a process $P$, we let $\initials(P)$ be the set of all the initially
available visible events of $P$, i.e. 
\begin{eqnarray*}
	\initials(P) & = & \set{ a | \trace{a} \in \traces(P)}.
\end{eqnarray*}
In addition, if $tr$ is a trace of $P$, then $P/tr$ (pronounced ``$P$ after
$tr$'') describes the behaviours of $P$ after it performs $tr$. So, in
particular,  
\begin{eqnarray*}
	\initials(P/tr) & = & \set{ a | tr \cat \trace{a} \in \traces(P)}.
\end{eqnarray*}

CSP specifications are expressed in the same formalism as implementations,
i.e.~as processes. An implementation $Impl$ is said to satisfy a specification
$\Spec$ if it \emph{refines} it, which we denote by writing $\Spec \refinedby
Impl$. Intuitively, process $Q$ refines process $P$ (or $P$ is \emph{refined
  by} $Q$) if $Q$ does not exhibit any behaviour that is not a behaviour
of~$P$. The type of behaviour that identifies a CSP process depends on the
denotational model that is used. 
In the traces model refinement is defined by:
\begin{eqnarray*}
	P \trefinedby Q & \iff & \traces(Q) \subseteq \traces(P).
\]
If $P \trefinedby Q$ and $Q \trefinedby P$, then we say that $P$ and $Q$ are
\emph{traces equivalent}, denoted $P \equiv_T Q$. 

In the \emph{stable failures model}, a process~$P$ is identified by the set of
its traces (as above) together with the set of its \emph{failures}
(written $\failures(P)$). A failure is a pair $(tr, X)$, where $tr \in
\traces(P)$ and $X \subseteq \Sigma$, and represents the behaviour where~$P$
performs trace~$tr$ to reach a stable state $P'$ (i.e.~$\tau$ is not
available in $P'$), in which it  refuses the whole of~$X$ (i.e.~none of the
events in~$X$ is available), denoted $P' \refuses X$. When refinement is
interpreted over the stable failures model, we get the notion of \emph{stable
  failures refinement}:
\begin{eqnarray*}
P \frefinedby Q & \iff & 
  \traces(Q) \subseteq \traces(P) \land \failures(Q) \subseteq \failures(P).
\end{eqnarray*}
If $P \frefinedby Q$ and $Q \frefinedby P$, then we say that $P$ and $Q$ are \emph{stable failures equivalent}, denoted $P \equiv_F Q$.

All denotational representations of a process $P$ (including $\traces(P)$ and
$\failures(P)$) can be obtained using the rules of denotational semantics,
which can be found, for example, in \cite[Chapter 8]{Roscoe:1997}. An
alternative approach (and the one we take most of the time in this paper) is
to extract denotational values from a labelled transition system
representing~$P$, obtained by applying an operational semantics. We describe
this method in more detail in Section~\ref{section:denotationalvaluesfromos}.

The FDR (Failures/Divergences Refinement) model checker~\cite{FDR:Manual}
allows
one to automatically perform refinement checks. When a CSP script with process
definitions, say $P$ and~$Q$, is loaded, FDR can automatically test for
refinement $P \refinedby_{\mathrm{M}} Q$ in a given denotational
model~$M$. 


\subsection{Example}
\label{sec:example}

We give here a simple example, to illustrate the problem we are addressing in
this paper.  

Consider a very simple token-based mutual exclusion protocol for a collection
of nodes.  Each node~$i$ obtains the token (event $getToken.i$),
enters and then leaves the
critical section (event $enterCS.i$, respectively, 
$leaveCS.i$), and returns the token (event $returnToken.i$):
\begin{eqnarray*}
Node(i) & = & getToken.i \then Entering(i), \\
Entering(i) & = & enterCS.i \then CS(i), \\
CS(i) & = & leaveCS.i \then Leaving(i), \\
Leaving(i) & = & returnToken.i \then Node(i).
\end{eqnarray*}
The nodes are interleaved; recall that we use the variable~$t$ to denote the
type of all node identities:
\begin{eqnarray*}
Nodes(t) & = & \Interleave i : t \spot Node(i).
\end{eqnarray*}
The nodes are combined with a controller that controls the token, repeatedly
giving it to a node and receiving it back.  The communications corresponding
to passing the token are considered internal so are hidden.
\begin{eqnarray*}
Controller(t) & = & 
  getToken?i\mathord:t \then returnToken?j\mathord:t \then Controller(t), \\
Impl(t) & = &
  \begin{align}
  ( Nodes(t) \parallel[\set{| getToken, returnToken |}] Controller(t) ) \\
  \qquad   \hide \set{| getToken, returnToken |}.
  \end{align}
\end{eqnarray*}

We would like to verify that at most a single node is in the critical section
at a time.  We can capture this using the specification process
\begin{eqnarray*}
Spec(t) & = & enterCS\$i\mathord:t \then leaveCS!i \then Spec(t).
\end{eqnarray*}
Our requirement, then, is
\begin{eqnarray}\label{eqn:requirement}
Spec(T) & \trefinedby & Impl(T) ,
  \qquad \mbox{for all instantiations~$T$ of~$t$.}
\end{eqnarray}


The approach we describe in~\cite{Tomasz-thesis, Mazur:2010} is to form an
abstraction of $Nodes(t)$ based on counter abstraction~\cite{Pnueli:2002}.  In
the process $NodesAbst(n,e,c,l)$, below, the four counter parameters $n$, $e$,
$c$ and~$l$ represent the number of nodes in the $Node$, $Entering$, $CS$ and
$Leaving$ states, respectively; however the counting is capped at some
value~$z$, where we take $z=2$ in this case; hence a counter value of~$z$
represents that there are $z$ \emph{or more} processes in the corresponding
state.  The definition of $NodesAbst$ is based on the transitions within a
single $Node$ process.  For most transitions, the counter for the prior
$Node$ state is decremented, and the counter for the new state is
incremented, but not beyond~$z$; we define the following function to perform
this:
\begin{eqnarray*}
inc(x) & = & \min (x+1, z). 
\end{eqnarray*}
However, if the counter for the prior state was at the cap~$z$, then there
might have been strictly more than~$z$ processes in this state before the
transition, so the counter should (nondeterministically) be able to stay
at~$z$. 
\[
\begin{align} 
NodesAbst(n,e,c,l)(t) = \\
\qquad
  \begin{align}
  (n>0 \mathbin\& getToken\$i\mathord:t \then \\
  \qquad  \If n<z \Then  NodesAbst(n-1,inc(e),c,l)(t) \\
  \qquad \Else NodesAbst(n-1,inc(e),c,l)(t) 
        \intchoice NodesAbst(n,inc(e),c,l)(t)) \\
  \extchoice \\
  (e>0 \mathbin\&  enterCS\$i\mathord:t \then \\
  \qquad   \If e<z \Then NodesAbst(n,e-1,inc(c),l)(t) \\
  \qquad \Else NodesAbst(n,e-1,inc(c),l)(t)
         \intchoice NodesAbst(n,e,inc(c),l)(t)) \\
  \extchoice \\
  (c>0 \mathbin\&  leaveCS\$i\mathord:t \then \\
  \qquad   \If c<z \Then NodesAbst(n,e,c-1,inc(l))(t) \\
  \qquad \Else NodesAbst(n,e,c-1,inc(l))(t) 
          \intchoice NodesAbst(n,e,c,inc(l))(t)) \\
  \extchoice \\
  (l>0 \mathbin\&  returnToken\$i\mathord:t \then \\
  \qquad   \If l<z \Then NodesAbst(inc(n),e,c,l-1)(t)  \\
  \qquad \Else NodesAbst(inc(n),e,c,l-1)(t) 
       \intchoice NodesAbst(inc(n),e,c,l)(t)) .
  \end{align}
\end{align}
\]
We can then build $Abst$ from $NodesAbst(z,0,0,0)$ in the same way that we
built $Impl$ from $Nodes$:
\begin{eqnarray*}
Abst(t) & = &
  \begin{align}
  ( NodesAbst(z,0,0,0)(t)
      \parallel[\set{| getToken, returnToken |}] Controller(t) ) \\
  \qquad   \hide \set{| getToken, returnToken |}.
  \end{align}
\end{eqnarray*}

In~\cite{Tomasz-thesis, Mazur:2010}, we show that the process built in this
way is an abstraction of the $Impl$ process in the following sense: for every
non-negative integer~$B$:
\begin{eqnarray}\label{eqn:abst}
Abst(\hatT) & \trefinedby & \phi(Impl(T))  ,
\end{eqnarray}
for all instantiations~$T$ of~$t$ with $\#T \ge B+z$,
where $\hatT = \set{0 \upto B}$, and $\phi$ is a $B$-collapsing function (see
Definition~\ref{definition:collapsing}).
We pick $B=1$ in this case.  We can then use
FDR to verify that  
\begin{eqnarray*}
Spec(\hatT) & \trefinedby & Abst(\hatT),
\end{eqnarray*}
and so deduce
\begin{eqnarray*}
Spec(\hatT) & \trefinedby & \phi(Impl(T)), 
  \qquad \mbox{for all instantiations~$T$ of~$t$  with $\#T \ge B+z = 3$,}
\end{eqnarray*}
by transitivity of refinement.
The results in this paper will allow us to deduce our requirement
(\ref{eqn:requirement}) from this.  We stress, though, that the results in
this paper can be used with any abstraction method that produces a process
$Abst$ such that (\ref{eqn:abst}) holds for all sufficiently large
instantiations~$T$ of~$t$. 

It is worth noting that the technique in~\cite{Tomasz-thesis, Mazur:2010} is
rather more general than the above example illustrates.  It allows node
processes to store the identities of other nodes, and to pass them on in
subsequent events; much of the difficulty of the theory concerns treating
these identities correctly. 

\section{Conditions on processes}
\label{sec:conditions}

In this section we define various conditions on processes that we will use
later.  In Section~\ref{section:tracesresults} we will mention tool support,
which is able to test for most of the conditions in this section. 

We mentioned above that we restrict our operational semantics to a fragment of
the CSP language when working with specifications. We aim to develop
mathematical machinery to prove (in Section~\ref{section:typereduction})
useful results about specifications that satisfy a certain normality
condition, which we define in Section~\ref{section:norm}. Earlier, in
Section~\ref{section:dataindep} we define data independence, a crucial part of
normality. We will strongly rely on our normality condition when defining our
Semi-Symbolic Operational Semantics (Section~\ref{section:ssos}) and when
deriving type reduction theory results in
Section~\ref{section:typereduction}. 

In Section~\ref{sec:TypeSym} we define the notion of type symmetry in the
type~$t$; our main theorems will require the implementation process to satisfy
this property.  Then in Section~\ref{sec:equalityTests} we define a property
concerning the use of equality tests; our main theorems will require the
specification process to satisfy this property.  


\subsection{Data independence}\label{section:dataindep}

Intuitively, we say that a process syntax treats type $t$ data independently
if it inputs and outputs values of type $t$, possibly storing them for later
use, but does not perform any operations on these values that could influence
either its control flow or the instantiations of type $t$ that can be
used. The following definition of a data independent process is based on the
one from \cite{Roscoe:1997}. 

\begin{defi}\label{def:dataindependence}
We say that a CSP process syntax is data independent with respect to type $t$
if it does not contain: 
\begin{romanenumerate}
\item replicated constructs indexed over any set depending on $t$,
   except for replicated nondeterministic choice ($\Intchoice$) indexed over
   the whole of $t$; however, we allow the use of deterministic and
   nondeterministic input selections, $?$ and $\$$;

\item conditional choices on $t$, except for equality and inequality
   tests; 

\item constants of type $t$;

\item functions whose domains or co-domains involve type $t$;

\item operations on $t$, including polymorphic operations (e.g.~tupling
  or lists);

\item selections from sets involving $t$, unless the selection is over
   the whole of $t$; and 

\item any operations that would extract information about $t$,
   e.g.\ $card(t)$. 
\end{romanenumerate}
\end{defi}

\begin{example}
The $Node(i)(T)$ processes from Section~\ref{sec:example} are data
independent in~$t$.  However, $Nodes(t)$ is not data independent because it
uses an indexed interleaving over~$t$. 
\end{example}




\begin{remark}\label{remark:tornott}
Clauses~(v) and (vi) of Definition~\ref{section:dataindep} together imply
that, for all constructs $c\S_1x_1\mathrm{:}X_1 \ldots \S_kx_k\mathrm{:}X_k$
of a given data independent process syntax, each $X_i$ is either  a
type not related to $t$ or precisely the type parameter $t$, unless $\S_i =
\mathord !$, in which case $X_i = \mathit{null}$.
\end{remark}


\subsection{The \textbf{Seq} condition}

In order to produce our Semi-Symbolic Operational Semantics, it is useful to
restrict the scope of processes considered.  


\begin{defi}\label{def:seq}
A process syntax $Proc(t)$ satisfies \textbf{Seq} if
\begin{enumerate}[(i)]
\item
it is data independent;

\item it is sequential and contains no renaming or hiding;

\item it contains no replicated external or nondeterministic choice
  (but we do allow nondeterministic selections through the use of the $\$$
  symbol);


\item all guards of conditional choices within $Proc(t)$ contain either
  only variables of type~$t$, or only variables and values of types other
  that $t$; 

\item in external and sliding choices, $Proc(t)$
  contains no name clashes between type $t$ nondeterministic-selection
  variables of one argument and free variables of another argument;
  e.g.~$c\$x\mathrm{:}t \then \STOP \extchoice d.x \then \STOP$ is not
  allowed;

\item constructs of $Proc(t)$ do not contain multiple occurrences of
  the same input variable of type $t$; e.g.~$c!x!x$, and $c?y\mathrm{:}Y!y$ for
  $Y$ not related to $t$ are allowed, but $c?x\mathrm{:}t!x$ is not. 
\end{enumerate}\smallskip
\end{defi}

\noindent \Seq\ may be seen as a rather strong condition.  However, in practice, almost
all useful specification processes can be easily re-written to meet its
requirements; we justify this below.  However, this condition does place
restrictions on the way the specifications are \emph{expressed}.  These
restrictions will make the production of the semi-symbolic operational
semantics easier, and also simplify subsequent proofs. 

We assume sequentiality (assumption~(ii)).  When a process is not sequential,
it can be rewritten into a sequential form using algebraic
equivalences~\cite{Roscoe:1997}.  Further, we forbid indexed choice operators,
since (for finite choices) such indexed operators can always be replaced by
binary ones.  Note that this means that \textbf{Seq} processes are taken from
processes with the following syntax:
\[ P ::=
\STOP | \alpha \then P | P \extchoice P | P \intchoice P | P \timeout P | 
\If b \Then P \Else P | X.
\]


Assumptions (iv)--(vi) have been introduced for technical reasons, to simplify
the production of the semi-symbolic operational semantics.  With the exception
of assumption~(vi), they do not reduce expressiveness.

Assumption~(iv) simplifies our treatment of conditionals when working with
symbolic representations of processes (see
Section~\ref{section:ssos}). Observe that the guard of every conditional can
be expressed using predicates that involve only types other than the
distinguished one, and predicates that involve only the distinguished type,
combined together using conjunction and disjunction.  The conjunctions and
disjunctions can be eliminated using the laws:
\begin{eqnarray*}
\If P \lor P' \Then Q \Else R & \equiv &
 \If P \Then Q \Else (\If P' \Then Q \Else R),
\\
\If P \land P' \Then Q \Else R & \equiv &
  \If P \Then (\If  P' \Then Q \Else R) \Else R.
\end{eqnarray*}
Hence any process can be rewritten to satisfy assumption~(iv).

We have introduced assumption~(v) as we will later store assignments of
values to variables explicitly; clashes of variables names could introduce
undesirable updates of values in such assignments. For example, consider the
syntax
\[
in_1\$x\mathrm{:}t \then (out.x \then \STOP \extchoice in_2\$x\mathrm{:}t \then \STOP).
\]
Then, the value of $x$ that is output using construct $out.x$ should be the
value that is assigned to variable $x$ at the time the nondeterministic
selection on channel $in_1$ is resolved. However, unless the output variable
$x$ is immediately substituted with the correct value, the nondeterministic
selection on channel $in_2$ can be resolved before the output is performed,
leading to the value of $x$ being overwritten. Using alpha-conversion, every
process definition that fails assumption~(v) can be easily rewritten into a
form that satisfies it.

Assumption~(vi) ensures that values of all outputs of type $t$ have to be
previously stored within a process's memory. This simplifies the semantics,
and does not greatly reduce expressiveness.

Thus, most  processes can be rewritten into a form that satisfies
\textbf{Seq}. 



\subsection{The \Norm{} condition}\label{section:norm}

When working with specification processes, it is desirable to ensure their
clarity and conformance to a certain standard (normality) to make analyses of
their behaviours easier. The \Norm{} condition, defined below, achieves this
without a major expressiveness reduction.  Its effect is to remove all
nondeterminism whose effect is not immediately observable.  In particular this
condition will allow us to deduce that a process reaches a unique state after
a particular trace (Proposition~\ref{prop:environmentuniqueness}), and that a
unique construct gives rise to each event following a given trace
(Proposition~\ref{corollaryA2}).

Given a sequential, data independent process syntax $P$, we define
$Channels(P)$ to be the set of the channel names of the initial constructs of
$P$. Formally, 
\begin{eqnarray*}
Channels(\STOP) & \defs & \set{},\\
Channels(\construct \then P) & \defs & \set{c},\\
Channels(P \extchoice Q) & \defs & Channels(P) \union Channels(Q),\\
Channels(P \intchoice Q) & \defs & Channels(P) \union Channels(Q),\\
Channels(P \timeout Q) & \defs & Channels(P) \union Channels(Q),\\
Channels(X) & \defs & Channels(P), \qquad \If E(X) = P  .
\end{eqnarray*}

\begin{defi}\label{def:seqnorm}
A process syntax $Proc(t)$ satisfies \Norm{} if it satisfies~\textbf{Seq}, and
in addition for all external choices $P(t) \extchoice Q(t)$, internal choices
$P(t) \intchoice Q(t)$ and sliding choices $P(t) \timeout Q(t)$ within
$Proc(t)$ we have that
\begin{iteMize}{$\bullet$}
\item $Channels(P(t)) \cap Channels(Q(t)) = \nullset$,
\item every conditional choice on $t$ in $P(t)$ and $Q(t)$ is after a
         prefix.
\end{iteMize}
\end{defi}

Our definition of \Norm{} is similar to definitions of \textbf{Norm} used in
the CSP literature \cite{Roscoe:1997,Lazic:1999}, except that it includes
\Seq, since we will always use \Norm\ with processes that satisfy \Seq.

The first clause does restrict expressiveness.  It bans processes such as $c!x
\then P \intchoice c!y \then Q$.  This is necessary to ensure that a unique
construct gives rise to each event (after a given trace), and that a process
reaches a unique state after a particular trace; for example, without this
condition, the above process could perform the event $c.0$ resulting from
either construct (assuming $x$ and~$y$ have value~0), and could reach either
state~$P$ or~$Q$ after this event.

If a particular process syntax fails \Norm{} because of the second subclause
of clause~(iv), then the following algebraic laws  can be used
to convert it to an equivalent process definition, satisfying this subclause: 
\begin{eqnarray*}
P \Join (\If\,b\,\Then Q \Else R)
  & \equiv &  \If\,b\,\Then (P \Join Q) \Else (P \Join R), \\
(\If\,b\,\Then Q \Else R) \Join P
  & \equiv  & \If\,b\,\Then (Q \Join P) \Else (R \Join P),
\end{eqnarray*}
where $\Join$ is one of $\extchoice, \intchoice$ or $\timeout$.

Thus, most processes can be rewritten into a form that satisfies
\textbf{SeqNorm}.  (A similar observation about the related \textbf{Norm}
condition is made in \cite[Section 15.2]{Roscoe:1997}.)  Indeed, we are not
aware of any specification used in practice that cannot.


\subsection{Type symmetry}
\label{sec:TypeSym}

In Section~\ref{section:dataindep} we defined the concept of data independence
which, undoubtedly, is a very useful property for studying parameterised
systems
\cite{Creese:1998,Roscoe:1999,Creese:1999a,Lowe:2004,Roscoe:2004}. However, in
practice it turns out to be too strong for the implementations we consider,
since we study parallel compositions of node processes indexed over the
parameter. Such compositions are banned by data independence. This is why we
define a weaker condition, which only requires all behaviours of a given
process to be \emph{symmetric} in the parameter.  A process syntax satisfies
the \textbf{TypeSym} condition if the behaviours of all its concretisations
are invariant under permutations of values of parameter instantiations.  Given
such a permutation~$\pi$, we write $\rename{\pi}$ for the renaming
$\rename{^{\pi(e)} / _e | e \in \Sigma}$.
\begin{defi}\label{def:typesym}\newnot{TypeSym}
A process syntax $Proc(t)$ satisfies the condition \textbf{TypeSym} if
$Proc(T)$ and $Proc(T) \rename{\pi}$ are bisimilar for every~$T$ and every
bijection $\pi: T \rightarrow T$.
\end{defi}

\begin{example}\label{example:typeSym-copy}
Consider the process
\begin{eqnarray*}
COPY(t) & = & in?x\mathord{:}t \then out!x \then COPY(t).
\end{eqnarray*}
This satisfies \textbf{TypeSym}, informally because it treats all elements
of~$t$ the same.  More formally, given an instantiation~$T$ of~$t$, and a
bijection~$\pi : T \fun T$, the relevant bisimulation is
\[
\begin{align}
\set{(COPY(t), COPY(t)\rename{\pi})} \union\null \\
\set{ (out!v \then COPY(t),\, (out!\pi^{-1}(v) \then COPY(t))\rename{\pi}) | 
       v \in T}.
\end{align}
\]
\end{example}

\begin{example}\label{typeSym-ring}
Consider a system of $N$ (where $N = \#T$) nodes that communicate using a ring
topology, where each node~$i$ can send messages only to the node~$(i+1) \bmod
N$.  For example (rather trivially):
\begin{eqnarray*}
\mathcal{N}_i(t) & = & 
  send!i!i\oplus1 \then \mathcal{N}_i(t) 
  \extchoice send!i\ominus1!i  \then \mathcal{N}_i(t),
\\
Nodes(t) & = &  
  \Parallel i \in t \spot [\set{send.i.i\oplus1, send.i\ominus1.i})]\ 
     \mathcal{N}_{i}(t),
\end{eqnarray*}
where $\oplus$ and $\ominus$ represent addition and subtraction mod $N$.  This
does \emph{not} satisfy \textbf{TypeSym}, which insists that the process is
\emph{fully} symmetric.  For example, if $T = \set{0 \upto 3}$ then $Nodes(T)$
has trace $\trace{send.1.2}$, but does not have the trace $\trace{send.1.3}$,
so \textbf{TypeSym} does not hold for $\pi = \set{0 \mapsto 2, 1 \mapsto 1, 2
  \mapsto 3, 3 \mapsto 0}$.
\end{example}

Semantic definitions, like Definition~\ref{def:typesym}, tend to be hard to
check efficiently, so we note here sufficient syntactic conditions for
\textbf{TypeSym}.
\begin{prop}\label{typesymproposition}
A process syntax $Proc(t)$ satisfies the
condition\ \mbox{\textbf{TypeSym}}\ if it uses no 
\begin{enumerate}[\em(i)]
\item constants of type $t$;

\item operations on type $t$, including polymorphic operations
  (e.g.~tupling or lists); 

\item functions whose domains or co-domains involve type $t$;

\item selections or indexing from sets involving $t$, unless the
  selection or indexing is over the whole of $t$, except this restriction does
  not apply to the alphabets of nodes in a parallel composition indexed
  over~$t$; and  

\item conditional choices on $t$, except for equality and inequality
  tests. 
\end{enumerate}
\end{prop}

%
\noindent
Note that the process of Example~\ref{example:typeSym-copy} satisfies the
conditions of this proposition, but the process of Example~\ref{typeSym-ring}
does not, because arithmetic operations are applied to type~$t$.
\proof[Proof sketch.] 
Let $CSP_t$ be the set of CSP syntaxes parameterised by~$t$, all of whose free
variables (other that $t$ itself) are of type $t$ and which satisfy the
conditions of the proposition. We use $[\Gamma]$ to denote the syntactic
substitution $[ \Gamma(x)/x | x \in \dom(\Gamma)]$ and $FV(P(t))$ to denote
the free variables of~$P(t)$. Then
\begin{eqnarray*}
\mathcal{B} & = &
   \left\{ 
     \left( P(T)[ \Gamma ], 
          (P(T)[ \pi^{-1}(\Gamma) ])\rename{\pi} \right) | 
      P(t) \in CSP_t,\, \Gamma \in FV(P(t)) \fun T \right\}
\end{eqnarray*}
is the required strong bisimulation relation.
The proof is  a structural induction on $P(t)$. 
\qed
\noindent
In practice, most systems where the nodes communicate using a fully connected
topology satisfy these conditions. 
In~\cite{moffat:2010}, Moffat proves a very similar result, for a larger
fragment of the machine-readable CSP language, including the underlying
functional language.

The syntactic definition of data independence
(Definition~\ref{def:dataindependence}) comprises a superset of the
requirements of Proposition~\ref{typesymproposition}, so we immediately have
the following result.

\begin{cor}\label{corollary:DIimpliesTypeSym}
Every data independent process satisfies \textbf{TypeSym}. 
\end{cor}



\begin{example}\label{example:system-typesym}
Consider a system built as the parallel composition of node
processes~$\mathcal{N}_{i}(t)$ for each~$i \in t$:
\begin{eqnarray*}
Nodes(t) & = & \mathop{\Parallel} i \in t \spot [A(i,t)]\ \mathcal{N}_{i}(t) .
\end{eqnarray*}
This process syntax satisfies \textbf{TypeSym} provided:
\begin{iteMize}{$\bullet$}
\item
the node process $\mathcal{N}_{i}(t)$ satisfies the conditions of
Proposition~\ref{typesymproposition}, so in particular it treats its
``identity'' parameter~$i$ polymorphically; informally, different nodes need
to be identical up to renaming of the identities;

\item
the alphabet $A(i,t)$ satisfies the conditions of
Proposition~\ref{typesymproposition}, so in particular no operations on
type~$t$ are applied;
informally, the different alphabets depend only on the identities~$i$,
polymorphically.
\end{iteMize}
Note, though, that $Nodes(t)$ does not satisfy data independence, since it
contains a replicated operator (parallel composition) that is indexed
over~$t$. 

Further, if we define the context $C_t[.]$ that composes its argument with a
controller process $Ctrl_t$ and hides some events:
\begin{eqnarray*}
C_t[X] & = & (X \parallel[A_t] Ctrl_t) \hide B_t
\end{eqnarray*}
then $C_t[ Nodes(t) ]$ satisfies \textbf{TypeSym} provided:
\begin{iteMize}{$\bullet$}
\item
the controller process $Ctrl_t$ satisfies the conditions of
Proposition~\ref{typesymproposition}; informally, it needs to treat different
nodes in the same way;

\item
the sets $A_t$ and~$B_t$ satisfy the conditions of
Proposition~\ref{typesymproposition}.
\end{iteMize}
Recall that this is the type of implementation process that we considered in
the Introduction.  In particular, the example from Section~\ref{sec:example}
meets this pattern. 
\end{example}

\begin{example}
The process 
\(
\Extchoice y\mathrm{:}t \spot c?x\mathrm{:}(t\setminus\set{y})!y 
  \then \STOP
\)
satisfies \textbf{TypeSym}.  However, it does not satisfy the conditions of
Proposition~\ref{typesymproposition}, in particular because $x$ is selected
from a proper subset of~$t$.
\end{example}

The following remark is a direct consequence of the \textbf{TypeSym}
condition.
\begin{remark}\label{symtraceremark}\label{symfailureremark}
Suppose that $Proc(t)$ satisfies \textbf{TypeSym}. Then, for all $T$:
\begin{iteMize}{$\bullet$}
\item
If $tr \in \tsem{Proc(T)}$ then for all bijections $\pi: T
\rightarrow T$,\, $\pi(tr) \in  \tsem{Proc(T)}$;

\item
If $(tr,X) \in \fsem{Proc(T)}$ then for all bijections $\pi: T
\rightarrow T$,\, $(\pi(tr), \pi(X)) \in  \fsem{Proc(T)}$.
\end{iteMize}
\end{remark}


\subsection{Equality tests}
\label{sec:equalityTests}

The syntactic condition \textbf{PosConjEqT}, formulated by Lazi\'{c}
in~\cite[Chapter 3]{Lazic:1998}, specifies that for a conditional choice with
an equality test on~$t$, the positive branch is a prefix and the negative
branch is simply $\STOP$. In~\cite{Roscoe:1998,Roscoe:1999}, a weaker version
of \textbf{Pos\-Conj\-EqT} is discussed, where no restriction on the process
in the positive branch is in place. Both of these definitions talk about the
condition only in relation to the traces model, but it is easy to extend it to
other models of CSP as the following definition shows.  

\begin{defi}\label{posconjeqt}
Given a CSP model $\mathcal{M}$ we say that a process syntax $Proc(t)$ satisfies \textbf{PosConjEqT$_\mathcal{M}$}\newnot{PosConjEqT} if for every conditional choice on $t$ of the form
\[\If cond \Then P(x_1,\ldots,x_k) \Else Q(x_1,\ldots,x_k)\]
within $Proc(t)$, we have that 
\begin{iteMize}{$\bullet$}
	\item $cond$ is a positive conjunction of equality tests on~$t$ (which gives rise to the name of the condition), and
 	\item $P(v_1,\ldots,v_k) \refinedby_\mathcal{M} Q(v_1,\ldots,v_k)$ for all values $v_1, \ldots, v_k$.
\end{iteMize}
\end{defi}

\noindent For technical reasons, it will be desirable in our work to assume
the opposite condition for specifications: that every positive branch of a
conditional choice is a refinement of the negative branch. This can be viewed
as a reversed version of \textbf{PosConjEqT$_\mathcal{M}$}. Hence, we have the
following definition.

\begin{defi}\label{revposconjeqt}\newnot{RevPosConjEqT}
Given a CSP model $\mathcal{M}$ we say that a process syntax $Proc(t)$
satisfies \textbf{RevPosConjEqT$_\mathcal{M}$} if for every conditional choice
on $t$ of the form 
\[\If cond \Then P(x_1,\ldots,x_k) \Else Q(x_1,\ldots,x_k)\]
within $Proc(t)$, we have that 
\begin{iteMize}{$\bullet$}
	\item $cond$ is a positive conjunction of equality tests on~$t$, and
 	\item $Q(v_1,\ldots,v_k) \refinedby_\mathcal{M} P(v_1,\ldots,v_k)$ for all values $v_1, \ldots v_k$.
\end{iteMize}
\end{defi}
Whenever $\mathcal{M}$ is clear from the context, we will simply
write \textbf{Rev\-Pos\-Conj\-EqT}. 

\begin{example}
The process syntax
\[
Proc(t) =
\align
	in?x\mathrm{:}t?y\mathrm{:}t?z\mathrm{:}t \then \If x = y\ 
	\align
		\Then out.x  \then out.y \then \STOP\\
		\Else out\$z \then (out.y \then \STOP \intchoice \STOP)\\
	\endalign
\endalign
\]
satisfies \textbf{RevPosConjEqT$_\mathrm{F}$}. However, the process syntax
\[
Proc(t) = in?x\mathrm{:}t?y\mathrm{:}t \then \If x = y\ 
\align
	\Then out.x \then \STOP\\
	\Else out.y \then \STOP\\
\endalign
\]
does not satisfy \textbf{RevPosConjEqT$_\mathrm{T}$}, because if $x$ and $y$
are two distinct values, then \mbox{$out.y \then \STOP \not \trefinedby out.x
  \then \STOP$}. 
\end{example}

Most specification processes that one tends to use in practice do not contain
conditionals, so vacuously satisfy both \textbf{PosConjEqT$_\mathcal{M}$} and
\textbf{RevPosConjEqT$_\mathcal{M}$} for all models~$\mathcal{M}$.  Further,
our experience is that many specifications that do contain conditionals
satisfy \textbf{RevPosConjEqT$_\mathcal{M}$}.

\section{Operational semantics}\label{section:operationalsemantics}

The main usefulness of a process algebra (like CSP) comes from the fact that
it allows us to reason about programs and processes rigorously. In this
section we look into the operational semantics for CSP\@.  An
operational semantics provides a precise step-by-step description of how
processes execute. It describes state changes as effects of events being
performed by representing processes using \emph{labelled transition systems},
defined as follows.

\begin{defi}
A \emph{labelled transition system (LTS)} is a tuple $\mathcal{L} = (S, s_0,
L,\trans)$, where $S$ is a set of states, $s_0 \in S$ is an initial state, $L$
is a set of labels, and $\mathord{\trans} \subseteq S \times L \times S$ is a
transition relation. We let $\hat{\mathcal{L}} = S$ denote the set of states
of $\mathcal{L}$.
\end{defi}

In Section~\ref{section:notation} we give some useful notation and definitions
that we will repeatedly use throughout the rest of this paper.

The various operational semantics we present in this paper do not aim to be
complete. Their main purpose is to formalise the foundations for the results
regarding specifications, presented in
Section~\ref{section:typereduction}.  This is why we describe only the minimal
operational semantics that allow us to generate transition graphs of the
processes that we consider in that section.  We therefore restrict ourselves
to processes that satisfy \textbf{Seq} throughout this section. 
As noted above, most CSP specifications one uses in practice lie within this
fragment of CSP, and others can be rewritten into this form using algebraic
laws.  We stress, though, that implementation processes can be written using
the full syntax of CSP. 

Operational semantics can be defined at different levels of abstraction. In
Section~\ref{section:os} we present a fairly standard operational semantics at
the lowest, implementation level. It generates LTSs from process syntax with
no free variables. This means that all parameters  must be substituted with
concrete values before the transition rules can be used.  When variables
become bound as the result of inputs or nondeterministic selections, the
binding is reflected by syntactic substitution.  

We introduce a running example, which we use to illustrate the different
styles of operational semantics.
\begin{example}\label{example:os}
Let
\begin{eqnarray*}
P(t) & = & 
  c\$x\mathord:\set{a,b}\$y\from t?z\from t \then 
  \If y=z \Then d!x \then STOP \Else STOP.
\end{eqnarray*}
In Figure~\ref{fig:os} we represent the standard operational semantics for
$P(T)$ where $T = \set{0,1}$.  We omit part of the semantics because of lack
of space.  In the figure, we write $Q_{x,y,z}$ as a shorthand for $\If y=z
\Then d!x \then STOP \Else STOP$.  For compatibility with the later semantics,
we choose to resolve all non-type-$t$ nondeterministic selections before the
type-$t$ nondeterministic selections: hence the $\tau$ transitions from the
initial states correspond to resolving the ``$\$x\mathord:\set{a,b}$''
selection, and the $\tau$ transitions from the subsequent states correspond to
resolving the~''$\$y\from t$'' selection.  The transitions labelled with
events on channel~$c$ also have the effect of resolving the subsequent
conditional (``$\If y=z \ldots$'').
\end{example}
\begin{figure}[htbp]
\begin{center}\small
\xymatrix @R=6.5mm @C=1mm{ 
& & & & P(T) \ar[dll]_\tau \ar[drr]^\tau
\\
& & {\align c!a\$y\from T?z\from T \\ \then Q_{a,y,z} \endalign} \ar[dl]_\tau \ar[dr]^\tau  
 &&&& {\align c!b\$y\from T?z\from T \\ \then Q_{b,y,z} \endalign}\ar[dl]_\tau \ar[dr]^\tau
\\
& {\align c!a!0?z\from T \\ \then Q_{a,0,z} \endalign} 
  \ar[dl]_{c.a.0.0} \ar[dr]^{c.a.0.1} & 
  & {\align c!a!1?z\from T \\ \then Q_{a,1,z} \endalign}
    \ar[dl]_{c.a.1.0} \ar[dr]^{c.a.1.1} 
  && {\align \ \ \ \ \ \ \ \ \ldots \\ \mbox{ } \endalign} 
  && {\align \ldots\  \\ \mbox{ } \endalign}
\\
d!a \then STOP \ar[rr]^{d.a} && STOP & & d!a \then STOP  \ar[ll]_{d.a}
}
\end{center}
\caption{Operational semantics for $P(T)$ from
  Example~\ref{example:os} with $T = \set{0,1}$.\label{fig:os}} 
\end{figure}

One of the main shortcomings of such an operational semantics, when working
with parameterised systems, is the need for repetitive application of the
transition rules for each instantiation of the parameters: this is reflected
in Figure~\ref{fig:os}, where the number of transitions depends on
the size of~$T$.  Lazi\'{c} addressed this problem in~\cite{Lazic:1999} by
defining a symbolic operational semantics (for a language similar to CSP, but
with an addition of certain lambda calculus terms), where the variables
related to the parameters are never instantiated, but rather left as symbols,
when an LTS is generated. The advantage of such an approach is that, given a
parameterised process syntax, a single symbolic LTS is generated and each of
the concrete LTSs can be easily obtained from it by an assignment of
values\footnote{In Lazi\'{c}'s work individual concrete LTSs are, in fact,
  never generated. Instead, the relationships with denotational semantics are
  established and the denotational values are derived directly from symbolic
  LTSs.}. Such a symbolic LTS can be viewed as a formal structure that
captures the essence of the behaviour of a process; it hides the details of
the data values, concentrating on the control states between which a process
can move by executing actions. This sort of symbolic structure is precisely
what we need for our work in Section~\ref{section:typereduction}. However, the
assumptions we make about the processes with which we work cause the
application of Lazi\'{c}'s work to be unnecessarily complex for our
needs. 

In Section~\ref{section:ssos} we define \emph{Semi-Symbolic Operational
  Semantics} (SSOS), a symbolic operational semantics similar to the one
from~\cite{Lazic:1999}.  
We explain the idea of SSOS via our running example. 
\begin{example}\label{example:ssos-intro}
Recall the following process from Example~\ref{example:os}:
\begin{eqnarray*}
P(t) & =  &
  c\$x\mathord:\set{a,b}\$y\from t?z\from t \then
  \If y=z \Then d!x \then STOP \Else STOP.
\end{eqnarray*}
In Figure~\ref{fig:ssosex} we represent the semi-symbolic operational semantics
for $P(t)$.  We again write $Q_{x,y,z}$ as a shorthand for $\If y=z
\Then d!x \then STOP \Else STOP$.  Note that transitions are \emph{symbolic}
in that they contain variables corresponding to type-$t$ selections; however,
non-type-$t$ values are treated concretely.  Further, we include transitions
corresponding to the conditional, labelled with the condition (``$y=z$'') and
its negation (``$\lnot y=z$'') respectively. 
\end{example}

\begin{figure}[tp]
\begin{center}
\xymatrix @R=6mm @C=1mm{ 
& & & P(t) \ar[dll]_\tau \ar[drr]^\tau
\\
& {\align c!a\$y\from t?z\from t \\ \then Q_{a,y,z} \endalign} 
  \ar[d]_{c!a\$y\from t?z\from t} 
 &&&& {\align c!b\$y\from t?z\from t \\ \then Q_{b,y,z} \endalign}
  \ar[d]_{c!b\$y\from t?z\from t}
\\
&  Q_{a,y,z} \ar[dl]_{y=z} \ar[dr]^{\lnot y=z}  
&&&& Q_{b,y,z} \ar[dl]_{y=z} \ar[dr]^{\lnot y=z}
\\
d!a \then STOP \ar[rr]_{d.a} && STOP\hspace*{11mm} 
&& d!b \then STOP \ar[rr]_{d.b} && STOP
}
\end{center}
\caption{Semi-symbolic operational semantics for $P(t)$ from
  Example~\ref{example:ssos-intro}.\label{fig:ssosex}} 
\end{figure}

The states of the resulting \emph{semi-symbolic LTSs} (SSLTSs) can be viewed
as the control states of families of concrete processes. In order to fully
concretise them, it is enough to provide a map of variable names to concrete
values; such a map will be called an \emph{environment}. In
Section~\ref{section:cose} we describe \emph{Concrete Operational Semantics
  with Environments} (COSE), a concrete operational semantics which, for a
fixed instantiation of the distinguished type, creates LTSs whose states are
triples consisting of a symbolic state (or a modification of such), an
environment giving values to the type-$t$ free variables, and the
instantiation of the distinguished type. The specification of COSE is provided
as a set of translation rules from SSOS, rather than a set of transition
rules.  We illustrate the idea via our running example.
\begin{example}
Recall the process $P(t)$ from above.  Figure~\ref{fig:cose} gives the COSE
semantics for $P(T)$ with $T = \set{0,1}$ (strictly speaking, each state
should also include~$T$ as a third term; we omit this due to lack of space).
The $\tau$ transitions from the initial states correspond to resolving the
``$\$x\mathord:\set{a,b}$'' selection; since $x$ is not of type~$t$, the
choice of~$x$ is reflected by syntactic substitution.  The $\tau$ transitions
from the subsequent states correspond to resolving the~``$\$y\from t$''
selection; the environment stores the resulting value for~$y$.  The subsequent
transitions  with events on channel~$c$ resolve the ``$?z\from t$''
choices; the environment stores the resulting value for~$z$.  Note also that
the operational semantics is strongly bisimilar to the semantics in
Figure~\ref{fig:os}.
\end{example}

\begin{figure}[tp]
\begin{center}\small
\xymatrix @R=6.5mm @C=1mm{ 
& & & (P(t), \set{}) \ar[dl]_\tau \ar[dr]^\tau
\\
& & {\align (c!a\$y\from t?z\from t \\ \then Q_{a,y,z}, \set{}) \endalign} 
    \ar[dl]_\tau \ar[dr]^\tau  
 && {\align \ldots \\ \mbox{\hspace{20mm}} \endalign}
\\
& {\align (c!a!y?z\from t  \then Q_{a,y,z},\\ \qquad\set{y \mapsto 0})\endalign} 
  \ar[dl]_{c.a.0.0} \ar[d]^{c.a.0.1} & 
 & {\align (c!a!y?z\from t  \then Q_{a,y,z}, \\ \qquad\set{y \mapsto 1})\endalign}
    \ar[d]_{c.a.1.0} \ar[dr]^{c.a.1.1} 
\\
{\align (Q_{a,y,z}, \\ \set{y \mapsto 0, z \mapsto 0})\endalign} \ar@/_3ex/[drr]_{d.a} 
  & {\align  (Q_{a,y,z}, \\  \set{y \mapsto 0, z \mapsto 1})\endalign}  &
  & {\align (Q_{a,y,z}, \\ \set{y \mapsto 1, z \mapsto 0})\endalign}  
 & {\align (Q_{a,y,z}, \\ \set{y \mapsto 1, z \mapsto 1})\endalign} \ar@/^3ex/[dll]^{d.a}
\\
&& (STOP,\set{}) & &
}
\end{center}
\caption{Concrete operational semantics with environments for $P(T)$  with $T
  = \set{0,1}$.\label{fig:cose}}  
\end{figure}

We show that the combination of SSOS and the translation rules of COSE is
always bisimilar to the standard one in
Section~\ref{section:congruence}. Finally, we define the relationship between
symbolic traces and concrete traces in Section~\ref{section:generates}.

\subsection{General definitions and notation}\label{section:notation}

We present some notation
and definitions that will often be used in the following sections. Additional
pieces of notation and local definitions will be introduced in the relevant
parts of this section.

We define $\Value$ to be the set of all values, and $\Var$ to be the
set of all variable names; we assume $\Var \inter \Value = \nullset$.

Prefix constructs may depend on the distinguished type, so, in theory, they
should be decorated with parameter $t$, e.g.~$Proc(t) = \alpha(t) \then
Proc'(t)$. However, for brevity, we omit the parameter (or its instantiation)
where it is clear from the context or indifferent.

For any construct $\alpha$ of the form $\construct$, we define functions that
return index sets of variables and values within $\alpha$, based on their type
and the kind of input or output they model:
\begin{eqnarray*}
\$^t(\alpha) & \defs & \set{i \in \set{1\upto k} | \S_i = \mathord \$ \land X_i =
  t},\vspace{0.5ex}
\\
\$^\nont(\alpha) & \defs & \set{i \in \set{1\upto k} | \S_i = \mathord \$ \land
  X_i \not = t},\vspace{0.5ex}
\\
\$(\alpha) & \defs & \mathord \$^t(\alpha) \union \mathord \$^\nont(\alpha),
\\[1ex]
?^t(\alpha) & \defs & \set{i \in \set{1\upto k} | \S_i = \mathord ? \land X_i =
  t},\vspace{0.5ex}
\\
?^\nont(\alpha) & \defs & \set{i \in \set{1\upto k} | \S_i = \mathord ? \land X_i
  \not = t},\vspace{0.5ex}
\\
?(\alpha) & \defs & \mathord ?^t(\alpha) \union \mathord ?^\nont(\alpha),
\\[1ex]
!^t(\alpha) & \defs & \set{i \in \set{1 \upto k} | \S_i = \mathord ! \land x_i \mbox{
    is of type $t$}},\vspace{0.5ex}
\\
!^\nont(\alpha) & \defs & \set{i \in \set{1 \upto k} | \S_i = \mathord ! \land x_i
  \mbox{ is not of type $t$}},\vspace{0.5ex}
\\
!(\alpha) & \defs & \mathord !^t(\alpha) \union \mathord !^\nont(\alpha).
\end{eqnarray*}

The following functions allow us to modify constructs. Let $\alpha =
\construct$ and let $\dagger$ be either $t$ or \textit{non-t}.  
\begin{iteMize}{$\bullet$}
 \item We define $\mathit{Replace}_{\$\mapsto!}^\dagger(\alpha)$ to be a
   construct like $\alpha$, but where for every $i$ in $\$^\dagger(\alpha)$
   the $\S_i$ symbol (which must be a $\$$) is replaced by a $!$ and $X_i$ is
   replaced by $\mathit{null}$;
 
 \item $\mathit{Replace}_{\$\mapsto!} \defs \mathit{Replace}_{\$\mapsto!}^t \circ \mathit{Replace}_{\$\mapsto!}^\nont$.
\end{iteMize}

\begin{example}
Let $\epsilon = c\$x_1\mathrm{:}t?x_2\mathrm{:}t\$x_3\mathrm{:}X!x_4$, where
$X$ it a type not related to $t$ and $x_4$ is some output variable. Then,
\begin{eqnarray*}
\mathit{Replace}_{\$\mapsto!}^t(\epsilon) & = &
  c!x_1?x_2\mathrm{:}t\$x_3\mathrm{:}X!x_4,
\\
\mathit{Replace}_{\$\mapsto!}^\nont(\epsilon) & = &
        c\$x_1\mathrm{:}t?x_2\mathrm{:}t!x_3!x_4, 
\\
\mathit{Replace}_{\$\mapsto!}(\epsilon) &  = &
        c!x_1?x_2\mathrm{:}t!x_3!x_4.
\end{eqnarray*}
\end{example}

Substitution will play an important role in defining the operational semantics
in the following sections. We use square brackets to denote substitution: for
a variable $x$ and a value $v$, $P[v/x]$ is like $P$, but with every free
occurrence of $x$ replaced with $v$ (here, $P$ can be a process, a definition
of a set, a definition of a relation, etc). Substitution is different from
renaming, since renaming is a function or relation from values to values,
while substitution is a function from variables to values.

\subsection{Standard CSP operational semantics}\label{section:os}


In this section we present a standard operational semantics for the fragment
of CSP corresponding to \Seq, i.e.~excluding parallel operators, hiding and
renaming.  (Rules for the remainder of the syntax can be found in,
e.g.,~\cite{Roscoe:1997}, but we do not need them in this paper.)  The
operational semantics generates LTSs from syntax without free variables. This
means that, when dealing with parameterised processes, all parameters have to
be assigned concrete values before the transitions rules can be applied.  We
let $T$ be a fixed instantiation of type $t$. 

We distinguish two types of transitions: \emph{visible} and
\emph{internal}. An internal transition, labelled with~$\tau$, represents an
event that can be performed by a process without any interaction from the
environment and which is not observable by the environment. A visible
transition, on the other hand, is labelled with an event that is observable by
the environment, and requires its synchronisation in order to be performed. We
write $P \trans[a] Q$ to mean that there is an $a$-labelled transition from
state $P$ to state~$Q$.




\subsubsection{Transition rules}\label{section:standardoperationalsemanticstransitionrules}

Most of the transition rules are given in Figure~\ref{fig:opsem}, and are
standard. 
We concentrate our discussion on the semantics for prefixing; a slightly
non-standard treatment is required due to the addition of nondeterministic
selections.


\begin{figure}[htbp]
\begin{center}


\noindent\begin{minipage}[b]{0.5\linewidth}
\infrule
P(T) \trans[\tau] P'(T)
\derive
P(T) \extchoice Q(T) \trans[\tau] P'(T) \extchoice Q(T)
\endinfrule
\end{minipage}
\begin{minipage}[b]{0.5\linewidth}
\infrule
Q(T) \trans[\tau] Q'(T)
\derive
P(T) \extchoice Q(T) \trans[\tau] P(T) \extchoice Q'(T)
\endinfrule
\end{minipage}

\noindent\begin{minipage}[b]{0.5\linewidth}
\infrule
P(T) \trans[a] P'(T)
\bigderive[a \not= \tau]
P(T) \extchoice Q(T) \trans[a] P'(T)
\endinfrule
\end{minipage}
\begin{minipage}[b]{0.5\linewidth}
\infrule
Q(T) \trans[a] Q'(T)
\bigderive[a \not= \tau]
P(T) \extchoice Q(T) \trans[a] Q'(T)
\endinfrule
\end{minipage}






\noindent\begin{minipage}[b]{0.5\linewidth}
\infrule
\mbox{ }
\derive
P(T) \intchoice Q(T) \trans[\tau] P(T)
\endinfrule
\end{minipage}
\begin{minipage}[b]{0.5\linewidth}
\infrule
\mbox{ }
\derive
P(T) \intchoice Q(T) \trans[\tau] Q(T)
\endinfrule
\end{minipage}





\noindent\begin{minipage}[b]{0.5\linewidth}
\infrule
\mbox{ }
\derive
P(T) \timeout Q(T) \trans[\tau] Q(T)
\endinfrule
\end{minipage}
\begin{minipage}[b]{0.5\linewidth}
\infrule
P(T) \trans[\tau] P'(T)
\derive
P(T) \timeout Q(T) \trans[\tau] P'(T) \timeout Q(T)
\endinfrule
\end{minipage}
 
\begin{minipage}[b]{0.5\linewidth}
\infrule
P(T) \trans[a] P'(T)
\bigderive[a \not = \tau]
P(T) \timeout Q(T) \trans[a] P'(T)
\endinfrule
\end{minipage}
\begin{minipage}[b]{0.5\linewidth}
\infrule
E(X) = P
\derive
X(T) \trans[\tau] P(T)
\endinfrule
\end{minipage}



\noindent
\infrule
P(T) \trans[a] P'(T)
\derive
\If \mathit{True} \Then P(T) \Else Q(T) \trans[a] P'(T)
\endinfrule
%
\infrule
Q(T) \trans[a] Q'(T)
\derive
\If \mathit{False} \Then P(T) \Else Q(T) \trans[a] Q'(T)
\endinfrule
\end{center}
\caption{Operational semantic rules for the choice operators and
  binding\label{fig:opsem}}
\end{figure}

Let $\alpha$ be a construct of the form $c\S_1x_1\mathrm{:}X_1\ldots
\S_kx_k\mathrm{:}X_k$.  In order to define the prefix transition rules for the
language with nondeterministic selections added in, we proceed in two
steps. Firstly, we deal with constructs with no nondeterministic selections.
\begin{Rule}{Prefix Rule 1} (Prefixes with no nondeterministic
  selections)\label{rule:prefixrule1} 
\infrule
\mbox{ }
\bigderive[c.v_1\ldots v_k \in \mathit{Comms}(\alpha) \land \#\$(\alpha) = 0]
\alpha \then P(T) \trans(5)[c.v_1\ldots v_k] P(T)[v_i / x_i | i \in \mathord ?(\alpha)]
\endinfrule
where $Comms(\alpha)$ is the set of concrete events that $\alpha$ describes;
formally:  
\[\label{def:comms}
\align
	\mathit{Comms}(c\S_1x_1\mathrm{:}X_1\ldots\S_kx_k\mathrm{:}X_k) =\\
	\qquad \{ c.v_1\ldots v_k | 
		\forall i \in \set{1 \upto k} \spot (\S_i = \mathord ? \land v_i \in X_i) \lor (\S_i = \mathord ! \land v_i = x_i) \}.
\endalign
\]
\end{Rule}

The second step involves deriving transitions from prefix constructs with at
least one nondeterministic selection, producing invisible transitions that
resolve the choices, and substituting the chosen values for the variables of
the choices. For reasons that will become clear later, we simultaneously
resolve all nondeterministic selections over types other than~$t$ before
simultaneously resolving all nondeterministic selections over type~$t$. 

The following rule resolve all nondeterministic selections over types other
than~$t$, replacing each variable $x_i \in \$^\nont(\alpha)$ with an
appropriate value $v_i \in X_i$.  (Here and subsequently we treat subscripting
as functional application, i.e.~$v_i$ is the result of applying function~$v$
to index~$i$.) 
\begin{Rule}{Prefix Rule 2a} (Prefixes with nondeterministic selections over non-$t$ types)\label{rule:prefixrule2a}
\infrule
\dom(v) = \mathord \$^\nont(\alpha) \land \forall i \in \mathord \$^\nont(\alpha) \spot v_i \in X_i
\bigderive[\#\$^\nont(\alpha) > 0]
\alpha \then P(T) \trans[\tau] \left(\mathit{Replace}_{\$\mapsto!}^\nont(\alpha) \then P(T)\right)[v_i / x_i | i \in \$^\nont(\alpha)]
\endinfrule
\end{Rule}

The following rule then resolve all nondeterministic selections over~$t$,
replacing each variable $x_i \in \$^t(\alpha)$ with an appropriate value $v_i
\in T$.
\begin{Rule}{Prefix Rule 2b} (Prefixes with nondeterministic selections only over
type $t$)\label{rule:prefixrule2b} 
\infrule
v \in \mathord \$^t(\alpha) \rightarrow T
\bigderive[\#\$^\nont = 0\\ \land \#\$^t(\alpha) > 0]
\alpha \then P(T) \trans[\tau] \left(\mathit{Replace}_{\$\mapsto!}^t(\alpha) \then P(T)\right)[v_i / x_i | i \in \$^t(\alpha)]
\endinfrule
\end{Rule}

The above two rules are consistent with defining $\alpha \then P(T)$
as 
\[
	\Intchoice \langle x_i\mathrm{:}X_i | i \in \mathord \$^\nont(\alpha)\rangle \spot \left(\Intchoice \langle x_i\mathrm{:}T | i \in \mathord \$^t(\alpha)\rangle \spot \mathit{Replace}_{\$\mapsto!}(\alpha) \then P(T)\right),
\]
where 
we use
$\Intchoice \langle x_i\mathrm{:}X_i | i \in \mathord\mathcal{I}\rangle \spot
P(x_{i_1}, \ldots, x_{i_n})$ as shorthand  for $\Intchoice (x_{i_1},
\ldots, x_{i_n}) \in X_{i_1} \times \cdots \times X_{i_n} \spot\linebreak[2] P(x_{i_1},
\ldots, x_{i_n})$.

\begin{example}
Recall our earlier running example:
\begin{eqnarray*}
P(t) & = & 
  c\$x\mathord:\set{a,b}\$y\from t?z\from t \then 
  \If y=z \Then d!x \then STOP \Else STOP.
\end{eqnarray*}
Figure~\ref{fig:os} represents the operational semantics for $P(T)$ where
$T = \set{0,1}$. 
The first $\tau$ transitions correspond to Prefix Rule~2a; the second
$\tau$ transitions correspond to Prefix Rule~2b; the visible transitions
correspond to Prefix Rule~1.
\end{example}



\subsubsection{Calculating denotational
  values}\label{section:denotationalvaluesfromos}

It is possible to calculate denotational values of processes without resorting
to operational semantics. Such a direct way, using denotational semantics, is
discussed in~\cite[Chapter 8]{Roscoe:1997}. However, since we will often work
with LTSs, it makes sense to derive these values directly from transition
graphs. Firstly, we need three definitions (from~\cite[Chapter
  7]{Roscoe:1997}). 
\begin{iteMize}{$\bullet$}
\item
Given two states $P(T)$ and $Q(T)$, and a sequence of events (visible or
invisible) $s = \trace{a_i | i \in \set{1\upto n}}$ for some $n \geq 0$, we
write $P(T) \mapstotrans[s] Q(T)$ if there exist states $P_0(T) = P(T),
P_1(T), \ldots, P_n(T) = Q(T)$ such that for all $i$ in $\set{0 \upto n-1}$ we
have that $P_i(T) \trans(3)[a_{i+1}] P_{i+1}(T)$.

\item
We write $P(T) \Trans[tr] Q(T)$ if there is $s$ such that $P(T)
\stackrel{s}{\longmapsto} Q(T)$ and $tr$ is the restriction of $s$ to visible
events.  

\item
We say that $Q(T)$ refuses $X$, written $Q(T) \refuses X$, if $Q(T)$ cannot
perform~$\tau$ (i.e.~it's stable) and cannot perform any event from~$X$:
\(
\forall x \in X \union \set{\tau} \spot Q(T) \not\trans(3)[x] .
\)
\end{iteMize}

Using the above, we have
\begin{eqnarray*}
\traces(P(T)) & = & 
  \set{tr \in \Sigma^* | \exists Q(T) \spot P(T) \Trans[tr] Q(T)},
\\
\failures(P(T)) & = & 
  \set{(tr, X) \in \Sigma^* \times \Sigma | \exists Q(T) \spot 
		P(T) \Trans[tr] Q(T) \land Q(T) \refuses X}.
\end{eqnarray*}
\subsection{Semi-Symbolic Operational Semantics}\label{section:ssos}

Symbolic representation of models is often used in model checking (see
e.g.~\cite{McMillan:1992,Burch:1992}). In most cases the approach taken is to
create a single, compact structure that represents the behaviour of multiple
instances of a given system. The specification check is then performed on the
symbolic model in order to deduce verification results for all the
concretisations this model corresponds to. 

In this section we present a symbolic operational semantics for CSP\@. Its
aim, however, is not to be used to perform abstract refinement checking of
processes. Given a process syntax $Proc(t)$, it generates a single structure,
which acts as a bridge between the different processes obtained from $Proc(t)$
by substituting different concrete values for the parameter~$t$. This will
allow us, in Section~\ref{section:typereduction}, to use known behaviours of a
given instance of the process to deduce facts about the behaviours of other
instances of the same process definition. In our work we will apply this
operational semantics only to specification processes.

One of the main characteristics of the symbolic operational semantics defined
in this section is that only the parts of systems that involve type $t$ are
left in their symbolic form. All other components are instantiated in a way
similar to that used in the standard operational semantics
(Section~\ref{section:os}). Therefore, the labels of transitions may contain
some symbolic parts and some concrete parts. This is why we call any resulting
transition graph a \emph{semi-symbolic labelled transition system}
(\emph{SSLTS}), and call this operational semantics \emph{Semi-Symbolic
  Operational Semantics} (\emph{SSOS}).

Throughout this section we assume that all processes satisfy \Seq.


\subsubsection{Symbolic transitions}

In order to be able to tell symbolic and standard transitions apart, the
symbolic transition relation is denoted by $\trans[][s]$, i.e.\, 
$P(t) \trans[\alpha][s] Q(t)$ denotes that there is an $\alpha$-labelled
transition from symbolic state $P(t)$ to symbolic state $Q(t)$.

We distinguish the following three types of symbolic transitions.
\begin{desCription}
\item\noindent{\hskip-12 pt\bf Internal:}\  The internal symbolic transitions, labelled~$\tau$, are in a
  direct correspondence with the standard internal transitions.

\item\noindent{\hskip-12 pt\bf Visible:}\ Visible symbolic transitions are similar to standard visible
  transitions. The main difference is that while the labels of
  standard visible transitions contain no input symbols and no variables,
  labels of visible symbolic transitions may contain nondeterministic
  selections of type $t$ (e.g.~$\$x\mathrm{:}t$), deterministic inputs of type
  $t$ (e.g.~$?x\mathrm{:}t$), outputs of type $t$ (e.g.~$!x$, where $x$
  is a variable of type $t$), or outputs of non-$t$ parts (e.g.~$!v$ where $v$
  is a value not of type~$t$).

  Formally, each visible symbolic transition is labelled with a \emph{visible
    symbolic event}, a construct of the form
  $c\S_1x_1\mathrm{:}X_1\ldots\S_kx_k\mathrm{:}X_k$, where
  \begin{iteMize}{$\bullet$}
	\item $c$ is a channel name,

	\item $\S_i \in \{\$,?,!\}$ is an input/output symbol,

	\item $x_i$ is a variable of type $t$ or a value of type other than
          $t$; it can be a value only if it is immediately preceded by the
          output symbol~$!$,

	\item $X_i$ is $t$ if and only if the preceding input/output symbol,
          $\S_i$, is either $\$$ or $?$; otherwise it is $null$. 
  \end{iteMize}
  For example, the process $c!a?x\mathord:t\$y\mathord:t \then STOP$ has an
  initial symbolic transition with label $c!a?x\mathord:t\$y\mathord:t$.  We
  let $\Visible$\newnot{visible} denote the set of all visible symbolic
  events.

\item\noindent{\hskip-12 pt\bf Conditional:}\ Since variables of type $t$ are not instantiated within
  SSLTSs, but left in their symbolic form, boolean conditions that contain
  such variables cannot, in general, be evaluated to either $\mathit{True}$ or
  $\mathit{False}$ at the time of generating a symbolic transition
  graph. Hence, in order to deal with processes with such conditional choices
  involving variables of type~$t$, we introduce conditional symbolic
  transitions. Each such transition is labelled with a \emph{conditional
    symbolic event}, a boolean expression obtained from the guard of a
  conditional choice on~$t$ or its negation. For example, the syntax ``$\If x=y
  \Then P \Else Q$'' gives raise to the conditional symbolic events ``$x=y$''
  and ``$\lnot x=y$''.
  We let $Cond$ denote the set of conditional symbolic events.  Without loss
  of generality, we assume that the process syntax contains no trivial
  condition such as ``$x=x$''.
\end{desCription}

\begin{remark}\label{remark:noNontPartsInSymbolicVisibleEvents}
If $\epsilon$ is a visible symbolic event, then $\$^\nont(\epsilon) = \mathord ?^\nont(\epsilon) = \nullset$.
\end{remark}


We will usually use $\alpha$ and its derivatives ($\alpha', \alpha_1$,
etc.)\ to denote labels whose kind is unknown or indifferent and $\epsilon$
and its derivatives ($\epsilon', \epsilon_1$, etc.)\ to denote visible
symbolic events.

\subsubsection{Transitions rules}\label{section:SSOSfiringrules}

We define the Semi-Symbolic Operational Semantics using the inference
rules below. Recall that we are considering only processes that satisfy
\Seq; therefore, we only provide transition rules for operators that the
condition allows.




We begin with prefixing.  Let $\alpha$ be a construct of the form
$c\S_1x_1\mathrm{:}X_1\ldots\S_kx_k\mathrm{:}X_k$. There are two transition
rules for prefix. The first one defines the initial symbolic events of
\mbox{$\alpha \then P(t)$} in the case when $\alpha$ contains no
nondeterministic selections over types other than~$t$. It is similar to
Prefix~Rule~1 from the standard operational semantics (see
Section~\ref{section:standardoperationalsemanticstransitionrules}), except
that variables of type~$t$ are left in their symbolic form when a transition
label is obtained from $\alpha$, so only deterministic selections over types
other than~$t$ are resolved.
\begin{Rule}{Symbolic Prefix Rule 1}\label{rule:symbolicprefixrule1}
\infrule
\mbox{ }
\bigderive[
  \epsilon = c\S'_1x'_1\mathrm{:}X'_1\ldots\S'_kx'_k\mathrm{:}X'_k \in
    Comms^\nont(\alpha) \\
  \land \#\$^\nont(\alpha) = 0]
\alpha \then P(t) \trans[\epsilon][s] 
  P(t)[x'_i / x_i | i \in \mathord ?^\nont(\alpha)]
\endinfrule
where $Comms^\nont(\alpha)$ is the set of events that $\alpha$ describes
(under the assumption that $\alpha$ contains no nondeterministic selections
over types other than $t$), with the parts involving type $t$ left in their
symbolic form; formally:  
\[\label{def:commsnont}
\align
Comms^\nont(c\S_1x_1\mathrm{:}X_1\ldots\S_kx_k\mathrm{:}X_k) = \\
\qquad \{ c\S'_1x'_1\mathrm{:}X'_1\ldots\S'_kx'_k\mathrm{:}X'_k | 
  \forall i \in \set{1\upto k} \spot\ \\
\qquad \qquad
	\align
	\S_i = \mathord ? \land X_i \not = t \land \S'_i = \mathord ! 
           \land x'_i \in X_i \land X'_i = \mathit{null}\\
	\lor \S_i = \S'_i \in \set{\$,?} \land X'_i = X_i = t \land x'_i = x_i\\
	\lor \S_i = \S'_i = \mathord ! \land X'_i = X_i \land x'_i = x_i \}.
	\endalign
\endalign
\]
\end{Rule}

The second transition rule of prefix deals with prefixes that contain at least
one nondeterministic selection over a type other than $t$. It is similar to
Prefix Rule 2a from the standard operational semantics (see
Section~\ref{section:standardoperationalsemanticstransitionrules}).
All the nondeterministic
selections over types other than $t$ are resolved simultaneously, the act of
which generates a single $\tau$ transition. The values~$v_i$  chosen are
substituted for the variables~$x_i$ of the selections. 
\begin{Rule}{Symbolic Prefix Rule 2}\label{rule:symbolicprefixrule2}
\infrule
\dom(v) = \mathord \$^\nont(\alpha) \land \forall i \in \mathord \$^\nont(\alpha) \spot v_i \in X_i
\bigderive[\#\$^\nont(\alpha) > 0]
(\alpha \then P(t)) \trans[\tau][s] \left(\mathit{Replace}_{\$\mapsto!}^\nont(\alpha) \then P(t)\right)[v_i / x_i | i \in \$^\nont(\alpha)]
\endinfrule
\end{Rule}

The transition rules for external, internal and sliding choice and for binding
are very similar to the standard rules, and are given in
Figure~\ref{fig:ssos}.  
One exception is the presence of conditional
symbolic transitions, which need to be taken into considerations
here. Conditional choices must be  resolved without any other
influence on the overall state of the system, which means that the members of
$\mathit{Cond}$ must be promoted by the $\extchoice$ and $\timeout$ operators
in the same way $\tau$'s are.

\begin{figure}[tp]
\begin{center}
\noindent\begin{minipage}[b]{\linewidth}
\infrule
P(t) \symtrans{\alpha} P'(t)
\bigderive[\alpha \in \mathit{Cond} \union \set{\tau}]
P(t) \extchoice Q(t) \symtrans{\alpha} P'(t) \extchoice Q(t)
\endinfrule
\end{minipage}
\\
\begin{minipage}[b]{\linewidth}
\infrule
Q(t) \symtrans{\alpha} Q'(t)
\bigderive[\alpha \in \mathit{Cond} \union \set{\tau}]
P(t) \extchoice Q(t) \symtrans{\alpha} P(t) \extchoice Q'(t)
\endinfrule
\end{minipage}


\noindent\begin{minipage}[b]{0.5\linewidth}
\infrule
P(t) \symtrans{\epsilon} P'(t)
\bigderive[\epsilon  \in Visible]
P(t) \extchoice Q(t) \symtrans{\epsilon} P'(t)
\endinfrule
\end{minipage}
\begin{minipage}[b]{0.5\linewidth}
\infrule
Q(t) \symtrans{\epsilon} Q'(t)
\bigderive[\epsilon  \in Visible]
P(t) \extchoice Q(t) \symtrans{\epsilon} Q'(t)
\endinfrule
\end{minipage}


\noindent\begin{minipage}[b]{0.5\linewidth}
\infrule
\mbox{ }
\derive
P(t) \intchoice Q(t) \symtrans{\tau} P(t)
\endinfrule
\end{minipage}
\begin{minipage}[b]{0.5\linewidth}
\infrule
\mbox{ }
\derive
P(t) \intchoice Q(t) \symtrans{\tau} Q(t)
\endinfrule
\end{minipage}


\vspace{1ex}\noindent\begin{minipage}[b]{0.4\linewidth}
\infrule
\derive
P(t) \timeout Q(t) \symtrans{\tau} Q(t)
\endinfrule
\end{minipage}
\begin{minipage}[b]{0.6\linewidth}
\infrule
P(t) \symtrans{\alpha} P'(t)
\bigderive[\alpha \in \mathit{Cond} \union \set{\tau}]
P(t) \timeout Q(t) \symtrans{\alpha} P'(t) \timeout Q(t)
\endinfrule
\end{minipage}
\\
\noindent\begin{minipage}[b]{0.5\linewidth}
\infrule
P(t) \symtrans{\epsilon} P'(t)
\bigderive[\epsilon  \in Visible]
P(t) \timeout Q(t) \symtrans{\epsilon} P'(t)
\endinfrule
\end{minipage}
\begin{minipage}[b]{0.5\linewidth}
\infrule
E(X) = P
\derive
X(t) \trans[\tau][s] P(t)
\endinfrule
\end{minipage}
\end{center}
\caption{Semi-symbolic operational semantics rules for external, internal and
  sliding choice, and for binding\label{fig:ssos}} 
\end{figure}

Clause~(iv) of the definition of \Seq{} (Definition~\ref{def:seq}) implies
that guards of conditional choices may not contain both variables of type $t$
and variables of non-$t$ types.  
The truth of any boolean condition that contains no variables of type $t$
(i.e.~every conditional not in $Cond$) can be fully evaluated at the time of
SSLTS generation.  Hence we have the following rules, similar to the standard
rules. 
\label{rule:symboliccondchoice} 
\begin{center}
\begin{minipage}[b]{0.5\linewidth}
\infrule
P(t) \symtrans{\alpha} P'(t)
\derive
\If \True \Then P(t) \Else Q(t) \symtrans{\alpha} P'(t)
\endinfrule
\end{minipage}
\begin{minipage}[b]{0.5\linewidth}
\infrule
Q(t) \symtrans{\alpha} Q'(t)
\derive
\If \mathit{False} \Then P(t) \Else Q(t) \symtrans{\alpha} Q'(t)
\endinfrule
\end{minipage}
\end{center}
Every conditional choice with a boolean condition $\mathit{cond}$ that
involves type $t$ 
(i.e.~every conditional in $Cond$) may either evolve to the
positive branch by following a conditional transition labelled with
$\mathit{cond}$ or it may evolve to the negative branch by following a
conditional transition labelled with the negation of $cond$. 
\infrule
\mbox{ }
\bigderive[\mathit{cond} \in \mathit{Cond}]
\If cond \Then P(t) \Else Q(t) \trans(3)[\mathit{cond}][s] P(t)
\endinfrule
\infrule
\mbox{ }
\bigderive[\mathit{cond} \in \mathit{Cond}]
\If cond \Then P(t) \Else Q(t) \trans(4)[\neg\mathit{cond}][s] Q(t)
\endinfrule



%

\begin{example}
\label{example:ssos}
Recall our running example:
\begin{eqnarray*}
P(t) & =  &
  c\$x\mathord:\set{a,b}\$y\from t?z\from t \then
  \If y=z \Then d!x \then STOP \Else STOP.
\end{eqnarray*}
Figure~\ref{fig:ssosex} represents the semi-symbolic operational semantics
for $P(t)$.  
The $\tau$ transitions correspond to Symbolic Prefix Rule~2; the visible
transitions correspond to Symbolic Prefix Rule~1.
\end{example}

\subsubsection{Symbolic traces}\label{section:symbolictraces}

Symbolic traces will play a vital role in the analysis of behaviour of process
families based on SSOS\@. They are similar to ordinary CSP traces
(Section~\ref{section:denotationalmodelsintro}), except they contain visible
symbolic events instead of ordinary visible events, and may contain both 
conditional and $\tau$ symbolic events.  

Formally, we define a symbolic trace as follows. Let $\mathcal{S} = (S, s_0,
L, \symtrans{})$ be the SSLTS obtained by applying the SSOS to process syntax
$Proc(t)$.  Given two symbolic states $P(t)$ and $Q(t)$ in $S$ and a sequence
of symbolic events $\sigma = \strace{\alpha_i | i \in \set{1\upto n}}$, we
write $P(t) \SymTrans{\sigma} Q(t)\newnot{mapstotranss}$ to mean that there
exist symbolic states $P_0(t) = P(t), P_1(t), \ldots, P_n(t) = Q(t)$ such that
for all~$i$ in $\set{0\upto n-1}$\,\ $P_i(t) \trans(3)[\alpha_{i+1}][s]
P_{i+1}(t)$;
$\sigma$ is
called a \emph{symbolic trace} of $P(t)$. Therefore, a symbolic trace of
$Proc(t)$ is a sequence of labels of symbolic events that form a path,
starting at $s_0$, through~$\mathcal{S}$. We let
$\stsem{Proc(t)}$ denote the set of all symbolic traces of
$Proc(t)$. Observe that symbolic traces are quite different from standard
traces as they may contain symbolic $\tau$ events and
conditional symbolic events, while ordinary traces contain only visible
events.  In Section~\ref{section:generates} we will study the relationship
between symbolic and concrete traces in more detail. We will usually use
$\sigma, \rho$ and their derivatives ($\sigma', \rho_1$, etc.) to denote
symbolic traces.

In Section~\ref{section:typereduction} we will work with symbolic traces that
are ``similar'' in the sense that their restrictions to conditional
and visible symbolic events are identical. 
\begin{defi}
Let $\sigma$ and $\sigma'$ be two symbolic traces. Then $\sigma$ and $\sigma$
are \emph{non-$\tau$ equivalent}, written $\sigma \nontauequiv \sigma'$, if
$\sigma \hide \set{\tau} = \sigma' \hide \set{\tau}$.
\end{defi}

\subsection{Concrete Operational Semantics with
  Environments}\label{section:cose}\newnot{cose}

So far we have presented a concrete and a semi-symbolic operational
semantics for CSP (see Section~\ref{section:os} and
Section~\ref{section:ssos}, respectively). In this section we present a
concrete operational semantics which joins the two together. We call it
\emph{Concrete Operational Semantics with Environments} (\emph{COSE}). The
states of an SSLTS correspond to the control states of a given process. In
order to link the symbolic and concrete states (where the latter contain
information not only about program state, but also about data of type $t$) we
need a mechanism for introducing concrete values into symbolic states. We do
this through the use of \emph{environments}. The environments defined in this
section are different from the environment with which processes communicate,
or the global map~$E$ of identifiers to process definitions that we introduced
in Section~\ref{section:cspsyntax}.  Intuitively, environments map free
variables within process syntaxes to concrete values that were previously
bound to such variables through  inputs. 
%
\begin{defi}
Let $\mathit{Env}(t) \defs \Var \pfun t$\newnot{pfun}\newnot{env}. Then an
\emph{environment} is a partial function $\Gamma \in Env(T)$ for some
instantiation $T$ of type~$t$. 
\end{defi}
For later convenience, we adopt the notational convention that for all $v$ in
$\Value$,\, ${\Gamma(v) = v}$.  We lift the application of environments to
various structures that we use (constructs, processes, sets, relations,
etc.)~in the natural way: if $X(T)$ is such a structure and $\Gamma$ is in
$\mathit{Env}(T)$, then $\Gamma(X(T))$ is a structure like $X(T)$ but with
every free variable $x$ of type $t$ replaced by $\Gamma(x)$ (assuming $x$ is
in $\dom(\Gamma)$). In particular, given a process definition $P(t)$ we define
the syntactic substitution
\begin{eqnarray*}
P(t)\envrename{\Gamma} & \defs & P(t)[ \Gamma(x) / x | x \in \dom(\Gamma)].
\end{eqnarray*}

Note that for all environments $\Gamma$, all symbolic events $\alpha$, and
$\dagger \in \set{t, \mbox{\textit{non-t}}}$, we have that
$\$^\dagger(\Gamma(\alpha)) = \mathord \$^\dagger(\alpha)$ and
$?^\dagger(\Gamma(\alpha)) = \mathord ?^\dagger(\alpha)$, which means that
$\$(\Gamma(\alpha)) = \mathord \$(\alpha)$ and ${\mathord ? (\Gamma(\alpha)) =
\mathord ?(\alpha)}$.

Let $T$ and $T'$ be two instantiations of type $t$. Then, given a function $f : T \rightarrow T'$ and an environment $\Gamma$ in $\mathit{Env}(T)$, we define
\begin{eqnarray*}
	f(\Gamma) &  \defs & \set{x \mapsto f(v) | \Gamma(x) = v}.
\end{eqnarray*}
Observe that $f(\Gamma)$ is an environment in $\mathit{Env}(T')$.

The states of the LTS $\mathcal{C}$ that COSE generates from a given process syntax $Proc(t)$ are configurations $(P(t),\Gamma,T)$, where:
\begin{iteMize}{$\bullet$}
	\item $P(t)$ is a symbolic state, equal to or slightly modified from a state of the SSLTS~$\mathcal{S}$ of $Proc(t)$,
	\item $\Gamma$ is an environment in $\mathit{Env}(T)$, and
	\item $T$ is a concrete instantiation of type $t$.
\end{iteMize}
Note that the inclusion of the type instantiation as the third element of a configuration means that each choice of $T$ gives rise to a different LTS\@.
Whenever the concrete type~$T$ is clear from the context or indifferent, we
omit it from the configurations and use pairs~$(P(t), \Gamma)$.


The initial state of $\mathcal{C}$ is defined to be the configuration
$(P_0(t),\nullset,T)$, where $P_0(t)$ is the initial state of $\mathcal{S}$;
we sometimes abbreviate this as $P_0(T)$.  To emphasise the fact that COSE is
a concrete operational semantics we denote the transition relation using the
same symbol ($\trans$) that we used in Section~\ref{section:os}.

We treat two configurations as identical if they describe exactly the same
process. Formally, $(P(t), \Gamma, T) = (P'(t), \Gamma', T')$ if and only if
$P(t)\envrename{\Gamma} \equiv_\alpha\newnot{alphaequivalence}
P'(t)\envrename{\Gamma'}$ and $T = T'$, where $\equiv_\alpha$ denotes
operational semantics alpha-equivalence, i.e.~equality of operational
semantics modulo renaming of bound variables. 


\begin{remark}\label{remark:minimalityofLTSs}
Observe that\footnote{$S \mathbin{\dres} \Gamma$ denotes $\Gamma$ restricted
  to domain~$S$, i.e.\ $\set{x \mapsto y | (x \mapsto y) \in \Gamma \land x
    \in S}$.}  $P(t)\envrename{\Gamma} = P(t)\envrename{\FV(P(t))
  \mathbin{\dres} \Gamma}$ for all symbolic states $P(t)$ and all environments
$\Gamma$, where $\FV(P(t))$ denotes the free variables of~$P(t)$.  Therefore,
configurations $(P(t), \Gamma, T)$ and $(P(t), \FV(P(t)) \mathbin{\dres}
\Gamma, T)$ are identical. From now on we always assume environment minimality
within configurations, which we achieve by restricting the environment
$\Gamma$ of every configuration $(P(t), \Gamma, T)$ to the free variables of
$P(t)$.
\end{remark}



\subsubsection{Translation rules}\label{section:translationrules}

We present the specification of COSE using translation rules that
translate transitions within SSLTSs into corresponding COSE transitions.  Let
$T$ be a fixed instantiation of the distinguished type parameter $t$.


Given a symbolic state $P(t)$, we let $Q(t) =
\mathit{Replace}_{\$\mapsto!}^t(c, P(t))$ be a symbolic state like $P(t)$,
except every transition from $P(t)$ labelled with a visible symbolic
transition $\epsilon$ on channel~$c$ is replaced with an identical transition
in $Q(t)$, but labelled with $\mathit{Replace}_{\$\mapsto!}^t(\epsilon)$
instead, i.e.\ if $P(t) \trans[\epsilon][s] P'(t)$ then $Q(t)
\trans(8)[Replace_{\$\mapsto!}^t(\epsilon)] P'(t)$.  We will see later
(Proposition~\ref{corollaryA2}) that
all such transitions over~$c$ result from the same construct.



Visible symbolic events that contain a nondeterministic selection over type
$t$ are translated into two concrete events: a $\tau$ that resolves the
nondeterminism; and a subsequent visible event.  The first translation rule
shows how the $\tau$ is produced; for each nondeterministically chosen
variable~$x_i$ in the symbolic event, the environment is updated to map $x_i$
to some value~$v_i$, and the nondeterministic choice in the subsequent
symbolic event is replaced by an output (to be dealt with later). 
%
\begin{Rule}{Translation Rule 1}\label{rule:translationrule1}
\infrule
P(t) \symtrans{\epsilon} Q(t)\\
\epsilon = \construct \land v \in \mathord \$^t(\epsilon) \rightarrow T\\
\bigderive[\#\$^t(\epsilon)>0][.]
(P(t),\Gamma,T) \trans[\tau]
  (\mathit{Replace}^t_{\$ \mapsto \mathord !} (c, P(t)),
   \Gamma \oplus \set{x_i \mapsto v_i | i \in \$^t(\epsilon)}, T)
\endinfrule
\end{Rule}

\noindent Clause~(v) of the definition of \Seq\ (Definition~\ref{def:seq})
implies that there is never a clash between a nondeterministic input variable
of type $t$ from one branch of an external or sliding choice and a free
variable present in the other branch. Without this assumption, Translation
Rule~1 could produce wrong answers, as demonstrated by the following example. 

\begin{example}\label{example:translationrule1}
Let $Proc(t) = c_1?x\mathrm{:}t \then (c_2\$x\mathrm{:}t?y\mathrm{:}t \then
\STOP \extchoice c_1!x \then \STOP)$ and $T = \set{0,1}$. Then, after
performing $c_1.0$ and a $\tau$ resolving the nondeterministic selection by
choosing $x=1$, the configuration $(Proc(t),\set{},T)$ evolves to
$(c_2!x?y\mathrm{:}t \then \STOP \extchoice c_1!x \then \STOP,\linebreak[2] \set{x \mapsto
  1}, T)$. Then, by Translation Rule~2 (see below), the event $c_1.1$ is
available, which clearly should not be the case.
\end{example}

Next, we show how visible symbolic events that contain no nondeterministic
selections of type $t$ get instantiated into concrete visible events by
substituting values from the environment for all the outputs of type $t$ and
choosing the values of all deterministic inputs of type $t$. 

\begin{Rule}{Translation Rule 2}\label{rule:translationrule2}\nopagebreak[4]
\infrule
P(t) \symtrans{\epsilon} Q(t)\\
\epsilon = \construct\\
\dom(v) = \set{1\upto k} \land (\forall i \in \mathord ?^t(\epsilon) \spot v_i \in T) \land \forall i \in \mathord !(\epsilon) \spot v_i = \Gamma(x_i)
\bigderive[\#\$^t(\epsilon) = 0][.]
(P(t),\Gamma,T) \trans(5)[c.v_1\ldots v_k] (Q(t),\Gamma \oplus \set{x_i \mapsto v_i | i \in \mathord ?^t(\epsilon)},T) 
\endinfrule
\end{Rule}

\begin{example}\label{example:COSE1}
Recall our running example
\begin{eqnarray*}
P(t) & = & 
  c\$x\mathord:\set{a,b}\$y\from t?z\from t \then 
  \If y=z \Then d!x \then STOP \Else STOP
\end{eqnarray*}
whose SSOS semantics appear in Figure~\ref{fig:ssosex}.  In particular,
consider the transition
\begin{eqnarray}
\label{eqn:replace}
c!a\$y\mathrm{:}t?z\mathrm{:}t \then Q_{a,y,z} & 
  \trans(7)[c!a\$y\mathrm{:}t?z\mathrm{:}t][s] & Q_{a,y,z}.
\end{eqnarray}
Translation Rule 1 implies 
that configuration $(c!a\$y\mathrm{:}t?z\mathrm{:}t \then Q_{a,y,z}, \nullset,
T)$, with $T = \set{0,1}$, can do a $\tau$ and become either of  
\begin{eqnarray*}
\label{eqn:replace1}
conf_0 & = & 
   (\mathit{Replace}^t_{\$ \mapsto \mathord !}
     (c, c!a\$y\mathrm{:}t?z\mathrm{:}t \then Q_{a,y,z}),  \set{y \mapsto 0},T),
\\
conf_1 & = &
 (\mathit{Replace}^t_{\$ \mapsto \mathord !} 
   (c, c!a\$y\mathrm{:}t?z\mathrm{:}t \then Q_{a,y,z}),  \set{y \mapsto 1},T).
\nonumber
\end{eqnarray*}
Now, from (\ref{eqn:replace}) and the definition of $\mathit{Replace}$:
\begin{eqnarray*}
\mathit{Replace}^t_{\$ \mapsto \mathord !} 
  (c, c!a\$y\mathrm{:}t?z\mathrm{:}t \then Q_{a,y,z})
  & \trans(6)[c!a!y?z\mathrm{:}t][s] & Q_{a,y,z}.
\end{eqnarray*}
Hence, using Translation Rule 2, we can deduce 
\begin{eqnarray*}
conf_0 & \trans(5)[c.a.0.0] &
  (Q_{a,y,z}, \set{y \mapsto 0, z  \mapsto 0}, T),
\\
conf_0 & \trans(5)[c.a.0.1] & 
  (Q_{a,y,z}, \set{y \mapsto 0, z  \mapsto 1}, T);
\end{eqnarray*}
and similarly for $conf_1$.  (In Figure~\ref{fig:cose}, the process
$\mathit{Replace}^t_{\$ \mapsto \mathord !}  (c,
c!a\$y\mathrm{:}t?z\mathrm{:}t \then Q_{a,y,z})$ is written as
$c!a!y\mathrm{:}t?z\mathrm{:}t \then Q_{a,y,z}$, for convenience.)
\end{example}

\begin{remark}\label{remark:translationrules1and2}
We can combine Translation Rules~1 and~2 to deduce that if
\[
\begin{align}
P(t) \symtrans{\epsilon} Q(t) \land
\epsilon = \construct \land\null\\
\dom(v) = \set{1\upto k} \land 
(\forall i \in \mathord \$^t(\epsilon) \union \mathord ?^t(\epsilon) \spot
   v_i \in T) \land 
\forall i \in \mathord !(\epsilon) \spot v_i = \Gamma(x_i),
\end{align}
\]
then
\begin{eqnarray*}
(P(t),\Gamma,T) & [\trans[\tau]]\trans(5)[c.v_1\ldots v_k] &
  (Q(t), 
   \Gamma \oplus \set{x_i \mapsto v_i | 
         i \in \mathord \$^t(\epsilon) \union \mathord ?^t(\epsilon)},
   T) ,
\end{eqnarray*}
where $[\trans[\tau]]$\newnot{optionaltransition} denotes an optional $\tau$
transition, present if and only if $\#\$^t(\epsilon) > 0$.
\end{remark}

The next translation rule says that when a concrete LTS is obtained from an
SSLTS, symbolic $\tau$~transitions are turned into standard
$\tau$~transitions.
\begin{Rule}{Translation Rule 3}\label{rule:translationrule3}
\infrule
P(t) \symtrans{\tau} Q(t)
\derive
(P(t),\Gamma,T) \trans[\tau] (Q(t),\Gamma,T)
\endinfrule
\end{Rule}

The final translation rule shows how conditional symbolic transitions
disappear when an SSLTS is instantiated into a concrete LTS using COSE; the
labels are evaluated in the environment, affecting the availability of the
subsequent transitions.
\begin{Rule}{Translation Rule 4}\label{rule:translationrule4}
\infrule
P(t) \trans(3)[\mathit{cond}][s] Q(t)\\
(Q(t),\Gamma,T) \trans[a] (R(t),\Gamma',T)
\bigderive[cond \in Cond \land \eval{cond}{\Gamma}][,]
(P(t),\Gamma,T) \trans[a] (R(t),\Gamma',T)
\endinfrule
where $\eval{cond}{\Gamma}$\newnot{evalcond} denotes the truth value of the
proposition obtained from $cond$ by substituting all free variables of type
$t$ with their corresponding values contained within the
environment~$\Gamma$. Note that if $(Q(t),\Gamma,T)$ is deadlocked, then so is
$(P(t),\Gamma,T)$. 
\end{Rule}

\begin{example}
Recall our running example
\begin{eqnarray*}
P(t) & = & 
  c\$x\mathord:\set{a,b}\$y\from t?z\from t \then 
  \If y=z \Then d!x \then STOP \Else STOP
\end{eqnarray*}
whose SSOS semantics appear in Figure~\ref{fig:ssosex}.  The COSE semantics is
given in Figure~\ref{fig:cose}.  The initial $\tau$-transitions follow from
Translation Rule~3.  The subsequent $\tau$-transitions and transitions with
events on~$c$ were explained in Example~\ref{example:COSE1}.  The left-hand
final transition with event~$d.a$ follows from Translation Rule 4, noting that
$\eval{y=z}{\set{y \mapsto 0, z \mapsto 0}}$, and using the fact that $(d!a
\then STOP, \set{y \mapsto 0, z \mapsto 0}, T) \trans[d.a] STOP$, by
Translation Rule 2; other transitions on~$d$ follow similarly.
\end{example}


\subsection{Congruence of COSE to the standard operational  semantics}
\label{section:congruence}

We will often work with concrete LTSs generated by COSE rather than by the standard operational semantics. It is therefore important that the two operational semantics are congruent so that any denotational values extracted from them are identical. The following theorem proves such a congruence. 

\begin{thm}\label{theorem:congruence}\textbf{\textup{(Congruence of COSE 
     to the standard operational semantics.)}} 
Suppose that $Proc(t)$ is some process syntax that satisfies \Seq. Let
$\mathcal{L}_1$ and $\mathcal{L}_2$ be the LTSs generated from $Proc(t)$, for
some fixed instantiation $T$ of type $t$, using COSE and the standard
operational semantics, respectively. Then $\mathcal{L}_1$ and $\mathcal{L}_2$
are strongly bisimilar.  
\end{thm}
\proof[Proof sketch.]
By showing that
\begin{eqnarray*}
\mathcal{B} & = & 
  \left\{((P(t),\Gamma),P(T)\envrename{\Gamma}) | 
     (P(t),\Gamma) \in \hat{\mathcal{L}}_1 \land
      P(T)\envrename{\Gamma} \in \hat{\mathcal{L}}_2 \right\}.
\end{eqnarray*}
is a strong bisimulation relation between $\hat{\mathcal{L}}_1$ and
$\hat{\mathcal{L}}_2$ (the states of $\mathcal{L}_1$ and $\mathcal{L}_2$),
using structural induction on $P(t)$.\qed

One implication of Theorem~\ref{theorem:congruence} is the fact that we can
express denotational values of configurations of LTSs obtained using COSE in
terms of the denotational values calculated from states of LTSs generated
using standard operational semantics; so:
\begin{eqnarray*}
\tsem{Proc(t),\Gamma,T} & = & \tsem{Proc(T)\envrename{\Gamma}}, \\
\fsem{Proc(t),\Gamma,T} & = & \fsem{Proc(T)\envrename{\Gamma}},
\end{eqnarray*}
for every process syntax $Proc(t)$,  instantiation $T$ of $t$ and environment
$\Gamma$ in $\mathit{Env}(T)$. 
\subsection{Relating symbolic and concrete traces}\label{section:generates}

In this section we define what it means for a concrete trace to be an
instantiation of a symbolic trace. We do this by using a ternary relation
$\generates{}$ that links symbolic traces
(Section~\ref{section:symbolictraces}), environments and concrete traces. The
environments are included in the relation, since, in order to relate a
symbolic trace to a concrete trace, concrete values need to be substituted for
the free variables that can occur within the symbolic trace; these concrete
values come from environments. 

Given a process syntax $Proc(t)$ and an instantiation $T$ of type~$t$, we
define a relation $\generates{}$
written using infix notation: $\sigma \generates{\Gamma} tr$, for a symbolic
trace $\sigma$, an environment $\Gamma$ and a concrete trace $tr$:
\begin{enumerate}[(i)]\label{def:generates}\newnot{generates}
\item $\strace{} \generates{\Gamma} \trace{}$,

\item $\sigma \generates{\Gamma} tr \iff \strace{\tau}\cat \sigma
  \generates{\Gamma} tr$,

\item $\sigma \generates{\Gamma} tr \land \eval{cond}{\Gamma} \iff
  \trace{cond}\cat \sigma \generates{\Gamma} tr$,

\item $e \in \mathit{Insts}_\Gamma(\epsilon) \land \sigma
  \generates{\Gamma \oplus \mathit{Match}(\epsilon,e)} tr \iff
  \trace{\epsilon}\cat \sigma \generates{\Gamma} \trace{e}\cat tr$, 
\end{enumerate}
where $\mathit{Insts}_\Gamma(\epsilon)$ gives all instantiations of~$\epsilon$
consistent with~$\Gamma$, and $\mathit{Match}(\epsilon,e)$ gives the extension
to the environment caused by instantiating~$\epsilon$ with~$e$; let $\epsilon$
be of the form $c\S_1x_1\mathrm{:}X_1\ldots\S_kx_k\mathrm{:}X_k$, and recall
that, by Remark~\ref{remark:noNontPartsInSymbolicVisibleEvents},
$\$^\nont(\epsilon) = \mathord ?^\nont(\epsilon) = \nullset$:
\begin{eqnarray*}
\label{def:insts}
\mathit{Insts}_\Gamma(\epsilon) & = &
  \set{c.v_1\ldots v_k | 
    \forall i \in \mathord \$^t(\epsilon) \union \mathord ?^t(\epsilon) \spot
      v_i \in T \land 
    \forall i \in \mathord !(\epsilon) \spot v_i = \Gamma(x_i)}\newnot{insts},
\\
\mathit{Match}(\epsilon,c.v_1\ldots v_k) & =  &
  \set{x_i \mapsto v_i | i \in \mathord \$^t(\epsilon) \union \mathord
    ?^t(\epsilon)}.
\end{eqnarray*}
Observe that rule~(ii) indicates that we essentially ignore any $\tau$'s
present in the symbolic traces. 
For brevity we write $\sigma, \sigma' \generates{\Gamma} tr$ to mean that both
$\sigma \generates{\Gamma} tr$ and $\sigma' \generates{\Gamma} tr$.  

Using the above definition we can describe what it means for a (concrete)
visible event to \emph{match} a visible symbolic event. In the following
definition we treat two concrete events as essentially different if they are
available after different traces. 
\begin{defi}\label{definition:matching1}
A visible event $e$ that is available in $Proc(T)$ immediately after a
trace~$tr$ \emph{matches} a visible symbolic event $\epsilon$ if there exists
a symbolic trace $\sigma$ such that $\sigma\cat\strace{\epsilon}$ is in
$\stsem{Proc(t)}$ and $\sigma\cat\strace{\epsilon} \generates{\nullset}
tr\cat\trace{e}$.
\end{defi}

\section{Regularity results}
\label{section:regularity}

In this section we present a series of regularity results that are consequences
of the \Norm{} condition.  These results show that specifications exhibit
certain clarity in their behaviour. 
Our main findings can be summarised as follows:  
\begin{enumerate}[(1)]
\item 
There is no ambiguity about what configuration a process reaches after
performing a sequence of concrete events that does not end with an internal
event (Proposition~\ref{prop:environmentuniqueness});

\item 
There is no ambiguity which construct gives rise to a given concrete event
that is available after a given trace (Proposition~\ref{corollaryA2}); and

\item 
Every event available in a process syntax instantiated with a collapsed type
is also an event available in the same process syntax instantiated with the
uncollapsed type, and the target configurations are the same except for the
underlying type (Proposition~\ref{proposition:adataindepimplication}).
\end{enumerate}
Regularity results will play a vital role in proving the main theorems of the
type reduction theory. 

In Section~\ref{section:generates} we defined what it means for a concrete visible event to match a visible symbolic event. In the following sections we will often need to relate concrete and symbolic visible events and syntax constructs that give rise to them. The following definition establishes such relationships formally.

\begin{defi}\label{definition:matching}
Given a sequential process syntax $Proc(t)$, let $\alpha_1, \alpha_2, \ldots$
be the prefix constructs of~$Proc(t)$ (where two prefix constructs are
regarded as different if they appear in different places in $Proc(t)$). Let
$T$ be a given instantiation of type~$t$. Then, for every pair of a trace $tr$
and a visible event $e$ such that $tr\cat\trace{e}$ is a trace of $Proc(T)$,
there must be at least one $\alpha_i$ that gives rise to $e$ immediately after
$tr$. We then say that $\alpha_i$ is \emph{matched} by~$e$ (or that $e$
\emph{matches} $\alpha_i$). We define visible symbolic events to match syntax
constructs in an analogous way. 
\end{defi}

\begin{example}
Let
\begin{eqnarray*}
P(t) & = &
  c?x\mathrm{:}t \then \STOP \extchoice c\$x\mathrm{:}t \then c.x \then \STOP.
\end{eqnarray*}
Then, for $T = \set{0,1}$, given trace $tr = \nil$, the event $e = c.1$
matches both the constructs $c?x\mathrm{:}t$ and $c\$x\mathrm{:}t$, but not
$c.x$ (as $c.x$ may give rise to $c.1$, but only after the trace $\trace{c.1}$
and not the empty trace).
\end{example}
\noindent
Note that the process in the above example does not satisfy \Norm.  We will
show (in Proposition~\ref{corollaryA2}) that for processes that do satisfy
\Norm, each event (after a given trace) matches a \emph{unique} construct.



An important property of normality is the lack of ambiguity about what state a
process reaches after performing a sequence of visible concrete events not
followed by a $\tau$. The following proposition establishes this formally.
\begin{prop}\label{prop:environmentuniqueness}
Suppose that $Proc(t)$ satisfies \Norm{}. Suppose further that
\[
(Proc(t),\Gamma_{init}) \stackrel{s}{\longmapsto} (P(t),\Gamma)
\mbox{\quad and \quad}
(Proc(t),\Gamma_{init}) \stackrel{s'}{\longmapsto} (Q(t),\Gamma'),
\]
where $s$, $s'$ do not end with a $\tau$, and $s \hide \tau = s' \hide
\tau$. Then $P(t) = Q(t)$ and $\Gamma = \Gamma'$.
\end{prop}


Further, \Norm{} implies that every two symbolic traces that give rise to the
same concrete trace, and are either the empty symbolic trace or both end in a
visible symbolic event, are identical up to internal actions.
\begin{prop}\label{propositionA2}
Suppose that $Proc(t)$ satisfies \Norm{}. Then, if $\sigma, \sigma'$ are
symbolic traces of $Proc(t)$ such that $\sigma \generates{\Gamma} tr$ and
$\sigma' \generates{\Gamma} tr$, and either $\sigma = \sigma' = \strace{}$ or
both $\sigma$ and $\sigma'$ end in a visible symbolic event, then $\sigma
\nontauequiv \sigma'$. 
\end{prop}



For a process that satisfies \Norm{}, for each visible event that is performed
after a given trace, there is never any ambiguity what construct this event
matches.
\begin{prop}\label{corollaryA2}
Suppose that  $Proc(t)$ satisfies \Norm{}. Then, if $tr\cat\trace{e}$ is a
trace of $(Proc(t), \Gamma, T)$ for some type $T$ and some environment
$\Gamma$, and $e$ matches constructs $\alpha$ and~$\alpha'$, then $\alpha =
\alpha'$. 
\end{prop}




The following lemma compares corresponding constructs in two processes, one of
which refines the other. 
\begin{lem}\label{lemmaD}
Suppose  that $P(t)$ and $Q(t)$ satisfy \Norm{}. Let $T$ be an instantiation
of type $t$. Suppose that $(Q(t),\Gamma_{init},T) \trefinedby
(P(t),\Gamma_{init},T)$. Then for all visible symbolic events $\epsilon,
\epsilon'$, symbolic traces $\sigma, \sigma'$ and traces $tr$ such that:
\begin{enumerate}[\em(i)]
\item $tr\cat \trace{e} \in \tsem{P(t),\Gamma_{init}, T}$;
\item $\sigma\cat \strace{\epsilon} \in \stsem{P(t)}
  \generates{\Gamma_{init}} tr\cat \trace{e}$;  and
\item $\sigma'\cat \strace{\epsilon'} \in \stsem{Q(t)}
  \generates{\Gamma_{init}} tr\cat \trace{e}$; 
\end{enumerate}
we have that  $!(\epsilon') \subseteq \mathop!(\epsilon)$.
\end{lem}


The following proposition and its corollary form our final consequence of
\mbox{\Norm{}}. The proposition says that every event available in a process
syntax instantiated with some type is also an event available in the same
process syntax instantiated with a larger type. In addition, the target
configurations are the same (except for the underlying
type). Corollary~\ref{corollary:adataindepimplication} then extends this
observation to traces. 
\begin{prop}\label{proposition:adataindepimplication}
Suppose that $Proc(t)$ satisfies \Norm{}. Let $T$ and $\shiftedhat{T}$ be
instantiations of type $t$ such that $\shiftedhat{T} \subseteq T$ and let
$\Gamma$ be an environment in $\mathit{Env}(T)$. Then, if 
\begin{eqnarray}\label{equation:adataindepimplicationeq1}
(Proc(t), \Gamma, \shiftedhat{T}) & \trans[a] &
  (Proc'(t), \Gamma', \shiftedhat{T})
\end{eqnarray}
then
\begin{eqnarray*}
(Proc(t), \Gamma, T) & \trans[a] & (Proc'(t), \Gamma', T).
\end{eqnarray*}
\end{prop}


\begin{cor}\label{corollary:adataindepimplication}
Suppose that $Proc(t)$ satisfies \Norm{}. Let $T$ and $\shiftedhat{T}$ be
instantiations of type $T$ such that $\shiftedhat{T} \subseteq T$. Then, if 
\begin{eqnarray*}
(Proc(t), \Gamma, \shiftedhat{T}) & \Trans[tr] & 
  (Proc'(t), \Gamma', \shiftedhat{T}),
\end{eqnarray*}
then
\begin{eqnarray*}
(Proc(t), \Gamma, T) & \Trans[tr] & (Proc'(t), \Gamma', T).
\end{eqnarray*}
\end{cor}

\section{Type reduction theory}\label{section:typereduction}

Recall that given an instantiation $T$ of type $t$ and a non-negative integer
$B$, we defined (Definition~\ref{definition:collapsing}) a 
\emph{$B$-collapsing function} $\phi$ to be a function from $T$ to $\set{0
  \upto   B}$ such that 
\begin{iteMize}{$\bullet$}
\item $\phi(v) = v$ for all $v$ in $\set{0 \upto B-1}$;
\item $\phi(v) = B$ for all $v$ in $\set{B \upto \#T-1}$.
\end{iteMize}
In other words, $\phi$ replaces all but a fixed finite number of members
of~$t$ by a single value.  Whenever $B$ is clear from context, we call $\phi$
a \emph{collapsing function}.

As described in the Introduction, our aim is develop a \emph{type reduction
  theory}, to show that
\begin{equation}\label{equation:typereductionprinciple}
\begin{align}
Spec(\shiftedhat{T}) \refinedby \phi(Impl(T))
\mbox{ implies that } 
Spec(T) \refinedby Impl(T),
\\
\mbox{for all $T$ such that $T \supseteq \shiftedhat{T}$, where $\phi : T
  \rightarrow \shiftedhat{T}$ is a collapsing function}
\end{align}
\end{equation}

The use of parameters in specifications and/or implementations leads to the
problem of having to decide infinitely many refinements in order to
deduce the answer to a verification problem.  Our technique of using
collapsing functions treats some values of type~$t$ as essentially identical.

Our two main results, Theorem~\ref{maintracesthm} and
Theorem~\ref{mainfailuresthm}, prove (\ref{equation:typereductionprinciple})
in the traces and stable failures models, respectively.  They require suitable
assumptions on $Spec$ (including \Norm) and $Impl$ (that it is symmetric
in~$t$); they give a lower bound on the size of~$\shiftedhat{T}$ based on the
syntax of~$Spec$.

A significant part of this section is devoted to showing how certain
behaviours (either in the traces or the stable failures model) of
specifications instantiated with uncollapsed types can be inferred from known
behaviours of the same specification instantiated with reduced types
(justifying the name of our theory). 


We present and prove type reduction theorems for both the traces and the
stable failures models in Sections~\ref{section:tracesresults}
and~\ref{section:failuresresults}, respectively.  Proofs of some subsidiary
results are in Appendix~\ref{sec:traces_proofs}.

First, we lift $\phi$ to various settings.
Given a boolean condition $cond$, we define $\phi(cond)$ to be like $cond$,
except that every value or variable $x$ of type $t$ is replaced by $\phi(x)$.
We adopt the notational convention that if $x$ is a value or a variable of a
type other than~$t$ or it is a type other than~$t$, then $\phi(x) = x$.  

We lift the application of $\phi$ to other common objects used in this paper
in the natural way (see Table~\ref{philifttable}).
\begin{table}
\renewcommand\arraystretch{1.1}
\begin{center}
\begin{tabular}{|c|c|c|}
\hline
 Object&Application&Meaning\\
\hline\hline
event&$\phi(c.v_1 \ldots v_k)$&$c.\phi(v_1) \ldots \phi(v_k)$\\
set/type&$\phi(S)$&$\set{\phi(x) | x \in S}$\\
trace&$\phi(tr)$&$\trace{\phi(e) | e \leftarrow tr}$\\
environment&$\phi(\Gamma)$&$\set{x \mapsto \phi(v) | \Gamma(x) = v}$\\
process&$\phi(P)$&$P\rename{^{\phi(e)} / _e | e \in \Sigma}$\\
\hline
\end{tabular}
\caption{Lifting the definition of \protect$\phi$.}
\label{philifttable}
\end{center}
\end{table}

Finally, given an instantiation $T$ of type $t$, a $B$-collapsing function
$\phi$, and a value $v$ in $\set{0 \upto B}$, we define 
\begin{eqnarray*}
\phi^{-1}(v) & = & 
  \left\{
    \begin{array}{ll}
      \set{ v' \in T | \phi(v') = v}, & \mbox{if $v \in \set{0 \upto B}$}, \\
      \set{v}, & \mbox{otherwise}.
    \end{array} \right.
\end{eqnarray*}
Also, given~$T$, we lift the definition of $\phi^{-1}$ to events:
\begin{eqnarray*}
\phi^{-1}(c.v_1\ldots v_k) & = & 
  \set{ c.v'_1 \ldots v'_k  | 
    \forall i \in \set{1 \upto k} \spot v_i' \in \phi^{-1}(v_i)},
\end{eqnarray*}
and to sets of events:
\begin{eqnarray*}
\phi^{-1}(S) & = & \Union\set{ \phi^{-1}(e) | e \in S}.
\end{eqnarray*}

\subsection{Threshold results for the traces model}
\label{section:tracesresults}

In this section we present the main results of our type reduction theory for
use within the traces model.  

We begin with a proposition that establishes that, provided $Proc(t)$
satisfies \Norm{} and $\textbf{RevPosConjEqT}_{\mathrm{T}}$ and given a
collapsing function $\phi$, if
\begin{iteMize}{$\bullet$}
\item $tr$ is a trace of $(Proc(t), \Gamma_{init},T)$ (for some sufficiently
large $T$),

\item $\phi(tr)\cat\trace{e}$ is a trace of $(Proc(t), \phi(\Gamma_{init}),
T)$, and

\item $e$ does not have outputs of type $t$ from outside of $\set{0 \upto
B-1}$,
\end{iteMize}
then every event that is like $e$, except with arbitrary values of inputs of
type $t$, is in $\initials(Proc(t), \Gamma_{init},T) / tr)$.
In both the statement and the proof of this proposition we take the
underlying type of all configurations to be the fixed type $T$.
\begin{prop}\label{bigtracesproposition}
Let $B$ be some natural number. Suppose that
\begin{iteMize}{$\bullet$}
\item $Proc(t)$ satisfies \Norm{} and $\textbf{RevPosConjEqT}_{\mathrm{T}}$;

\item $\phi$ is a $B$-collapsing function; and

\item $T$ is an instantiation of type $t$ of size at least $B+1$.
\end{iteMize}
Suppose further that
\begin{enumerate}[\em(i)]
\item $tr \in  \tsem{Proc(t), \Gamma_{init}}$;

\item $\phi(tr)\cat \trace{e} \in \tsem{Proc(t), \phi(\Gamma_{init})}$
with $e= c.v_1\ldots v_k$;

\item $\sigma\cat \strace{\epsilon} \in \stsem{Proc(t)}$ and
$\sigma\cat \strace{\epsilon} \generates{{\phi(\Gamma_{init})}} \phi(tr)\cat 
\trace{e}$, where $\epsilon$ is a visible symbolic event of the form
$\construct$; and 

\item $\forall i \in \mathord !^t(\epsilon) \spot v_i \in \set{0 \upto
B-1}$. 
\end{enumerate}
Then\footnote{The notation $\forall x \in X | P(x) \spot Q(x)$ is equivalent
to $\forall x \in X \spot P(x) \implies Q(x)$.}
\[
\align
\forall v' \in \set{1 \upto k} \rightarrow \Value | 
  (\forall i \in \mathord \$^t(\epsilon) \union \mathord ?^t(\epsilon) \spot
      v'_i \in T) \land 
  (\forall i \in \mathord !(\epsilon) \spot v'_i = v_i) \spot \\
\qquad tr\cat \trace{c.v'_1\ldots v'_k} \in \tsem{Proc(t),\Gamma_{init}}.
\endalign
\]
\end{prop}
\proof[Proof sketch]
By a structural induction on $Proc(t)$.  The details are in
Appendix~\ref{ssec:bigtracesproposition_proof}. 
\qed

The following example illustrates some aspects of
Proposition~\ref{bigtracesproposition}.
\begin{example}\label{example:bigtracesproposition}
Let 
\begin{eqnarray*}
Proc(t) & = & c!x\$y\from t?z \from t \then 
              \If y=z \Then d!x \then STOP \Else d\$w \from t \then STOP.
\end{eqnarray*}
Note in particular that $Proc(t)$ satisfies
$\textbf{RevPosConjEqT}_{\mathrm{T}}$.  Let $T = \set{0,1,2}$,\, $B = 1$ and
let $\phi$ be the appropriate $1$-collapsing function.  We consider four
instances.
\begin{enumerate}[(1)]
\item
Let $\Gamma_{init}(x) = 0$,\, $tr = \trace{}$,\, $e = c.0.1.2$, $\sigma =
\trace{}$ and $\epsilon = c!x\$y\from t?z \from t$.  It is easy to check
that conditions (i)--(iv) of the proposition hold.  The proposition then
implies
\[\qquad
\forall v_2', v_3' \in T \spot 
  \trace{c.0.v_2'.v_3'} \in traces(Proc(t),\Gamma_{init}),
\]
which is clearly true.

\item
Now suppose $\Gamma_{init}(x) = 2$ and again $tr = \trace{}$ and $\epsilon =
c!x\$y\from t?z \from t$.  Then condition~(ii) implies $e$ is of the form
$c.1.v_2.v_3$ for some $v_2, v_3$.  But now condition~(iv) does not hold, so
no conclusion can be reached from the proposition; and indeed
$traces(Proc(t),\Gamma_{init})$ does not include traces of the form
$\trace{c.1.v_2'.v_3'}$.  Condition~(iv) ensures that all the values $v_i$ for
$i \in \mathord!(\epsilon)$ are not collapsed within $\phi(tr) \cat
\trace{e}$.

\item
Now consider $\Gamma_{init}(x) = 0$,\, $tr = \trace{c.0.0.2}$, so $\phi(tr) =
\trace{c.0.0.1}$, and $e = d.2$,\, $\sigma = \trace{c!x\$y\from t?z
  \from t, \lnot y=z}$ and $\epsilon = d\$w \from t$.  It is easy to
check that conditions (i)--(iv) of the proposition hold.  The proposition then
implies
\[\qquad
\forall v_1' \in T \spot 
  \trace{c.0.0.2,\, d.v_1'} \in traces(Proc(t),\Gamma_{init}),
\]
which is clearly true, since the process reaches the ``$\Else$'' branch after
$\trace{c.0.0.2}$. 

\item
Now suppose $tr = \trace{c.0.1.2}$, so $\phi(tr) = \trace{c.0.1.1}$, and $e =
d.0$,\, $\sigma = \trace{c!x\$y\from t?z \from t,\linebreak[1] {y=z}}$ and
$\epsilon = d!x$.  It is easy to check that conditions (i)--(iv) of the
proposition hold.  The proposition then implies
\[\qquad
\trace{c.0.1.2,\, d.0} \in traces(Proc(t),\Gamma_{init}),
\]
which is clearly true.  This case shows the importance of
$\textbf{RevPosConjEqT}_{\mathrm{T}}$: the ``$\Else$'' branch must be able to
perform (at least) the same events as the ``$\Then$'' branch. 
\end{enumerate}
\end{example}


We will need the following definition in order to define our threshold.
\begin{defi}
We say that visible symbolic events $\epsilon$ and $\epsilon'$ are non-$t$
equivalent, written $\epsilon \nontequiv \epsilon'$, if they agree on all the
fields not of type~$t$.  For example, $c!a?t\from T \nontequiv c!a\$t\from T$,
but $c!a?t \from T \not\nontequiv c!b?t\from T$ where $a$ and~$b$ are not of
type~$t$.

We lift the relation to sequences of visible symbolic events by pointwise
application.

Finally we say that symbolic traces $\sigma$ and $\sigma'$ are non-$t$
equivalent, written $\sigma \nontequiv \sigma'$, if their restrictions to
visible symbolic events are non-$t$ equivalent.
\end{defi}

The following function returns the indices of all output variables of type~$t$
in all constructs of~$P(t)$ corresponding to a symbolic trace that is non-$t$
equivalent to $\sigma \cat \trace\epsilon$.
\begin{eqnarray*}
!^t(\sigma, \epsilon)(P(t)) & = & 
  \Union \set{!^t(\epsilon') | 
     \begin{align}
     \sigma' \cat \strace{\epsilon'} \in  \stsem{P(t)} \land 
     \sigma \cat \trace\epsilon \nontequiv \sigma' \cat \strace{\epsilon'} } .
     \end{align}
\end{eqnarray*}

\begin{example}
\label{example:traces-count}
Consider
\begin{eqnarray*}
P(t) & = & 
  \begin{align}
  in?x\from t?y\from t?z\from t \then \\
  \qquad
    \begin{align}
    \If x=y \Then (\If x=z \Then STOP \Else out!x\$w\from t \then STOP) \\
    \Else (\If x=z \Then out\$w\from t!y \then STOP 
      \Else out\$v \from t\$w\from t \then STOP) .
    \end{align}
  \end{align}
\end{eqnarray*}
Then $!^t(\strace{in?x\from t?y\from t?z\from t, \lnot x=y, \lnot x=z}, out\$v
\from t\$w\from t)(P(t)) = \set{1,2}$, since all the constructs using $out$
can be reached on a trace that is non-$t$ equivalent to $\strace{in?x\from
  t?y\from t?z\from t,\linebreak[1] {\lnot x=y},\linebreak[1] {\lnot x=z},
  out\$v \from t\$w\from t}$.

Now consider
\begin{eqnarray*}
Q(t) & = & 
  in?x\from \set{a,b}?i\from t \then 
  \If x=a \Then c?j\from t!i \then STOP \Else c!i?j \from t \then STOP,
\end{eqnarray*}
where $a$ and~$b$ are not of type~$t$.  Then $!^t(\trace{in!a?i\from t},
c?j\from t!i)(Q(t)) = \set{2}$, since the construct in the else branch cannot
be reached after a symbolic trace that is non-$t$ equivalent to
$\trace{in!a?i\from t}$.
\end{example}


We now present the first of our two main results of this paper. The following
theorem establishes a threshold $Thresh_{\mathrm{T}}$ such that if $\Spec(t)$
and $Impl(t)$ fulfil certain requirements, then, for all $B \geq
Thresh_{\mathrm{T}}$, if $\phi$ is a $B$-collapsing function, then for all
$n \geq B$
\[
\begin{align}
\mbox{if }	\Spec(\set{0\upto B}) \trefinedby \phi(Impl(\set{0\upto n}),\
\mbox{ then }	\Spec(\set{0\upto n}) \trefinedby Impl(\set{0 \upto n}).
\end{align}
\]
In Section~\ref{section:failuresresults} we will present an analogous result
for the stable failures model.


\begin{thm}[Extendibility of traces refinement of systems with replicated
components]\label{maintracesthm} 
Suppose that
\begin{enumerate}[\em(i)]
\item $\Spec(t)$ satisfies \Norm{} and
$\textbf{RevPosConjEqT}_{\mathrm{T}}$; 

\item $Impl(t)$ satisfies \textbf{TypeSym};

\item 
 $\Thresh_{\mathrm{T}}$ is the maximum number of output positions reachable on
  non-$t$ equivalent symbolic traces of $\Spec(t)$, i.e.\
\begin{eqnarray*}
\Thresh_{\mathrm{T}} & = & 
  \max\set{ \# !^t(\sigma,\epsilon)(\Spec(t)) | 
     \sigma \cat \trace\epsilon \in \stsem{\Spec(t)} };
\end{eqnarray*}

\item $B \geq \Thresh_{\mathrm{T}}$;

\item $T$ is an instantiation of type $t$ of size at least $B+1$; and

\item $\phi$ is a $B$-collapsing function.
\end{enumerate}
Then if $\Spec(\phi(T)) \trefinedby \phi(Impl(T))$, then $\Spec(T) \trefinedby
Impl(T)$. 
\end{thm}

\proof Suppose that $\Spec(\phi(T)) \refinedby_T \phi(Impl(T))$ and assume for
a contradiction that $\Spec(T) \not \refinedby_T Impl(T)$.  Consider a
shortest trace that demonstrates this non-refinement; this trace is
necessarily non-empty, so of the form $tr \cat \trace{e}$ such that
\begin{eqnarray*}
tr \cat \trace{e} & \in & \tsem{Impl(T)}, \\
tr & \in & \tsem{\Spec(T)}, \\
tr \cat \trace{e} & \not \in & \tsem{\Spec(T)}.
\end{eqnarray*}
Suppose that $e = c.v_1\ldots v_k$.  Suppose $tr$ is generated by symbolic
trace $\sigma_1$ of $Spec(t)$.  We can construct a symbolic event $\epsilon_1
= c\S x_1\mathord{:}X_1 \ldots \S x_k\mathord{:}X_k$ to generate~$e$ (although
$\sigma_1 \cat \trace{\epsilon_1}$ might not be a symbolic trace
of~$Spec(t)$): if $v_i$ is of type~$t$ we set $\S x_i\mathord{:}X_i$ to
$\$x_i\mathord{:}t$; otherwise we set $\S x_i\mathord{:}X_i$ to $!v_i
\mathord{:} null$.



By assumptions~(iii) and~(iv), 
$\# \mathord!^t(\sigma_1, \epsilon_1)(Spec(t)) \le B$,
so let $\pi:
T \rightarrow T$ be a bijection that maps $\set{v_i |
i \in \mathord!^t(\sigma_1, \epsilon_1)(Spec(t))}$ 
into $\set{0 \upto B-1}$.
%
%
By Corollary~\ref{corollary:DIimpliesTypeSym} $\Spec(t)$ satisfies
\textbf{TypeSym}, and by assumption $Impl(t)$ satisfies \textbf{TypeSym}, so
by Remark~\ref{symtraceremark} we have that
\begin{eqnarray}
\pi(tr)\cat \trace{\pi(e)} & \in & \tsem{Impl(T)}, \nonumber
\\
\pi(tr) & \in & \tsem{\Spec(T)}, \label{eqn:condi}
\\
\pi(tr)\cat \trace{\pi(e)} & \not \in & \tsem{\Spec(T)}. \nonumber
\end{eqnarray}
Hence,
\begin{eqnarray*}
\phi(\pi(tr))\cat \trace{\phi(\pi(e))} & \in & \tsem{\phi(Impl(T))}.
\end{eqnarray*}
But $\Spec(\phi(T)) \trefinedby \phi(Impl(T))$, so
\begin{eqnarray*}\label{equation:maintracesthmeq1}
\phi(\pi(tr))\cat \trace{\phi(\pi(e))} & \in & \tsem{\Spec(\phi(T))}.
\end{eqnarray*}
However, by Corollary~\ref{corollary:adataindepimplication}, $\Spec(T) \trefinedby \Spec(\phi(T))$, so
\begin{eqnarray}\label{equation:maintracesthmeq2}
\phi(\pi(tr))\cat \trace{\phi(\pi(e))} & \in & \tsem{\Spec(T)}.
\end{eqnarray}

We can now apply Proposition~\ref{bigtracesproposition} to $Spec(T)$, with
$\pi(tr)$ in place of~$tr$, and $\phi(\pi(e))$ in place of~$e$: condition~(i)
is satisfied, by (\ref{eqn:condi}); condition~(ii) is satisfied,
by~(\ref{equation:maintracesthmeq2}); condition~(iii) is satisfied by taking a
suitable choice of~$\sigma$ to generate $\phi(\pi(tr))$, and taking $\epsilon$
to be the symbolic event that generates $\phi(\pi(e))$; condition~(iv) is
satisfied since
$!^t(\epsilon) \subseteq  \mathord!^t(\sigma_1, \epsilon_1)(Spec(t))$ (since
$\sigma \cat \trace\epsilon \nontequiv \sigma_1 \cat \trace{\epsilon_1}$),
and by
construction, all the corresponding fields of
$\phi(\pi(e))$ are in $\set{0 \upto B-1}$.  
Considering the valuation $v'$
such that $\pi(e) = c.v_1' \ldots v_k'$ then  allows us to deduce that
\begin{eqnarray*}
\pi(tr)\cat \trace{\pi(e)} & \in & \tsem{\Spec(T)}.
\end{eqnarray*}
This is a contradiction, which completes our proof. 
\qed


\begin{remark}
For every verification problem, the value of $\Thresh_{\mathrm{T}}$ in
Theorem~\ref{maintracesthm} depends only on the specification.  
\end{remark}

\begin{example}
Recall the process syntax $P(t)$ from Example~\ref{example:traces-count}.  We
argued there that $!^t(\strace{in?x\from t?y\from t?z\from t, \lnot x=y,
  \lnot x=z}, out\$v \from t\$w\from t)(P(t)) = \set{1,2}$.  Clearly 
$!^t(\sigma, \epsilon)(P(t))$ has fewer elements for other values of~$\sigma$
and~$\epsilon$.  Hence $\Thresh_{\mathrm{T}} = 2$ in this case.
\end{example}


If $\Spec(t)$ contains no conditional choices, then we can obtain a simpler
expression for the threshold.  
\begin{prop}\label{prop:noEqT-threshold}
If $\Spec(t)$ contains no conditional choices then
\begin{eqnarray*}
\Thresh_{\mathrm{T}} & \le & 
 \max\set{\# !^t(\alpha) | \mbox{$\alpha$ is a construct of $\Spec(t)$}}.
\end{eqnarray*}
\end{prop}
The proof is in Appendix~\ref{sec:traces_proofs}, and shows that in this case
there is a \emph{unique} construct that contributes towards the calculation of
each $!^t(\sigma,\epsilon)(\Spec(t))$. 

If $\Spec(t)$ uses a conditional, then there may be two such constructs, but
by Lemma~\ref{lemmaD},
$!^t(\alpha_{then}) \supseteq \mathord!^t(\alpha_{else})$, where
$\alpha_{then}$ and $\alpha_{else}$ are the constructs in the ``then'' and
``else'' branches, respectively; hence the above equality still holds.  It's
only when $Spec(t)$ contains nested conditionals, as in
Example~\ref{example:traces-count}, that one needs to consider multiple
constructs together.


\begin{remark}\label{remark:finitecheckingforthresholdtraces}
For all specifications with a finite SSLTS, the value of
$\Thresh_{\mathrm{T}}$ in Theorem~\ref{mainfailuresthm} can be calculated in a
finite amount of time.  All states that can be reached by non-$t$ equivalent
traces need to be considered together; this can be performed by a process
similar to normalisation~\cite[Appendix~C]{Roscoe:1997}.
\end{remark}


\begin{example}\label{example:CA-traces}
Recall the example from Section~\ref{sec:example}.  Earlier, we explained how
to use counter abstraction techniques from~\cite{Tomasz-thesis, Mazur:2010} to
show 
\begin{eqnarray*}
Spec(\phi(T)) & \trefinedby & \phi(Impl(T)), 
  \qquad \mbox{for all instantiations~$T$ of~$t$  with $\#T \ge 3$,}
\end{eqnarray*}
where
\begin{eqnarray*}
Spec(t) & = & enterCS\$i\mathord:t \then leaveCS!i \then Spec(t),
\end{eqnarray*}
and where we took $B = 1$.  We can now apply Theorem~\ref{maintracesthm}.
It's clear that condition~(i) holds.  Condition~(ii) holds from the discussion
in Example~\ref{example:system-typesym}.  From condition~(iii) we obtain
$Thresh_{\mathrm{T}} = 1$, essentially because $Spec(t)$ contains a
single~``$!$''; hence condition~(iv) holds.  Condition~(v) gives a lower bound
of~2 on the size of~$T$, which is a weaker condition than we have already
imposed.  Finally condition~(vi) holds by construction.  Hence we can apply
the theorem to deduce
\begin{eqnarray*}
Spec(T) & \trefinedby & Impl(T), 
  \qquad \mbox{for all instantiations~$T$ of~$t$  with $\#T \ge 3$.}
\end{eqnarray*}
Smaller values of~$T$ can be verified directly. 
\end{example}

In \cite{Mazur:2010}, we describe tool support, called TomCAT, for our counter
abstraction techniques.  In particular, the tool checks the conditions of
Theorem~\ref{maintracesthm}, and calculates the
threshold~$Thresh_{\mathrm{T}}$.  This part of the tool could easily be
adapted to other abstraction techniques that build on the type reduction
theory of this paper.

\subsection{Threshold results for the stable failures model}
\label{section:failuresresults}

In this section we present type reduction theory results analogous to those in
Section~\ref{section:tracesresults}, but extended to the stable failures
model. 

We begin with a proposition that shows that, provided $Proc(t)$
satisfies \Norm{} and $\textbf{RevPosConjEqT}_{\mathrm{F}}$ and given a collapsing function
$\phi$, if $tr$ is a trace of $(Proc(t), \Gamma_{init},T)$ (for some
sufficiently large $T$), $(\phi(tr), X)$ is a failure of
$(Proc(t), \linebreak[1]\phi(\Gamma_{init}), T)$ and events in
$\initials(Proc(t), \Gamma_{init}, T) / tr$ do not have outputs of type $t$
from outside $\set{0 \upto B-1}$, then $(tr, X)$ is a failure of
$(Proc(t), \Gamma_{init},T)$. 
In this proposition we assume that the underlying type of all configurations
is the fixed type $T$.
\begin{prop}\label{bigfailuresproposition}
Let $B$ be some natural number. Suppose that 
\begin{iteMize}{$\bullet$}
\item $Proc(t)$ satisfies \Norm{} and $\textbf{RevPosConjEqT}_{\mathrm{F}}$;

\item $\phi$ is a $B$-collapsing function; and

\item $T$ is an instantiation of type $t$ of size at least $B+1$.
\end{iteMize}
Suppose further that
\begin{enumerate}[\em(i)]
\item $tr \in  \tsem{Proc(t), \Gamma_{init}}$;

\item $(\phi(tr), X) \in \fsem{Proc(t), \phi(\Gamma_{init})}$; and

\item if $P$ is a configuration such that
$(Proc(t), \Gamma_{init}) \Trans[tr] P$, then every output value of type~$t$
of every event  in $\initials(P)$ is in $\set{0 \upto B-1}$.  
\end{enumerate}
Then $(tr, X) \in \fsem{Proc(t), \Gamma_{init}}$.
\end{prop}
\proof[Proof sketch]
By a structural induction on $Proc(t)$.  The details are in
Appendix~\ref{ssec:bigfailuresproposition_proof}. 


The following example illustrates some aspects of
Proposition~\ref{bigfailuresproposition}.
\begin{example}
Recall the following process from Example~\ref{example:bigtracesproposition}: 
\begin{eqnarray*}
Proc(t) & = & c!x\$y\from t?z \from t \then 
              \If y=z \Then d!x \then STOP \Else d\$w \from t \then STOP.
\end{eqnarray*}
Note in particular that $Proc(t)$ satisfies
$\textbf{RevPosConjEqT}_{\mathrm{F}}$.  Let $T = \set{0,1,2}$,\, $B = 1$ and
let $\phi$ be the appropriate $1$-collapsing function.  We consider four
instances.
\begin{enumerate}[(1)]
\item
Let $\Gamma_{init}(x) = 0$,\, $tr = \trace{}$ and $X = \eset{c} -
\eset{c.0.1}$.  Condition~(i) of the proposition clearly holds.
Condition~(ii) holds, considering the case that the nondeterministic selection
picks $y = 1$.  Condition~(iii) holds since the only output value in an event
after~$tr$ is the value~0 for~$x$.  The proposition then implies $(tr, X) \in
\fsem{Proc(t), \Gamma_{init}}$, which is clearly true, considering the case
that the nondeterministic selection again picks $y = 1$.

\item
Now suppose $\Gamma_{init}(x) = 2$,\, $tr = \trace{}$ and $X = \eset{c.2}$.
Condition~(i) clearly holds; condition~(ii) holds since the environment
$\phi(\Gamma_{init})$ maps $x$ to~$1$.  However, condition~(iii) does not
hold, since the initial configuration can output~2 for~$x$.  And indeed
$(tr,X) \nin \fsem{Proc(t), \Gamma_{init}}$, since for some $y \in T$ the
event $c.2.y.0$ will be available.  Condition~(iii) ensures that the output
values in initial events after~$tr$ are not collapsed. 

\item
Now consider $\Gamma_{init}(x) = 0$,\, $tr = \trace{c.0.0.2}$, so $\phi(tr) =
\trace{c.0.0.1}$, and $X = \eset{c} \union \set{d.v | v \ne 2}$.
Conditions~(i) and~(iii) clearly hold.  Condition~(ii) holds, since
after~$\phi(tr)$ the process takes the ``$\Else$'' branch and can select $w =
2$.  The proposition then implies $(tr, X) \in \fsem{Proc(t), \Gamma_{init}}$,
which is clearly true, since after~$tr$ the process again takes the
``$\Else$'' branch and can select $w = 2$.

\item
Now suppose $tr = \trace{c.0.1.2}$, so $\phi(tr) = \trace{c.0.1.1}$, and $X =
\eset{c} \union \set{d.v | v \ne 0}$.  It is easy to check the three
conditions; note that for condition~(ii), the failure corresponds to the
``$\Then$'' branch.  The proposition then implies $(tr, X) \in \fsem{Proc(t),
  \Gamma_{init}}$, which is clearly true, since after~$tr$ the process
takes the ``$\Else$'' branch and can select $w = 0$.  This case shows the
importance of $\textbf{RevPosConjEqT}_{\mathrm{F}}$: the ``$\Else$'' branch
must have (at least) all the failures of the ``$\Then$'' branch.
\end{enumerate}
\end{example}


The following theorem is our second key result of this paper. It extends
Theorem~\ref{maintracesthm} to the stable failures model by establishing a
threshold~$\Thresh$ such that if $\Spec(t)$ and $Impl(t)$ fulfil certain
requirements, then, for all $B \geq \Thresh$, if $\Spec(\set{0\upto
B}) \frefinedby \phi(Impl(\set{0\upto n})$ then $\Spec(\set{0\upto
n}) \frefinedby Impl(\set{0 \upto n})$ for all $n \geq B$. 


Recall that, given a symbolic conditional event $cond$ and an
environment~$\Gamma$, $\eval{cond}{\Gamma}$ denotes the truth value of the
proposition obtained from $cond$ by substituting all free variables of type
$t$ with their corresponding values contained within~$\Gamma$.  We lift the
definition of $\eval{\cdot}{\Gamma}$ to symbolic traces without visible
symbolic events in the following way. Given a symbolic trace $\sigma$ in
$(Cond \union \set{\tau})^*$ we let $\eval{\sigma}{\Gamma}$ be equal to
$\bigwedge \set{\eval{cond}{\Gamma} | cond \In \sigma, cond \in  Cond}$. 

\begin{thm}[Extendibility of stable failures refinement of systems with
replicated components]\label{mainfailuresthm} 
Suppose that
\begin{enumerate}[\em(i)]
\item $\Spec(t)$ satisfies \Norm{} and
$\textbf{RevPosConjEqT}_{\mathrm{F}}$, and is divergence-free and has a finite
alphabet for every finite instantiation of type $t$;

\item $Impl(t)$ satisfies \textbf{TypeSym};

\item no construct $\alpha$ in $\Spec(t)$ combines nondeterministic
inputs of type $t$ and deterministic input of any type, i.e.\ if
$\#\$^t(\alpha) > 0$, then $\#?(\epsilon) = 0$;

\item $T$ is an instantiation of type $t$ such that $\#T \geq B+1$,
where $B$ is as below;

\item 
$\align 
\Thresh =  \max\{
  \begin{align}
  Thresh_{\mathrm{T}}, \\
  \max\{
    \align
    \Thresh_{!^t}(\sigma,\Gamma) + \Thresh_{?^t}(\sigma,\Gamma) |\\
    \qquad \align
           \sigma \in \stsem{Spec(t)}\\
	   \land (\sigma = \strace{} \lor \mathit{last}(\sigma) \in \Visible)
           \land \Gamma \in \mathit{Env}(T)\} \},
	   \endalign
    \endalign
  \end{align}
\endalign$ \\
where
\begin{iteMize}{$\bullet$}
\item 
$\Thresh_{!^t}(\sigma,\Gamma)$ counts the number of unique output variables of
type $t$ in all the visible symbolic events $\epsilon$ available in $\Spec(t)$
immediately after~$\sigma$ such that all conditionals between the last
symbolic event of $\sigma$ and $\epsilon$ evaluate to $\True$, i.e. 
\[\hspace*{\mathindent} 
\align
\Thresh_{!^t}(\sigma,\Gamma) =\\
\qquad \# \{x_i | 
  \align
  \Spec(t) \SymTrans{\sigma}\SymTrans{\rho}\symtrans{\epsilon}\ \land
     \rho \in (Cond \union \set{\tau})^* \land \eval{\rho}{\Gamma} = \True \\
  \land \epsilon = \construct \in \Visible \land 
     i \in \mathord !^t(\epsilon)\};
  \endalign
\endalign
\]

\item 
$\Thresh_{?^t}(\sigma,\Gamma)$ counts the number of (not necessarily unique)
input variables of type $t$ in all the visible symbolic events $\epsilon$
available in $\Spec(t)$ immediately after $\sigma$ such that all conditionals
between the last symbolic event of $\sigma$ and $\epsilon$ evaluate to
$\True$, i.e. 	
\[ 
\hspace*{\mathindent}
\Thresh_{?^t}(\sigma,\Gamma)  \;=\; \sum \{ \#?^t(\epsilon) | 
\align
	\Spec(t) \SymTrans{\sigma}\SymTrans{\rho}\symtrans{\epsilon}\ \land \rho \in (Cond \union \set{\tau})^*\\
	\land \epsilon \in \Visible \land \eval{\rho}{\Gamma} = \True\};
\endalign
\] 

\item
$Thresh_{\mathrm{T}}$ is as in Theorem~\ref{maintracesthm};  
\end{iteMize}

\item $B \geq \Thresh$; and

\item $\phi$ is a $B$-collapsing function.
\end{enumerate}
Then, if $\Spec(\phi(T)) \frefinedby \phi(Impl(T))$, then
$\Spec(T) \frefinedby Impl(T)$. 
\end{thm}

\proof
Suppose that the refinement $\Spec(\phi(T)) \frefinedby \phi(Impl(T))$ holds
and assume for a contradiction that $\Spec(T) \not \frefinedby
Impl(T)$. Refinement in the stable failures model implies refinement in the
traces model, so $\Spec(\phi(T)) \trefinedby \phi(Impl(T))$. Then, by
Theorem~\ref{maintracesthm} (which is applicable since its assumptions are
weaker than those of this theorem), $\Spec(T) \trefinedby Impl(T)$. 

Consider a
minimal counterexample $(tr,X)$ to the refinement $\Spec(T) \frefinedby
Impl(T)$, i.e.\ 
\begin{eqnarray}
\label{secondassumption}
(tr,X) & \in & \fsem{Impl(T)},
\\
\label{thirdassumption}
(tr,X) & \not \in & \fsem{\Spec(T)},
\\
\label{Xminimality}
\forall e \in X  \spot (tr,X\setminus\set{e}) & \in & \fsem{\Spec(T)}.
\end{eqnarray}
Observe that there is such a minimal counterexample since we have assumed that
specifications are divergence-freedom and have finite alphabets.  

Combining (\ref{thirdassumption}) and (\ref{Xminimality}) we obtain that for all events $e$ in $X$ there exists a state $P_e(T)$ such that
\begin{equation}\label{refuseandaccept}
\Spec(T) \Trans[tr] P_e(T) \;\land\; 
P_e(T) \refuses (X) \setminus \set{e} \;\land\; P_e(T)\trans[e].
\end{equation}
This also means that every event in $X$ is accepted in some stable state of
$\Spec(T)$ after $tr$. Hence,
\begin{eqnarray}\label{Xunionainitials}
X & \subseteq & \initials (\Spec(T)/tr).
\end{eqnarray}

We now aim to show that $X$ is dependent upon at most $Thresh$
values from $T$, in a sense that we make precise below.  We begin with two
properties of~$X$. 
\begin{enumerate}[(1)]
\item 
\label{clause:clause1}
Firstly, we prove that $X$ is closed under type $t$ nondeterministic
inputs of the specification, i.e.\ we suppose that $e = c.v_1\ldots v_k \in X$
matches a construct $\alpha$ of $\Spec(t)$ (uniqueness follows from
Proposition~\ref{corollaryA2}) with $\#\$^t(\alpha) > 0$  (which, by
assumption~(iii), implies that $\#?(\alpha) = 0$) and show that  
\begin{equation}\label{nondeterministicinputsclosure}
\hspace{\mathindent}
\align
\forall v' : \set{1 \upto k} \rightarrow \Value | \\
\qquad (\forall i \in \mathord \$^t(\alpha) \spot v'_i \in T) 
    \land (\forall i \in \set{1 \upto k} \setminus \mathord \$^t(\alpha) \spot
      v'_i = v_i) \spot\\
\qquad \qquad c.v'_1\ldots v'_k \in X.
\endalign
\end{equation}
Let $v'$ be as in (\ref{nondeterministicinputsclosure}) and let $e' =
c.v'_1\ldots v'_k$. Assume for a contradiction that $e'$ is not an event in
$X$. Consider the same behaviour that leads to the stable state $P_e(T)$
of~(\ref{refuseandaccept}) where $X\setminus\set{e}$ is refused and $e$ is
accepted, except that the nondeterministic selections of $\alpha$ are resolved
in a way such that the values $v'_i$ are chosen instead of $v_i$ for all
$i \in \mathord \$^t(\alpha)$; call this stable state $P_{e'}(T)$ (see
Figure~\ref{figure:mainfailurestheoremexample} for an example). The initials
of $P_{e'}(T)$ are the same\footnote{This would not be true if specifications
could contain constructs that combine nondeterministic selections over type
$t$ and deterministic inputs over any type; see
Example~\ref{example:mainfailurestheoremexample1} and
Example~\ref{example:mainfailurestheoremexample2} below.} as those
of~$P_e(T)$, except they contain $e'$ instead of~$e$. Therefore, since
$X \setminus \set{e}$ is refused in $P_e(T)$, $X \setminus \set{e'}$ must be
refused in $P_{e'}(T)$. However, $e' \not \in X$ by assumption, so $X$ is
refused in $P_{e'}(T)$, which contradicts (\ref{thirdassumption}).

\begin{figure}[t]
\centering
\psfrag{tau}{\small$\tau$}
\psfrag{e=chn.a.0}{\small$e=chn.a.0$}
\psfrag{e'=chn.a.1}{\small$e'=chn.a.1$}
\psfrag{chn.a.2}{\small$chn.a.2$}
\psfrag{chn.b.0}{\small$chn.b.0$}
\psfrag{chn.b.1}{\small$chn.b.1$}
\psfrag{chn.b.2}{\small$chn.b.2$}
\psfrag{s}{\scriptsize\hspace{-0.2em}\raisebox{-0.3ex}{$P_e(T)$}}
\psfrag{s'}{\scriptsize\hspace{-0.4em}\raisebox{0.4ex}{$P_{e'}(T)$}}
\includegraphics[width=0.5\textwidth]{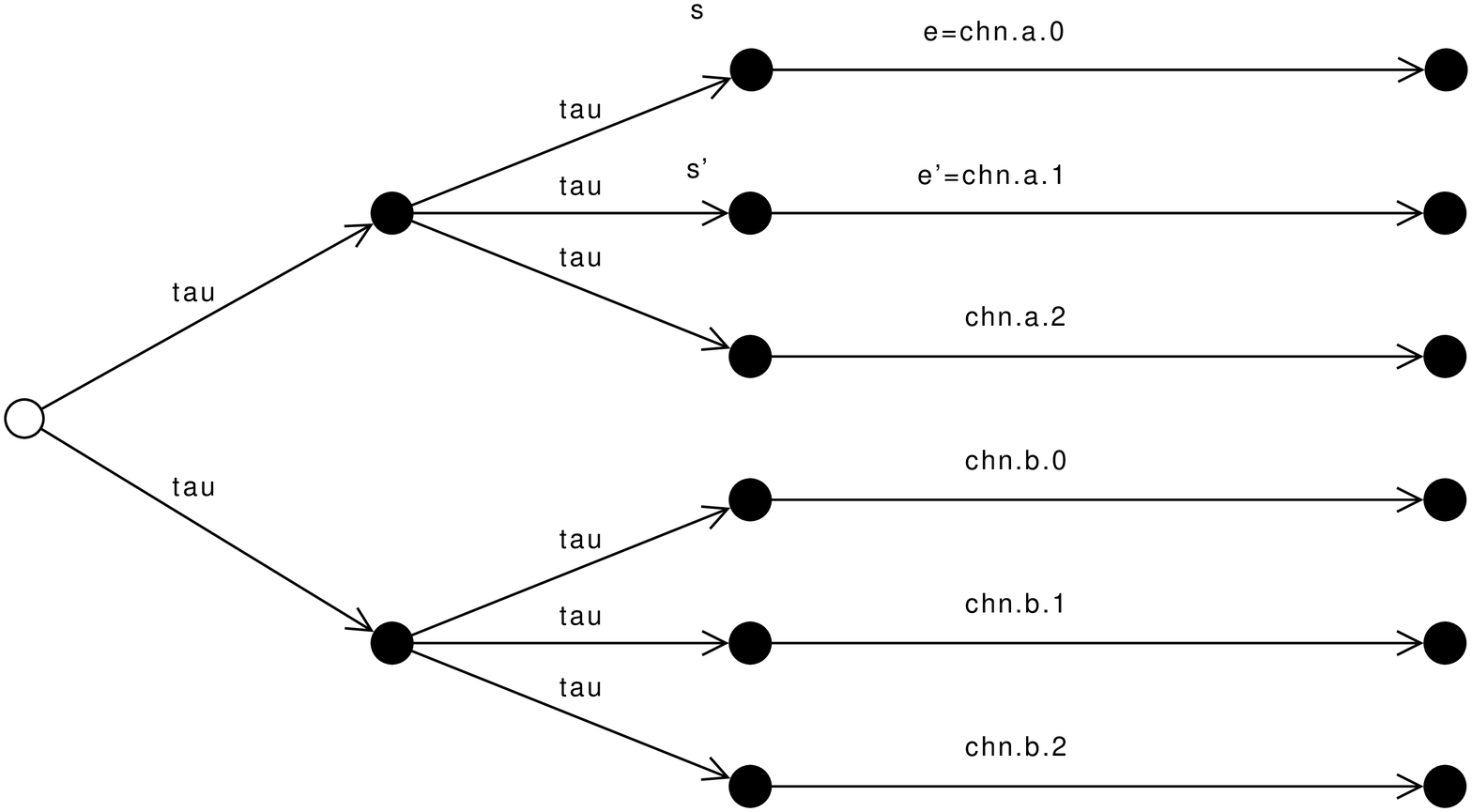}
\caption{The LTS of \protect $Proc(\set{0,1,2})$, where \protect $Proc(t) =
chn\$x\mathrm{:}\set{a,b}\$y\mathrm{:}t \then
STOP$.\label{figure:mainfailurestheoremexample}}  
\end{figure}

\item
\label{clause:clause2} Secondly, we show that $X$ contains no pairs of
events that differ only in values of deterministic inputs of any type, i.e.\
we suppose that $e = c.v_1\ldots v_k \in X$ is an event matching a 
construct $\alpha$ of $\Spec(t)$ (uniqueness guaranteed by
Proposition~\ref{corollaryA2}) with $\#?(\alpha) > 0$ (which, by
assumption~(iii), implies that $\#\$^t(\alpha) = 0$) and show that 
\begin{equation}\label{noduplicatesofinputs}
\hspace{\mathindent}\align
\forall v': \set{1 \upto k} \rightarrow \Value |\\
\qquad 
  (\forall i \in \set{1 \upto k} \setminus \mathord ?(\alpha) \spot
    v'_i = v_i) 
  \land (\exists i \in \mathord ?(\alpha) \spot v'_i \not = v_i) \spot\\
\qquad \qquad c.v'_1\ldots v'_k \not \in X.
\endalign
\end{equation}
Let $v'$ be as in (\ref{noduplicatesofinputs}) and assume for a contradiction that $e' = c.v'_1\ldots v'_k$ is an event in $X$. Consider the state $P_e(T)$ of (\ref{refuseandaccept}) where $X \setminus \set{e}$ is refused and $e$ is available. Clearly $e'$ is refused in this state, since we assumed it to be in $X$ and hence in $X\setminus\set{e}$ (since $e \not = e'$). This is a contradiction since $e$ and $e'$ differ only in the values of deterministic inputs and hence $e$ is available if and only if $e'$ is.
\end{enumerate}

Recall that $tr$ is a trace of $\Spec(T) = (\Spec(t), \nullset, T)$; let
$(Spec'(t),\Gamma,T)$ be the unique resulting configuration (with uniqueness
following from Proposition~\ref{prop:environmentuniqueness}); i.e.\
\begin{eqnarray}\label{equation:specperformstr'sequence} 
	(\Spec(t),\nullset,T) & \mapstotrans[s] & (Spec'(t),\Gamma,T),
\end{eqnarray}
for some sequence of concrete events $s$ such that $s$ does not end with a
$\tau$ and $s \hide \set{\tau} = tr$.  Observe that 
\begin{eqnarray*}
(\Spec(t),\nullset,T) & \Trans[tr] & (\Spec'(t),\Gamma,T).
\end{eqnarray*}
So, thanks to the uniqueness of $\Spec'(t)$ and $\Gamma$, and since $s$ does
not end with a $\tau$, 
\begin{eqnarray*}\label{equation:initialsofspecaftertr}
\initials(\Spec(T) / tr) & = & \initials(\Spec'(t), \Gamma, T).
\end{eqnarray*}

Let $S = S_1 \union S_2$, where
\begin{eqnarray}
\label{equation:definitionofs1}
S_1 & = & \{v_i |
  \align
  e = c.v_1\ldots v_k \in \initials(\Spec(T)/tr)\\
  \land e\mbox{ matches  construct }\alpha\mbox{ of }\Spec(t)
    \land i \in \mathord !^t(\alpha)\},
\endalign
\\
\label{equation:definitionofs2}
S_2 & = & \{v_i |
  \align
  c.v_1\ldots v_k \in X \land i \in \set{1\upto k} \land v_i \in T\\
  \land \exists v' \in T \spot 
    c.v_1\ldots v_{i-1}.v'.v_{i+1}\ldots v_k \not \in X\}.
  \endalign
\end{eqnarray}
We now show that $B$ is an upper bound on $\#S$. 

Firstly, we deduce from the closure of~$X$ under type $t$ nondeterministic
inputs of the specification (clause 1, above) that $S_2$ contains no values of
type $t$ that come from nondeterministic selections of constructs of
$\Spec(t)$. Hence and by (\ref{Xunionainitials}), $S_2$ is a subset of the set
of values of type $t$ in the events in $\initials(\Spec(T)/tr) =
\initials(\Spec'(t), \Gamma, T)$ that come from deterministic inputs and
outputs only. Observe that if a concrete event of the specification is
obtained from a symbolic event using the translation rules of COSE, then all
the preceding conditional symbolic events have to evaluate to $\True$ in
appropriate environments. Also, all conditional symbolic events occurring
between any two visible symbolic events are always evaluated within the same
environment. Therefore, when working out $\initials(Spec'(t), \Gamma, T)$, we
can ignore those initial visible symbolic events of $Spec'(t)$ that are
preceded by a conditional symbolic event that evaluates to $\False$ in
$\Gamma$. Finally, from (\ref{equation:specperformstr'sequence}) and the
translation rules of COSE (see Section~\ref{section:translationrules}) we have
that there exists $\sigma \in \stsem{\Spec(t)}$ such that either $\sigma =
\strace{}$ or $\mathit{last}(\sigma) \in \Visible$ and $\Spec(t)
\SymTrans{\sigma} \Spec'(t)$ (so $\sigma \generates{\set{}} s$). Hence, the
number of type~$t$ values matched by deterministic inputs in the events in
$S_2$ is at most
\begin{eqnarray*}
\Thresh_{?^t}(\sigma,\Gamma) & = &
  {\textstyle\sum} \{ \#?^t(\epsilon) | 
	\align
	\Spec(t) \SymTrans{\sigma}\SymTrans{\rho}\symtrans{\epsilon}\ \land \rho \in (Cond \union \set{\tau})^*\\
	\land \epsilon \in \Visible \land \eval{\rho}{\Gamma} = \True\}.
	\endalign
\end{eqnarray*}

Now, since we assumed no constants of type $t$ in the definition of
$\Spec(t)$, any type~$t$ output value in $S$ must come from some
environment. Therefore, the total number of output values of type $t$ in $S$
can never be greater than the total number of different output variable names
used in the constructs of $\Spec(t)$ that are matched by the members of
$\initials(\Spec(T)/tr)$, i.e.~the total number of output values of type~$t$
in~$S$ is at most 
\begin{eqnarray*}
\Thresh_{!^t}(\sigma,\Gamma) & = &
  \# \{x_i |
 	\align
	\Spec(t) \SymTrans{\sigma}\SymTrans{\rho}\symtrans{\epsilon}\ 
          \land \rho \in (Cond \union \set{\tau})^* 
          \land \eval{\rho}{\Gamma} = \True \\
	\land \epsilon = \construct \in \Visible
           \land i \in \mathord !^t(\epsilon)\}.
	\endalign
\end{eqnarray*}

Summarising the last two paragraphs: all elements of $S$ either match
deterministic inputs in the events in $S_2$ (at most
$\Thresh_{?^t}(\sigma,\Gamma)$ such values), or match outputs in either~$S_1$
or $S_2$ (at most $\Thresh_{!^t}(\sigma,\Gamma)$ such values).  Therefore,
\[
\begin{array}{cl}
& \#S\\
\leq & 
\Thresh_{!^t}(\sigma,\Gamma) + \Thresh_{?^t}(\sigma,\Gamma)\\
\leq & \max\{ \Thresh_{!^t}(\sigma',\Gamma') + \Thresh_{?^t}(\sigma',\Gamma') |\
	\align
		\sigma' \in \stsem{Proc(t)}\\
		\land (\sigma' = \strace{} \lor last(\sigma') \in \Visible)\\
		\land \Gamma' \in \mathit{Env}(T)\}
	\endalign\\
\leq & \Thresh\\
\leq & B.\\
\end{array}
\]

Let $\pi : T \rightarrow T$ be a bijection that maps $S$ into $\set{0 \upto
B-1}$. Then, using Remark~\ref{symfailureremark}, we can infer from
(\ref{secondassumption}) and (\ref{thirdassumption}) that  
\begin{eqnarray}
\label{secondassumption'}
(\pi(tr),\pi(X)) & \in & \fsem{Impl(T)},
\\
\label{thirdassumption'}
(\pi(tr),\pi(X)) & \not \in & \fsem{\Spec(T)}.
\end{eqnarray}

Now, by the denotational semantics of renaming~\cite{Roscoe:1997}, 
\begin{equation}\label{equation:defoffailuresofphiofimpl}
\align
\fsem{\phi(Impl(T))} = 
  \set{(\phi(tr'),Y) | (tr', \phi^{-1}(Y)) \in \fsem{Impl(T)}}.
\endalign
\end{equation}
Let $c.v_1\ldots v_k$ be an event in $X$. Let $i$ be in $\set{1 \upto k}$ such
that $v_i$ is of type $t$. Our construction of $S$ implies that either
(1)~$v_i$ is in $S$, in which case $\pi(v_i)$ is in $\set{0 \upto B-1}$, or
(2)~$v_i$ matches a nondeterministic input of type $t$, in which case the
closure of $X$ under nondeterministic inputs of the specification (clause 1 on
p.~\pageref{clause:clause1}) implies that for all values $v'$ in $T$,
$c.v_1 \ldots v_{i-1}.v'.v_{i+1} \ldots v_k$ is in $X$. Therefore, if
$c.w_1 \ldots w_k$ is an event in $\pi(X)$, then for every $i$ in
$\set{1 \upto k}$ such that $w_i$ is of type~$t$, we have that either
(1)~$w_i$ is in $\set{0 \upto B-1}$, or (2)~$c.w_1 \ldots
w_{i-1}.w'.w_{i+1} \ldots w_k$ is in $\pi(X)$ for all values $w'$ in
$T$. This, thanks to the definition of $\phi$, means that
\(\forall e \in \phi(\pi(X)) \spot \phi^{-1}(e) \subseteq \pi(X) \).
This implies that
\(
	\phi^{-1}(\phi(\pi(X))) \subseteq \pi(X)
\),
which trivially implies that
\[
	\phi^{-1}(\phi(\pi(X))) = \pi(X).
\]
Combining with (\ref{secondassumption'}) and
(\ref{equation:defoffailuresofphiofimpl}),
we get that  
\begin{eqnarray*}
(\phi(\pi(tr)),\phi(\pi(X))) & \in & \fsem{\phi(Impl(T))}.
\end{eqnarray*}
However, $\Spec(\phi(T))\frefinedby \phi(Impl(T))$, so
\begin{eqnarray}\label{phifaileqn} 
(\phi(\pi(tr)),\phi(\pi(X))) & \in & \fsem{\Spec(\phi(T))}.
\end{eqnarray}

We now show that
\begin{eqnarray}\label{failureslemmaequation1}
(\phi(\pi(tr)),\pi(X)) & \in & \fsem{\Spec(T)}.
\end{eqnarray}
Firstly, (\ref{phifaileqn}) implies that there exists a configuration $(P(t),\Gamma, \phi(T))$ such that
\begin{equation}\label{equation:mainfailurestheoremeq0.1}
\Spec(\phi(T)) \; = \; (\Spec(t), \set{}, \phi(T)) 
  \; \Trans(5)[\phi(\pi(tr))] \; (P(t),\Gamma, \phi(T))
  \; \refuses \; \phi(\pi(X)).
\end{equation}
Let $e = c.v_1 \ldots v_k$ be in $\phi(\initials(P(t), \Gamma, T))$ and let
$\alpha$ be the unique construct of $\Spec(t)$ that $e$ matches (with
uniqueness following from Proposition~\ref{corollaryA2}). Then there exists a
function $v' : \set{1 \upto k} \rightarrow \Value$ such that 
\begin{equation}\label{equation:mainfailurestheoremeq1}
	e' \; = \; c.v'_1 \ldots v'_k \; \in \; \initials(P(t), \Gamma, T)
\end{equation}
and $\phi(e') = e$, i.e.
\begin{equation}\label{equation:mainfailurestheoremeq2}
	\forall i \in \set{1 \upto k} \spot \phi(v'_i) = v_i.
\end{equation}
We know that $(P(t),\Gamma, \phi(T))$ must be a stable state, as otherwise it
would not be able to refuse $\phi(\pi(X))$. This means that all values of
nondeterministic selections of constructs that generate the events in
$\initials(P(t),\Gamma, \phi(T))$ had been chosen before this state was
reached (and are necessarily in $\phi(T)$). Hence, and since
$\Gamma \in \mathit{Env}(\phi(T))$ implies 
$\Gamma \in \mathit{Env}(T)$, we have that $\initials(P(t), \Gamma, \phi(T))$
and $\initials(P(t), \Gamma, T)$ are the same, except for values of
deterministic inputs of type $t$. Formally,
\begin{equation}\label{equation:mainfailurestheoremeq3}
\begin{align}
\initials(P(t),\Gamma, \phi(T)) = \\
\qquad  \{c.w_1\ldots w_k | 
  \align
  c.w'_1\ldots w'_k \in \initials(P(t),\Gamma,T)
    \mbox{ matches construct $\alpha'$} \land \\
   w \in \set{1\upto k} \rightarrow \Value \land 
   (\forall i \in \set{1\upto k} \setminus \mathord ?^t(\alpha') \spot
       w_i = w'_i) \land \\
   \forall i \in \mathord ?^t(\alpha') \spot w_i \in \phi(T)\}.
   \endalign
\end{align}
\end{equation}
Also, since $\Gamma \in Env(\phi(T))$, all values of type $t$ used in the
events of $\initials(P(t), \Gamma, T)$ and that match nondeterministic
selections or outputs, are in $\phi(T) = \set{0 \upto B}$:
\[
	\forall i \in \mathord \$^t(\alpha) \union \mathord !^t(\alpha) \spot v'_i \in \set{0 \upto B},
\]
which, thanks to the properties of $\phi$, 
and combined with the fact that for all non-$t$ values $val$, $\phi(val) =
val$, gives us that 
\[
\forall i \in \set{1 \upto k} \setminus \mathord ?^t(\alpha) \spot
   \phi(v'_i) = v'_i.
\]
Hence and by the definition of $v'$ (\ref{equation:mainfailurestheoremeq2}),
\begin{equation}\label{equation:mainfailurestheoremeq5}
	\forall i \in \set{1 \upto k} \setminus \mathord ?^t(\alpha) \spot v_i = v'_i.
\end{equation}
In addition, since $c.v_1 \ldots v_k$ is in $\phi(initials(P(t), \Gamma, T))$,
it must be that
\begin{equation}\label{equation:mainfailurestheoremeq6}
	\forall i \in \mathord ?^t(\alpha) \spot v_i \in \phi(T). 
\end{equation}
Combining (\ref{equation:mainfailurestheoremeq1}),
(\ref{equation:mainfailurestheoremeq3}),
(\ref{equation:mainfailurestheoremeq5}) and
(\ref{equation:mainfailurestheoremeq6}) we get that $e$ is in
$\initials(P(t), \Gamma, \phi(T))$. Hence 
\begin{eqnarray}\label{equation:mainfailurestheoremeq7}
\phi(\initials(P(t), \Gamma, T)) & \subseteq & 
   \initials(P(t), \Gamma, \phi(T)).
\end{eqnarray}

Conversely, let $e = c.v_1 \ldots v_k$ be in
$\initials(P(t), \Gamma, \phi(T))$. From
Proposition~\ref{proposition:adataindepimplication} we can infer that 
\begin{eqnarray*}
\initials(P(t), \Gamma, \phi(T)) & \subseteq & \initials(P(t), \Gamma, T),
\end{eqnarray*}
so $e$ is in $\initials(P(t), \Gamma, T)$. Hence, $\phi(e)$ is in
$\phi(\initials(P(t), \Gamma, T))$. However, for all $i$ in $\set{1 \upto k}$,
$v_i$ is either a value of a non-$t$ type or it is a value in
$\phi(T)$. Hence, 
\[
\forall i \in \set{1 \upto k} \spot \phi(v_i) = v_i,
\]
so $\phi(e) = e$ and therefore $e \in \phi(\initials(P(t), \Gamma), T)$. Hence,
\begin{eqnarray*}
\initials(P(t), \Gamma, \phi(T)) & \subseteq &
   \phi(\initials(P(t), \Gamma, T)).
\end{eqnarray*}
Combining the above with (\ref{equation:mainfailurestheoremeq7}) we have that 
\begin{eqnarray}\label{initialsphiinitials}
\phi(\initials(P(t),\Gamma,T)) & = & \initials(P(t),\Gamma,\phi(T)).
\end{eqnarray}

We now aim to show that 
\begin{eqnarray}\label{PTGammaref}
(P(t),\Gamma, T) & \refuses & \pi(X).
\end{eqnarray}
Suppose for a contradiction that there exists an event $x$ in $\pi(X) \cap
\initials(P(t),\Gamma, T)$. Then (\ref{initialsphiinitials}) implies that
$\phi(x) \in \phi(\pi(X)) \cap\, \initials(P(t),\Gamma,
\phi(T))$. This in turn means
that ${\phi(\pi(X)) \cap\, \initials(P(t),\Gamma, \phi(T))}$ is non-empty,
contradicting (\ref{equation:mainfailurestheoremeq0.1}). Hence,
(\ref{PTGammaref}) holds. 
Finally, applying Corollary~\ref{corollary:adataindepimplication} to
(\ref{equation:mainfailurestheoremeq0.1}),  we have that 
\begin{eqnarray*}
(\Spec(t), \set{}, T) & \Trans(5)[\phi(\pi(tr))] & (P(t),\Gamma,T).
\end{eqnarray*}
This, combined with (\ref{PTGammaref}) implies that
(\ref{failureslemmaequation1}) holds. 

We now seek to apply Proposition~\ref{bigfailuresproposition} to $\pi(tr)$ and
$\pi(X)$.   
From equation~(\ref{secondassumption'}),
\(	\pi(tr) \in \tsem{Impl(T)} \).
However, we have already shown that $\Spec(T) \trefinedby Impl(T)$, so
\[
\pi(tr) \; \in \; \tsem{\Spec(T)} \; = \; \tsem{\Spec(t), \nullset, T}.
\]
This gives us condition~$(i)$ of Proposition~\ref{bigfailuresproposition}.
Equation~(\ref{failureslemmaequation1}) (and the fact that
$\phi(\nullset) = \nullset$) gives us condition~(ii). In
addition, our definition of $S_1$ (\ref{equation:definitionofs1}), combined
with the definition of $\pi$, implies that every output value of type $t$ of
every event in $\initials(Spec(T)/\pi(tr))$ is in $\set{0 \upto B-1}$, which
gives us condition (iii). Hence, we can infer  that 
\[
(\pi(tr),\pi(X)) \; \in \; \fsem{Spec(t), \nullset, T} \; = \; \fsem{\Spec(T)}.
\]
This is a contradiction to (\ref{thirdassumption'}), which completes our
proof.\qed 

Some observations related to Theorem~\ref{mainfailuresthm} are now in order.
\begin{remark}
For every verification problem, the value of $\Thresh$ in
Theorem~\ref{mainfailuresthm} depends only on the specification.  
\end{remark}

\begin{remark}\label{remark:finitecheckingforthreshold}
For all specifications with a finite SSLTS, the value of $\Thresh$ in
Theorem~\ref{mainfailuresthm} can be calculated in a finite amount of time.
The term $Thresh_{\mathrm{T}}$ can be calculated as in
Remark~\ref{remark:finitecheckingforthresholdtraces}.  The other term can be
obtained by calculating $\Thresh_{!^t}(\sigma,\Gamma)
+ \Thresh_{?^t}(\sigma,\Gamma)$ for each symbolic state that is either the
initial state or that has an incoming visible transition.
\end{remark}

\begin{example}
Recall the example from Section~\ref{sec:example}.  In
Example~\ref{example:CA-traces} we showed how to apply
Theorem~\ref{maintracesthm} to deduce results in the traces model.  We now,
similarly, show how to apply Theorem~\ref{mainfailuresthm} to deduce results
in the stable failures model.  We can use the counter abstraction techniques
to verify 
\begin{eqnarray*}
Spec(\phi(T)) & \frefinedby & \phi(Impl(T)), 
  \qquad \mbox{for all instantiations~$T$ of~$t$  with $\#T \ge 3$,}
\end{eqnarray*}
where again
\begin{eqnarray*}
Spec(t) & = & enterCS\$i\mathord:t \then leaveCS!i \then Spec(t),
\end{eqnarray*}
and where we took $B = 1$.  It is clear that conditions~(i), (ii) and~(iii) of
Theorem~\ref{mainfailuresthm} hold.  From condition~(v) we obtain $Thresh =
1$, essentially because $Spec(t)$ contains a single~``$!$'' and no~``$?$'';
hence condition~(vi) holds.  Condition~(iv) gives a lower bound of~2 on the
size of~$T$, which is a weaker condition than we have already imposed.
Finally condition~(vii) holds by construction.  Hence we can apply the theorem
to deduce
\begin{eqnarray*}
Spec(T) & \frefinedby & Impl(T), 
  \qquad \mbox{for all instantiations~$T$ of~$t$  with $\#T \ge 3$.}
\end{eqnarray*}
Smaller values of~$T$ can be verified directly. 
\end{example}

As with the theorem for the traces model, the tool TomCAT can be used to
verify the conditions of Theorem~\ref{mainfailuresthm} and to calculate the
threshold.

At first, condition~(iii) of Theorem~\ref{mainfailuresthm} ---that no
construct of the specification combines nondeterministic inputs of type $t$
and deterministic inputs of any type--- may seem somewhat arbitrary. However,
without it, there are specifications $\Spec(t)$ and implementations $Impl(t)$
such that no threshold exists: for all values of $B$, there exists  an
instantiation~$T$ of 
type~$t$ such that $\Spec(\phi(T)) \frefinedby \phi(Impl(T))$ and yet
$\Spec(T) \not \frefinedby Impl(T)$.  The following examples illustrate such
pairs of processes, where a nondeterministic input of type $t$ is combined
with a deterministic input of type $t$
(Example~\ref{example:mainfailurestheoremexample1}) and with a deterministic
input of a non-$t$ type (Example~\ref{example:mainfailurestheoremexample2}).
\begin{example}\label{example:mainfailurestheoremexample1}
Let
\begin{eqnarray*}
\Spec(t) & = & c\$x\mathrm{:}t?y\mathrm{:}t \then \STOP, \\
Impl(t) & = & 
  \Extchoice y\mathrm{:}t \spot c?x\mathrm{:}(t\setminus\set{y})!y 
  \then \STOP.
\end{eqnarray*}
Note, in particular, that $Impl(t)$ satisfies \textbf{TypeSym}.

Let $B$ be an arbitrary positive number, and let $\phi$ be as in the statement
of Theorem~\ref{mainfailuresthm}. Let $T = \set{0 \upto N}$ where $N \geq
B+1$.  It is easy to see that $\traces(\phi(Impl(T))) \subseteq
traces(\Spec(\phi(T)))$.
Further, whatever value $Impl(T)$ chooses for~$y$,\, 
$(T \setminus \set{y}) \cap \set{B \upto N} \not = \nullset$; hence
$(\nil, \set{c.B.B}) \not \in \fsem{\phi(Impl(T))}$. This helps to see that 
\[
\begin{array}{rl}
	& \fsem{\phi(Impl(T))}\\
= &\set{(\nil, X) | X \subseteq \set{c.x.x | x \in \set{0 \upto B-1}}}\\
&	\union \set{(\trace{c.x.y}, X) | y \in \set{0 \upto B} \land x \in \set{0 \upto B}\setminus\set{y} \land X \subseteq \Sigma}\\
\subseteq &\set{(\nil, X) | X \subseteq \set{c.x.y | x \in \set{0 \upto B} \setminus \set{p} \land y \in \set{0 \upto B}} \land p \in \set{0 \upto B}}\\
&	\union \set{(\trace{c.x.y}, X) | x,y \in \set{0 \upto B} \land X \subseteq \Sigma}\\
=	& \fsem{\Spec(\phi(T))}.\\
\end{array} 
\]
Hence
\( \Spec(\phi(T)) \frefinedby \phi(Impl(T)) \).

However,
\[
(\nil,\set{c.x.x | x \in T}) \;\in\; \fsem{Impl(T)} \setminus \fsem{Spec(T)},
\]
so $\Spec(T) \not \frefinedby Impl(T)$.
\end{example}


\begin{example}
\label{example:mainfailurestheoremexample2}
Let $B$ be an arbitrary positive integer, and $Y = \set{y_1,y_2}$ a type other
than~$t$ of size~2.  Let
\begin{eqnarray*}
\Spec(t) & = & c\$x\mathrm{:}t?y\mathrm{:}Y \then STOP, \\
Impl_B(t) & = & 
  \Extchoice y\mathrm{:}Y \spot \left( 
    \Intchoice X \subseteq t \land \#X = B + 1 \spot 
      c?x\mathrm{:}X!y \then STOP 
  \right).
\end{eqnarray*}
Note, in particular, that $Impl(t)$ satisfies \textbf{TypeSym}.  

Let $\phi$ be
as in the statement of Theorem~\ref{mainfailuresthm}, and let $T
= \set{0 \upto N}$ where $N \geq 2B+1$.  It is easy to see that
$\traces(\phi(Impl_B(T))) \subseteq traces(\Spec(\phi(T)))$.  Further, whatever
value $Impl_B(T)$ chooses for~$X$,\, $X \inter \set{B\upto N} \not
= \nullset$; hence $(\nil, \set{c.B.y}) \not \in \fsem{\phi(Impl_B(T))}$
for every $y \in Y$. This helps to see that
\[
\begin{array}{rl}
& \fsem{\phi(Impl_B(T))}\\
\subseteq &\set{(\nil, R) | R \subseteq \set{c.x.y | x \in \set{0\upto B-1} \land y \in Y}}\\
&	\union \set{(\trace{c.x.y}, R) | x \in \set{0\upto B} \land y \in Y \land R \subseteq \Sigma}\\
\subseteq &\set{(\nil, R) | R \subseteq \set{c.x.y | x \in \set{0\upto B} \setminus \set{p} \land y \in Y} \land p \in \set{0\upto B}}\\
&	\union \set{(\trace{c.x.y}, R) | x \in \set{0\upto B} \land y \in Y \land R \subseteq \Sigma}\\
=	& \fsem{\Spec(\phi(T))}.\\
\end{array} 
\]
Hence
\( \Spec(\phi(T)) \frefinedby \phi(Impl_B(T)) \).

However, suppose the two values chosen for~$X$ when $y=y_1$ and $y=y_2$ are
disjoint (this is possible since $\#T \ge 2B+2$).  Then $Impl_B(T)$ has an
initial failure $(\trace{}, R)$ such that for each $x \in T$, there is some
$z$ such that $c.x.z \in R$; this is not a failure allowed by $Spec(T)$, so
$\Spec(T) \not \frefinedby Impl_B(T)$. 
\end{example}

\section{Conclusions}
\label{section:conclusions}

Given a specification $\Spec(t)$ and an implementation $Impl(t)$, direct model
checking can help us to find bugs in the implementation for a finite (and
small) number of instantiations~$T$ of parameter $t$. However, one is often
interested in uniform verification, i.e.~in proving correctness for all $T$. 

Lazi\'{c}'s theory of data independence~\cite{Lazic:1999} (see
Section~\ref{section:dataindep}) for the CSP process algebra solves the
problem of uniform verification of parameterised systems with the parameter
being a datatype. Inspired by these results, we have developed a type
reduction theory (with the key results captured by Theorem~\ref{maintracesthm}
and Theorem~\ref{mainfailuresthm}), which establishes the size of a fixed
type~$\shiftedhat{T}$ and a collapsing function $\phi$ that maps all types $T$
larger than $\shiftedhat{T}$ to~$\shiftedhat{T}$, and such that for all $T$
such that $\shiftedhat{T} \subseteq T$,
\begin{equation}\label{equation:conclusionssummaryeq3}
Spec(\shiftedhat{T}) \refinedby \phi(Impl(T))
\quad\mbox{implies that}\quad
Spec(T) \refinedby Impl(T)
\end{equation}
with both refinements in either the traces or the stable failures model. In
order for the above to hold, the processes have to satisfy certain
conditions, the most important of which include a normality condition, \Norm{}
(see Definition~\ref{def:seqnorm}), for specifications and a type symmetry
condition, \textbf{TypeSym} (see Definition~\ref{def:typesym}), for
implementations. 

Our type reduction theory makes extensive use of symbolic representation of
process behaviour, which allows us to use known behaviours of one
instantiation of a specification to deduce behaviours of another one. In
Section~\ref{section:operationalsemantics} we presented a symbolic operational
semantics for CSP processes that satisfy \Seq, and we provided a set of
translation rules that allow us to concretise symbolic transition graphs. We
also showed that, crucially, the combination of the symbolic operational
semantics and the translation rules is congruent to a fairly standard
operational semantics. 

Since the process $\phi(Impl(T))$ used in
(\ref{equation:conclusionssummaryeq3}) still depends on $T$, the type
reduction theory, on its own, does not resolve the problem of an infinite
number of refinement checks needed to solve a given verification
problem. However, the usefulness of the theory comes from the fact that it can
be combined with an abstraction method that produces models $Abstr$ such that
for all sufficiently large $T$,
\begin{eqnarray}\label{equation:conclusionssummaryeq4}
	Abstr & \refinedby & \phi(Impl(T)).
\end{eqnarray}
We can then test, using a model checker,  that $\Spec(\shiftedhat{T})
\refinedby Abstr$.  This allows us to deduce,
from transitivity of refinement and (\ref{equation:conclusionssummaryeq3}),
that $\Spec(T) \refinedby Impl(T)$ holds for all sufficiently large $T$ (and
the verification problem can be solved directly for all smaller $T$). One
suitable abstraction technique (based on ideas of counter abstraction) can be
found in~\cite{Tomasz-thesis, Mazur:2010}. 


\subsection{Automation}

We can automatically check process syntaxes for the syntactic requirements of
data independence (Definition~\ref{def:dataindependence}) and \Norm{}
(Definition~\ref{def:seqnorm}). 
Checking for the semantic requirements of the \textbf{TypeSym}
condition (Definition~\ref{def:typesym}) is difficult in practice due to the
universal quantification over all instantiations of the type parameter
$t$. However, we can automatically verify implementation definitions as to
whether they satisfy the five simple  syntactic conditions of
Proposition~\ref{typesymproposition} and infer \textbf{TypeSym}. 

Checking \textbf{RevPosConjEqT}
(Definition~\ref{revposconjeqt}) is the most problematic when it comes to
automation. The problem lies in the universal quantification over all
instantiations of the parameter variables of the arguments of conditional
choices. Currently, it is left to the user to provide a proof that for every
conditional choice of the form ``$\If cond \Then P(x_1,\ldots,x_k) \Else
Q(x_1,\ldots,x_k)$'' in a given specification, where $cond \in Cond$, $cond$
is a positive conjunction of equality tests on~$t$ and $Q(v_1,\ldots,v_k)
\refinedby P(v_1,\ldots,v_k)$ for all values $v_1, \ldots, v_k$. In general,
the problem of \textbf{RevPosConjEqT} satisfiability is undecidable, since a
general (undecidable) PMCP problem of the form $Spec(x) \refinedby\verproblem
Impl(x)$, where $x$ is a parameter, can be reduced to checking whether 
\( in?i\mathrm{:}x?j\mathrm{:}x \then \If i = j \Then Impl(x) \Else \Spec(x) \)
satisfies \textbf{RevPosConjEqT}. However, in most practical situations it is
not too difficult to provide a convincing proof that, regardless of
parameters, the ``then'' branch of every conditional choice on $t$ forms a
refinement of its ``else'' branch, as often the branches are similar, except
for the use of operators that introduce different levels of nondeterminism
(e.g.~using $\intchoice$ in the positive branch versus $\extchoice$ in the
negative one).

As noted in Remarks~\ref{remark:finitecheckingforthresholdtraces}
and~\ref{remark:finitecheckingforthreshold}, the calculation of the thresholds
in Theorems~\ref{maintracesthm} and~\ref{mainfailuresthm} can be fully
automated.

\subsection{Multiple distinguished types}

Throughout this paper we assumed the presence of a single distinguished type
$t$. It is easy to extend our techniques to any finite number of distinguished
types, say $t_1, t_2, \ldots, t_n$, provided all of them are pairwise
independent. All requirements are extended in the natural way, e.g.\ each
specification $\Spec(t_1, t_2, \ldots, t_n)$ must now be data independent in
each of the $n$ types, and each implementation $Impl(t_1, \ldots t_n)$ must
satisfy \textbf{TypeSym} with respect to each of $t_1$, $t_2, \ldots,
t_n$. The threshold in each of Theorems~\ref{maintracesthm}
and~\ref{mainfailuresthm} is then replaced by a tuple of values $(\Thresh_1,
\Thresh_2, \ldots, \Thresh_n)$, where each $\Thresh_i$ is a threshold for the
collapsing of the values of type $t_i$.


\subsection{Related work}

We are not aware of any other, similar type reduction theory for parameterised
systems with the parameter describing the number of node processes forming a
network. Ideas closest to ours are those of data
independence. In~\cite{Lazic:1999} Lazi\'{c} provides results similar to ours,
except that allows us to deduce that
\[
Spec(\shiftedhat{T}) \refinedby Impl(\shiftedhat{T})
\quad\mbox{implies that}\quad
Spec(T) \refinedby Impl(T)
\]
instead of (\ref{equation:conclusionssummaryeq3}). This makes data
independence theory applicable without the need for abstraction techniques,
but, since both $Spec$ and $Impl$ are assumed to be data independent, it does
not allow the use of replicated operators indexed over the distinguished type,
which is a key part of all the implementations we consider.


\subsection{Future work}


The operational semantics that we presented in Section~\ref{section:ssos}
served an important purpose in proving the results of our type reduction
theory. However, our type reduction theory assumes that processes satisfy the
\Seq condition. Therefore, for brevity, we provided operational semantics
rules only for those operators that \Seq allows. To increase the
generality, it would be desirable to formalise symbolic transition rules for
parallel compositions, renaming, hiding and replicated choices.

We would like to extend our type reduction theory to the failures/divergences
model of CSP (see e.g.~\cite{Roscoe:1997}). However, usually the only
divergences property one is interested in is full divergence-freedom. In
practice, this might be an easier problem to verify for all instantiations of
the distinguished type than verifying failures/divergences refinement. Once a
system is shown to be divergence-free, a refinement check in the stable
failures model implies refinement in the failures/divergences model.

Finally, we presented our type reduction techniques for processes modelled
using the CSP process algebra. It would be desirable to research how well
these ideas map across to other formalisms.

\subsection*{Acknowledgements}

We would like to thank Bill Roscoe, Marta Kwiatkowska, Ranko Lazi{\'c} and the
anonymous referees for very useful comments.  This work was partially funded
by EPSRC and ONR.

\appendix
\section{Proofs for Section \ref{section:regularity}}

In this appendix, we prove the results from Section~\ref{section:regularity}.
We start with a few lemmas that are used in the proofs of those results. 

In some proofs by structural induction in this and subsequent appendices, some
of the cases are straightforward and are omitted; they can be found
in~\cite{Tomasz-thesis}.

The following lemma shows that (for a process that satisfies \Norm), each
visible or conditional symbolic event leads to a unique symbolic state.
\begin{lem}\label{lemmaX}
Suppose that $Proc(t)$ satisfies \Norm{}. Let $\epsilon$ be a visible or
conditional symbolic event and suppose that 
\[
Proc(t) \SymTrans{\sigma_1} Proc'_1(t)
\mbox{\quad and \quad}
Proc(t) \SymTrans{\sigma_2} Proc'_2(t),
\]
where $\sigma_1 = \tau^a \cat \strace{\epsilon}$ for $a \geq 0$, and $\sigma_2
= \tau^b \cat\strace{\epsilon}$ for $b \geq 0$. Then $Proc'_1(t) =
Proc'_2(t)$. 
\end{lem}

\proof
We prove the result by a structural induction on $Proc(t)$. 
We give just the cases for prefix and external choice.


\para{Prefix} 

Suppose $Proc(t) = \alpha \then Proc'(t)$ for some construct $\alpha =
\construct$.  Clearly $\epsilon$ must be a visible symbolic event matching
$\alpha$, say $c\S'_1x'_1\mathrm{:}X'_1\ldots \S'_kx'_k\mathrm{:}X'_k$. We
consider two different cases, corresponding to the number of nondeterministic
selections over non-$t$ types of $\alpha$.

\subpara{Case 1.} Suppose that $\#\$^\nont(\alpha) = 0$.  Then, by Symbolic
Prefix Rule 1 (p.~\pageref{rule:symbolicprefixrule1}), it must be that 
$\sigma_1 = \sigma_2 = \strace{\epsilon}$
with $\epsilon$ in $\mathit{Comms}^\nont(\alpha)$ and
\[
Proc'_1(t) \;=\; Proc'_2(t) 
  \;=\; Proc'(t)[x'_i / x_i | i \in \mathord ?^\nont(\alpha)].
\]

\subpara{Case 2.} Suppose that $\#\$^\nont(\alpha) > 0$.
Then, Symbolic Prefix Rule~2 (p.~\pageref{rule:symbolicprefixrule2}) implies
that the only symbolic transitions in $Proc(t)$ are 
\begin{eqnarray*}
Proc(t) & \symtrans{\tau} & 
  \left( \mathit{Replace}_{\$\mapsto!}^\nont(\alpha) \then Proc'(t) \right)
    [v_i / x_i | i \in  \mathord \$^\nont(\alpha)] ,
\end{eqnarray*}
for each function $v$ with $\dom(v) = \set{1\upto k}$ and such that if $i$ is
in $\$^\nont(\alpha)$, then $v_i$ is in~$X_i$. We are guaranteed that  
$\#\$^\nont(\mathit{Replace}_{\$\mapsto!}^\nont(\alpha)) = 0$,
so, Symbolic Prefix Rule~1 (p.~\pageref{rule:symbolicprefixrule1}) implies
that there are two functions $v$ as above, say $v^1$ and $v^2$, such that for
$j \in \set{1,2}$:
\begin{eqnarray*}
Proc'_j(t) & = & 
  (Proc'(t)[v^j_i / x_i | i \in \$^\nont(\alpha)])
     [x'_i / x_i | i \in \mathord ?^\nont(\alpha)],
\\
\epsilon & \in & 
  \mathit{Comms}^\nont((\mathit{Replace}_{\$\mapsto!}^\nont(\alpha))
    [v^j_i / x_i | i \in \mathord \$^\nont(\alpha)]).
\end{eqnarray*}
Then, thanks to the definition of $Comms^\nont$, $v^1$ and $v^2$ are equal
under domain restriction to $\$^\nont(\alpha)$. Therefore, $Proc'_1(t) =
Proc'_2(t)$.

\para{External choice}

Suppose that $Proc(t) = P(t) \extchoice Q(t)$ for some process syntaxes $P(t)$
and~$Q(t)$. Since $Proc(t)$ satisfies \Norm{}, we know that neither $P(t)$ nor
$Q(t)$ contains a conditional choice on $t$ before a prefix. Therefore
$\epsilon$ cannot be a conditional symbolic event, so must be a
visible symbolic event. \Norm{} implies that the channels of the initial
visible symbolic events of $P(t)$ and $Q(t)$ are disjoint, so we have that
either 
$P(t) \symtransstar{\tau}\symtrans{\epsilon}$ or
$Q(t) \symtransstar{\tau}\symtrans{\epsilon}$,
but not both. Without loss of generality we assume the former. Then the inductive hypothesis implies that there is a unique symbolic state $P'(t)$ such that
\begin{eqnarray*}
P(t) & \symtransstar{\tau}\symtrans{\epsilon} & P'(t).
\end{eqnarray*}
Even though there may be some $\tau$'s, contributed by $Q(t)$, in the symbolic
trace of $P(t)$ leading to $Proc'_1(t)$, the uniqueness of $P'(t)$ implies that
$Proc'_1(t) = Proc'_2(t) = P'(t)$.
\qed









The following corollary lifts the previous lemma to traces.
\begin{cor}\label{corollaryX}
Suppose that $Proc(t)$ satisfies \Norm{}. Suppose further that
\[
Proc(t) \SymTrans{\sigma} P(t)
\mbox{\quad and \quad}
Proc(t) \SymTrans{\sigma'} Q(t).
\]
Then if neither $\sigma$ nor $\sigma'$ ends with a $\tau$ and $\sigma
\nontauequiv \sigma'$, then $P(t) = Q(t)$. 
\end{cor}
\proof
By induction on the  number of visible and conditional symbolic events
of $\sigma$ and $\sigma'$ (which must be equal,
since $\sigma \nontauequiv \sigma'$), and using Lemma~\ref{lemmaX}. 
\qed


The following lemma relates two initial visible symbolic events on the same
channel.
\begin{lem}\label{lemmaB}
Suppose that $Proc(t)$ satisfies \Norm{}, and $\sigma,\sigma' \in (Cond \union
\set{\tau})^*$ are such that $\sigma \nontauequiv \sigma'$. Then, if $Proc(t)
\SymTrans{\sigma}\symtrans{\epsilon}$ and $Proc(t)
\SymTrans{\sigma'}\symtrans{\epsilon'}$, where $\epsilon =
c\S_1x_1\mathrm{:}X_1\ldots \S_kx_k\mathrm{:}X_k$ and $\epsilon' =
c'\S'_1x'_1\mathrm{:}X'_1\ldots \S'_lx'_l\mathrm{:}X'_l$ are visible symbolic
events, then:
\begin{enumerate}[\em(i)]
\item if the channels of $\epsilon$ and $\epsilon'$ are identical
  (i.e.\ $c = c'$), then the parts of $\epsilon$ and $\epsilon'$ involving
  type $t$ are equal, i.e.\  
  \[
  \begin{align}
  \$^t(\epsilon) \union \mathord ?^t(\epsilon) \union \mathord !^t(\epsilon) =
   \mathord \$^t(\epsilon') \union \mathord ?^t(\epsilon') 
   \union \mathord !^t(\epsilon') \land \\
  \forall i \in \$^t(\epsilon) \union \mathord ?^t(\epsilon)
      \union \mathord !^t(\epsilon) \spot 
        \S_i = \S'_i \land x_i = x'_i \land X_i = X'_i ;
  \end{align}
  \]
	
\item if $\mathit{Insts}_\Gamma(\epsilon) \inter
  \mathit{Insts}_\Gamma(\epsilon') \not = \nullset$ for some environment
  $\Gamma$, then $\epsilon = \epsilon'$. 
\end{enumerate}
\end{lem}

\proof Firstly, observe that if $\mathit{Insts}_\Gamma(\epsilon) \inter
\mathit{Insts}_\Gamma(\epsilon') \not = \nullset$, then the channels of
$\epsilon$~and $\epsilon'$ must be the same. Hence in both cases $c =
c'$. Since every channel has a fixed structure of the communication along it,
the number of components of $\epsilon$ and $\epsilon'$ must be identical,
i.e.\ $k = l$. We can prove both clauses using a structural induction on
$Proc(t)$.  We give just the case for prefix, since it is the most
interesting. 


Suppose that $Proc(t) = \alpha \then Proc'(t)$ for some construct $\alpha =
\construct$ and some process syntax $Proc'(t)$. We perform a case analysis on
the number of nondeterministic selections over non-$t$ types of $\alpha$.

\subpara{Case 1.} Suppose that $\#\$^\nont(\alpha) = 0$.  Then, by Symbolic
Prefix Rule~1 (p.~\pageref{rule:symbolicprefixrule1}) (observe that the other
symbolic firing rules are not applicable in this case), it must be that
$\sigma = \sigma' = \strace{}$, and that both $\epsilon$ and $\epsilon'$ are
in $\mathit{Comms}^\nont(\alpha)$. By the definition of $\mathit{Comms}^\nont$
(p.~\pageref{def:commsnont}), $\epsilon$ and $\epsilon'$ may differ only in
the values of deterministic inputs of non-$t$ types of $\alpha$. Hence, clause
(i) of the lemma holds. To prove clause (ii), we let $c.v_1 \ldots v_k$ be a
common member of $\mathit{Insts}_\Gamma(\epsilon)$ and
$\mathit{Insts}_\Gamma(\epsilon')$. Then the definition of
$\mathit{Insts}_\Gamma$ (p.~\pageref{def:insts}) implies that
\begin{equation}\label{equation:partialidentityeq1a}
\forall i \in \mathord !(\epsilon) \spot
   \S_i = \S'_i = \mathord ! \land x_i = x'_i \land v_i = \Gamma(x_i) 
   \land X_i = X'_i = \mathit{null}.
\end{equation}
The definition of $\mathit{Comms}^\nont$ implies that
$?^\nont(\alpha) \subseteq \mathord !(\epsilon)$, so 
\[
\forall i \in \mathord ?^\nont(\alpha) \spot \S_i = \S'_i = \mathord ! 
  \land x_i = x'_i \land X_i = X'_i = \mathit{null}.
\]
This, combined with clause (i) of the lemma,
(\ref{equation:partialidentityeq1a}) and the fact that $\$^\nont(\alpha) = 0$,
implies that $\epsilon = \epsilon'$.

\subpara{Case 2.} 

Suppose that $\#\$^\nont(\alpha) > 0$.  Then, by Symbolic Prefix Rule~2
(p.~\pageref{rule:symbolicprefixrule2}) (observe that the other symbolic
firing rules are not applicable in this case), the only transitions in
$Proc(t)$ are
\begin{eqnarray*}
Proc(t) & \trans[\tau][s] &
  \left(Replace_{\$\mapsto !}^\nont(\alpha) \then Proc'(t)\right)
    [v_i/x_i | i \in \$^\nont(\alpha)] \\
&  = & Replace_{\$\mapsto !}^\nont(\alpha) [v_i/x_i | i \in \$^\nont(\alpha)]
	 \then Proc'(t)[v_i/x_i | i \in \$^\nont(\alpha)]
\end{eqnarray*}
for functions $v$ such that $\dom(v) = \mathord \$^\nont(\alpha)$, and if $i$
is in $\$^\nont(\alpha)$ then $v_i$ is in~$X_i$. Clearly 
\begin{eqnarray*}
\#\$^\nont \left(
  Replace_{\$\mapsto !}^\nont(\alpha) [v_i/x_i | i \in \$^\nont(\alpha)]
   \right) &  = & 0
\end{eqnarray*}
for every such function $v$, so Symbolic Prefix Rule~1 (p.~\pageref{rule:symbolicprefixrule1}) implies that
\begin{eqnarray*}
\epsilon & \in &
   Comms^\nont\left(\left(\mathit{Replace}_{\$\mapsto !}^\nont(\alpha)\right)
      [v_i/x_i | i \in \$^\nont(\alpha)]\right),\\
\epsilon' & \in &
   Comms^\nont\left(\left(\mathit{Replace}_{\$\mapsto !}^\nont(\alpha)\right)
      [v'_i/x_i | i \in \$^\nont(\alpha)]\right),
\end{eqnarray*}
for some functions $v$ and $v'$ such that $\dom(v) = \dom(v') = \mathord
\$^\nont(\alpha)$ and if $i$ is in $\$^\nont(\alpha)$, then $v_i$ and $v'_i$
are in $X_i$. The definition of $\mathit{Comms}^\nont$
(p.~\pageref{def:commsnont}) implies that the parts of $\epsilon$ and
$\epsilon'$ that involve type $t$ are identical, which proves clause (i) of
the lemma. To prove clause (ii), we let $c.v_1 \ldots v_k$ be a common member
of $\mathit{Insts}_\Gamma(\epsilon)$ and
$\mathit{Insts}_\Gamma(\epsilon')$.  Then, the definition of
$\mathit{Insts}_\Gamma$ (p.~\pageref{def:insts}) implies that  
\begin{equation}\label{equation:partialidentityeq2a}
\forall i \in \mathord !(\epsilon) \spot 
  \S_i = \S'_i = \mathord ! \land x_i = x'_i \land v_i = \Gamma(x_i) 
  \land X_i = X'_i = \mathit{null}.
\end{equation}
However, the definition of $\mathit{Comms}^\nont$ implies that
\[
\begin{array}{rcccl}
?^\nont(\alpha) & = &
   \mathord ?^\nont\left(
      \left( \mathit{Replace}_{\$\mapsto !}^\nont(\alpha) \right)
         [v_i/x_i | i \in \$^\nont(\alpha)] \right)
  & \subseteq & \mathord !(\epsilon),
\\
\$^\nont(\alpha) & \subseteq & 
  \mathord !^\nont\left(
    \left( \mathit{Replace}_{\$\mapsto !}^\nont(\alpha) \right)
       [v_i/x_i | i \in \$^\nont(\alpha)] \right)
  & \subseteq & \mathord !(\epsilon).
\end{array}
\]
Hence,
\[
\forall i \in \mathord \$^\nont(\alpha) \union \mathord ?^\nont(\alpha) \spot
   \S_i = \S'_i \land x_i = x'_i \land X_i = X'_i.
\]
This, combined with clause (i) of the lemma and
(\ref{equation:partialidentityeq2a}), implies that $\epsilon = \epsilon'$.
\qed

The following lemma shows that if a process can perform a conditional event
initially (after only $\tau$s), then \emph{all} its initial events (after
$\tau$s) must be that conditional or its negation. 
\begin{lem}\label{lemma:lonelyconditionals}
Suppose that $Proc(t)$ satisfies \Norm{}. Then, if $Proc(t)
\SymTrans{\sigma}\trans(3)[cond][s]$ and $Proc(t)
\SymTrans{\sigma'}\symtrans{\alpha}$, where $\sigma, \sigma' \in
\set{\tau}^*$, $cond \in Cond$ and $\alpha \not = \tau$, then $\alpha \in
\set{cond,\neg cond}$. 
\end{lem}
\proof
Since $Proc(t)$ satisfies \Norm{}, we know that there are no conditionals
before prefixes in branches of external, internal and sliding choices. Hence
one of the following must hold: 
\begin{enumerate}[(i)]
\item $Proc(t)$ is a conditional choice on $t$, where the boolean
  condition is equal to $cond$ or $\neg cond$;

\item $Proc(t)$ is a process identifier bound by the global
  environment~$E$ to a conditional choice like that in clause~(i)~or~(iii); or

\item $Proc(t)$ is a conditional choice whose boolean condition
  immediately evaluates to $\True$ or $\False$, and the appropriate branch is
  a process syntax as in clause~(i)~or~(ii). 
\end{enumerate}
This means that the only transitions available in $Proc(t)$ are $Proc(t)
\starit{\trans}[\tau][s]\trans(3)[cond][s]$ and $Proc(t)
\starit{\trans}[\tau][s]\trans(4)[\neg cond][s]$. 
\qed


The following lemma shows that if two symbolic traces each contain a single
visible symbolic event, and each trace can be instantiated in the same
environment, then they contain the same conditional events before the visible
event, essentially because those conditionals must evaluate to $True$ in the
initial environment.
\begin{lem}\label{lemmaA1}
Suppose that $Proc(t)$ satisfies \Norm{}. Let $\sigma\cat\strace{\epsilon},
\sigma'\cat\strace{\epsilon'}$ be symbolic traces of $Proc(t)$ such that
$\sigma, \sigma' \in (Cond \union \set{\tau})^*$, $\epsilon, \epsilon' \in
\Visible$, $\sigma\cat\strace{\epsilon} \generates{\Gamma} \trace{e}$ and
$\sigma'\cat\strace{\epsilon'} \generates{\Gamma} \trace{e'}$ for some
environment $\Gamma$ and some visible events $e$ and $e'$. Then $\sigma
\nontauequiv \sigma'$. 
\end{lem}

\proof
Let $\kappa : \stsem{Proc(t)} \rightarrow \Nats$ be a function that returns
the number of conditional symbolic events within a given symbolic trace. We
prove the result using an induction on $\kappa(\sigma)$. 

\para{Base case.} Suppose that $\kappa(\sigma) = 0$. Then $\sigma \in
\set{\tau}^*$, so $Proc(t)
\starit{\trans}[\tau][s]\trans[\epsilon][s]$. Suppose, for a contradiction,
$\kappa(\sigma') > 0$. Then $\sigma' = \tau^a\cat\strace{cond}\cat\rho$ for
some $a \geq 0$, some $cond \in Cond$ and some symbolic trace~$\rho \in (Cond
\union \set{\tau})^*$. Therefore $Proc(t)
\starit{\trans}[\tau][s]\trans(3)[cond][s]$. By
Lemma~\ref{lemma:lonelyconditionals}, $\epsilon \in \set{cond, \neg
  cond}$. This is a contradiction as $\epsilon \in \Visible$. Therefore
$\kappa(\sigma') = 0$, which means that $\sigma' \in \set{\tau}^*$. Hence
$\sigma \nontauequiv \sigma'$.

\para{Inductive case.}

Suppose that the result holds for all process syntaxes and all their symbolic
traces with exactly $k$ conditional symbolic events. Consider $\kappa(\sigma)
= k+1$. Then $\sigma = \tau^a \cat\strace{cond}\cat\rho$ for some $a \geq 0$,
some $cond \in Cond$ and some $\rho \in (Cond \union \set{\tau})^*$ with
$\kappa(\rho) = k$. Arguing
similarly in the base case, $\kappa(\sigma') > 0$. Therefore, $\sigma' =
\tau^b\cat\strace{cond'}\cat\rho'$ for some $b \geq 0$, some $cond' \in Cond$
and some $\rho' \in (Cond \union \set{\tau})^*$. By
Lemma~\ref{lemma:lonelyconditionals}, $cond' \in \set{cond, \neg cond}$. Since
$\sigma\cat\strace{\epsilon} \generates{\Gamma} \trace{e}$ and
$\sigma'\cat\strace{\epsilon'} \generates{\Gamma} \trace{e'}$, $cond$ and
$cond'$ must both evaluate to $\True$ within $\Gamma$, because there are no
visible symbolic events within $\sigma$ and $\sigma'$ before $cond$ and
$cond'$, respectively, that could modify the environment~$\Gamma$. So it must
be that $cond = cond'$. By Lemma~\ref{lemmaX}, there is a unique state $P(t)$
such that $Proc(t) \starit{\trans}[\tau][s]\trans(3)[cond][s] P(t)$. Hence,
$\rho\cat\strace{\epsilon}$ and $\rho'\cat\strace{\epsilon'}$ are both
symbolic traces of $P(t)$ such that $\rho\cat\strace{\epsilon}
\generates{\Gamma} \trace{e}$ and $\rho'\cat\strace{\epsilon'}
\generates{\Gamma} \trace{e'}$. Therefore, by the inductive hypothesis, $\rho
\nontauequiv \rho'$, which implies that $\sigma \nontauequiv \sigma'$. 
\qed


\subsection{Proofs of main results}

We can now prove the results stated in Section~\ref{section:regularity}.  In
order to prove Proposition~\ref{prop:environmentuniqueness}.  We will need the
following lemma.
\begin{lem}\label{environmentuniquenesssinglestep}
Suppose that $Proc(t)$ satisfies \Norm{}. Suppose further that
\begin{eqnarray*}
(Proc(t),\Gamma_{init}) & \starit{\trans}[\tau]\trans[e] & (P(t),\Gamma), \\
(Proc(t),\Gamma_{init}) & \starit{\trans}[\tau]\trans[e] & (Q(t),\Gamma'),
\end{eqnarray*}
where $e$ is a visible event. Then $P(t) = Q(t)$ and $\Gamma = \Gamma'$.
\end{lem}

\proof
By the translation rules of COSE (see Section~\ref{section:translationrules}),
there must exist symbolic traces~$\sigma, \sigma'$ and visible symbolic events
$\epsilon = \construct$ and $\epsilon' = c'\S'_1x'_1\mathrm{:}X'_1\ldots
\S'_lx'_l\mathrm{:}X'_l$ such that 
\begin{iteMize}{$\bullet$}
\item $Proc(t) \SymTrans{\sigma}\symtrans{\epsilon} P(t)$ and $Proc(t)
  \SymTrans{\sigma'}\symtrans{\epsilon'} Q(t)$; and
 
\item $\sigma\cat\strace{\epsilon} \generates{\Gamma_{init}} \trace{e}$ and
  $\sigma'\cat\strace{\epsilon'} \generates{\Gamma_{init}} \trace{e}$. 
\end{iteMize}
Then it must be that $\sigma,\sigma' \in (Cond \union \set{\tau})^*$. So, by
Lemma~\ref{lemmaA1}, $\sigma \nontauequiv \sigma'$. From the definition of the
$\generates{}$ relation (p.~\pageref{def:generates}) we have that $e \in
\mathit{Insts}_{\Gamma_{init}}(\epsilon) \inter
\mathit{Insts}_{\Gamma_{init}}(\epsilon')$, so Lemma~\ref{lemmaB} implies that
$\epsilon = \epsilon'$. Hence, $\sigma\cat\strace{\epsilon} \nontauequiv
\sigma'\cat\strace{\epsilon'}$ and so we can infer, using
Corollary~\ref{corollaryX}, that $P(t) = Q(t)$. In addition, the translation
rules of COSE (see Section~\ref{section:translationrules}) imply that if $e=
c.v_1 \ldots v_k$, then 
\begin{eqnarray*}
\Gamma & = & 
  \Gamma_{init} \oplus \set{x_i \mapsto v_i | 
     i \in \mathord \$^t(\epsilon) \union \mathord ?^t(\epsilon)},
\\
\Gamma' & = &
  \Gamma_{init} \oplus \set{x'_i \mapsto v_i | 
    i \in \mathord \$^t(\epsilon') \union \mathord ?^t(\epsilon')}.
\end{eqnarray*}
However, $\epsilon = \epsilon'$, so $\Gamma = \Gamma'$, as required.
\qed

\proof[Proof of Proposition~\ref{prop:environmentuniqueness}]
By a straightforward induction on the length of $s \hide \set{\tau}$ and using
Lemma~\ref{environmentuniquenesssinglestep}. 
\qed


\proof[Proof of Proposition \ref{propositionA2}]
We prove the result by an induction on the length of~$tr$.

\para{Base case.} Suppose that $tr = \trace{}$. Then $\sigma
\generates{\Gamma} \trace{}$ and $\sigma' \generates{\Gamma}
\trace{}$. Therefore, by the definition of the $\generates{}$ relation
(p.~\pageref{def:generates}), $\sigma$ and $\sigma'$ cannot contain visible
symbolic events. Hence, by the assumptions of this proposition, $\sigma =
\sigma' = \strace{}$, which means that $\sigma \nontauequiv \sigma'$. 

\para{Inductive case.} Suppose that the result holds for all traces of length
$k$. Consider a trace $tr_{k+1}$ of length $k+1$. Then there exists a trace
$tr_k$ of length $k$ and a visible event $e$ such that $tr_{k+1} =
tr_k\cat\trace{e}$. There must also exist symbolic traces $\sigma_1,
\sigma'_1, \sigma_2$, and $\sigma'_2$ such that 
\begin{iteMize}{$\bullet$}
\item either $\sigma_1 = \sigma'_1 = \strace{}$ or both $\sigma_1$ and
  $\sigma'_1$ end in a visible symbolic event;

\item $\sigma = \sigma_1\cat\sigma_2$ and $\sigma' = \sigma'_1\cat\sigma'_2$;
  and 

\item $\sigma_1 \generates{\Gamma} tr_k$ and $\sigma'_1 \generates{\Gamma}
  tr_k$. 
\end{iteMize}
Then, by the inductive hypothesis, $\sigma_1 \nontauequiv \sigma'_1$. Hence,
if $P(t)$ and $Q(t)$ are such that $Proc(t)
\SymTrans{\sigma_1} P(t)$ and $Proc(t) \SymTrans{\sigma'_1} Q(t)$, then, by
Corollary~\ref{corollaryX}, $P(t) = Q(t)$. 

Let $\Gamma$ be the environment reached after $tr_k$, i.e.~such that
$(Proc(t),\Gamma) \mapstotrans[s] (P(t),\Gamma)$ for some $s$ such that $s
\hide \tau = tr_k$; by Proposition~\ref{prop:environmentuniqueness}, $\Gamma$
is unique.  We now have that $\sigma_2 \generates{\Gamma} \trace{e}$ and
$\sigma'_2 \generates{\Gamma} \trace{e}$. Hence, $\sigma_2, \sigma'_2 \not =
\strace{}$. Therefore, both $\sigma_2$ and~$\sigma'_2$ must end in a visible
symbolic event (since they are suffixes of $\sigma$ and $\sigma'$).  So,
$\sigma_2 = \rho\cat\strace{\epsilon}$ and $\sigma'_2 =
\rho'\cat\strace{\epsilon'}$ for some symbolic traces $\rho, \rho' \in (Cond
\union \set{\tau})^*$ and some visible symbolic events $\epsilon$ and
$\epsilon'$. Hence, by Lemma~\ref{lemmaA1}, $\rho \nontauequiv \rho'$. Also
from the definition of $\generates{}$ we have that $e \in
\mathit{Insts}_\Gamma(\epsilon) \inter \mathit{Insts}_\Gamma(\epsilon')$, so,
by Lemma~\ref{lemmaB}, $\epsilon = \epsilon'$. Therefore, $\sigma_2
\nontauequiv \sigma'_2$, and hence $\sigma \nontauequiv \sigma'$.  \qed


\proof[Proof of Proposition \ref{corollaryA2}]
Suppose for contradiction that $\alpha \not =
\alpha'$. Definition~\ref{definition:matching} implies that $\alpha$ and
$\alpha'$ give rise to $e$ immediately after $tr$. By
Proposition~\ref{propositionA2}, if $\sigma\cat\strace{\epsilon}$ and
$\sigma'\cat\strace{\epsilon'}$ are both symbolic traces of $Proc(t)$ such
that $\sigma\cat\strace{\epsilon}, \sigma'\cat\strace{\epsilon'}
\generates{\nullset} tr\cat\trace{e}$ and where $\epsilon$ and $\epsilon'$ are
visible, then $\sigma \nontauequiv \sigma'$ and $\epsilon = \epsilon'$. We now
have that $\epsilon$ matches both $\alpha$ and $\alpha'$. Then, the firing
rules of SSOS (see Section~\ref{section:SSOSfiringrules}) imply that $\alpha$
and $\alpha'$ must be constructs in different branches of an external,
internal or sliding choice. Since both of these constructs give rise to the
same concrete event, $e$, their channels must be identical. Hence, $Proc(t)$
contains a binary choice with branches sharing a common channel name. This
contradicts the fact that $Proc(t)$ satisfies \Norm{}, so it must be that
$\alpha = \alpha'$. 
\qed


\proof[Proof of Lemma~\ref{lemmaD}]
Suppose for a contradiction that there is some $j \in \mathord !(\epsilon')
\setminus \mathord !(\epsilon)$. Then $j \in \mathord \$^t(\epsilon) \union
\mathord ?^t(\epsilon)$ (since $\$^\nont(\epsilon) = \mathord
?^\nont(\epsilon) = \nullset$ by
Remark~\ref{remark:noNontPartsInSymbolicVisibleEvents}). This means that the
$j$-th variable or value of $\epsilon'$ is of type $t$, so $j \in \mathord
!^t(\epsilon')$.  Suppose $e = c.v_1\ldots v_k$ and let $e' = c.v_1 \ldots
v_{j-1}.v'_j.v_{j+1}\ldots v_k$, where $v'_j \in T \setminus \set{v_j}$. By
Remark~\ref{remark:tornott}, if a process can perform a given event, then it
can also perform every other event that differs only in the values of
inputs. Therefore, $tr\cat \trace{e'} \in \tsem{P(t),\Gamma_{init},T}$,
i.e.~$e'$ matches $\epsilon$. 

Since $(Q(t),\Gamma_{init},T) \trefinedby (P(t),\Gamma_{init},T)$, we must
have $tr\cat \trace{e'} \in \tsem{Q(t),\Gamma_{init},T}$. Clause~(iii),
combined with the fact that $v'_j$ is an output for~$Q(t)$, different from
$v_j$, implies that $\sigma'\cat \strace{\epsilon'}$ cannot generate
$tr\cat\trace{e'}$ within $\Gamma_{init}$. So let $\rho\cat
\strace{\epsilon''} \in \stsem{Q(t)}$ be such that $\rho\cat
\strace{\epsilon''} \generates\Gamma_{init} tr\cat\trace{e'}$. Also, let
$\sigma' = \sigma_1\cat \sigma_2$  and  $\rho = \rho_1\cat \rho_2$,
where $\sigma_1$ and $\rho_1$ are either both the empty symbolic trace or both
end with visible symbolic events, and $\sigma_2, \rho_2 \in (Cond \union
\set{\tau})^*$. Then, clause~(iii) implies that $\sigma_1
\generates{\Gamma_{init}} tr$ and $\rho_1 \generates{\Gamma_{init}} tr$, so by
Proposition~\ref{propositionA2}, $\sigma_1 \nontauequiv \rho_1$. Hence, if
$Q'(t)$ and $Q''(t)$ are symbolic states such that 
$Q(t) \SymTrans{\sigma_1} Q'(t)$ and $Q(t) \SymTrans{\rho_1} Q''(t)$,
then, thanks to Corollary~\ref{corollaryX}, we have that $Q'(t) = Q''(t)$. 

\begin{figure}[htbp]
\SelectTips{cm}{}
\begin{center}\ 
\xymatrix @R=0mm {
& \save[] \hspace{-9em} \sigma_1 \nontauequiv \rho_1 \restore
& \save[] \hspace{-1em} \sigma_2 \cat \strace{\epsilon'} 
    \generates{\Gamma} \trace{e} \restore
\\ \\
&& {\bullet} \ar@{|->}_>{s}[r]^{\strace{\epsilon'}} & {\bullet}
\\
Q(t) \ar@{|->}@/^2pc/_>{s}[r]^{\sigma_1}
 \ar@{|->}@/_2pc/_>{s}[r]^{\rho_1}
& *+++{\hspace{4mm}Q'(t) =  Q''(t)\hspace{-2mm}} \ar@{|->}^(0.7){\sigma_2}_>{s}[ur] 
    \ar@{|->}_>{s}[dr]^(0.7){\rho_2} 
\\
&& {\bullet} \ar@{|->}_>{s}[r]^{\strace{\epsilon''}} & {\bullet}
\\ \\
& \save[] \hspace{-8em} \sigma_1, \rho_1 \generates{\Gamma_{init}} tr \restore
& \save[] \hspace{-1em} \rho_2 \cat \strace{\epsilon''} 
    \generates{\Gamma} \trace{e'} \restore
}\ 
\end{center}
\caption{Illustration of the proof of Lemma \ref{lemmaD}}
\end{figure}

Let $\Gamma$ be the environment reached after $tr$, i.e.\ such that
$(Q(t),\Gamma_{init},T) \stackrel{s}{\longmapsto} (Q'(t),\Gamma,T)$ for some
$s$ such that $s \hide \tau = tr$; by
Proposition~\ref{prop:environmentuniqueness}, $\Gamma$ is unique.  Then, we
have that
\begin{iteMize}{$\bullet$}
\item$\sigma_2\cat\strace{\epsilon'}, \rho_2\cat\strace{\epsilon''} \in
  \stsem{Q'(t)}$; 

\item $\trace{e}, \trace{e'} \in \tsem{Q'(t),\Gamma,T}$;

\item $\sigma_2\cat\strace{\epsilon'} \generates{\Gamma} \trace{e}$; and

\item $\rho_2\cat\strace{\epsilon''} \generates{\Gamma} \trace{e'}$.
\end{iteMize}
Hence, we can deduce from Lemma~\ref{lemmaA1} that $\sigma_2 \nontauequiv
\rho_2$. Let $\epsilon' =
c\S'_1x'_1\mathrm{:}X'_1\ldots\S'_kx'_k\mathrm{:}X'_k$ and $\epsilon'' =
c\S''_1x''_1\mathrm{:}X''_1\ldots\S''_kx''_k\mathrm{:}X''_k$. Then, since the
channels of $\epsilon'$ and $\epsilon''$ are the same, Lemma~\ref{lemmaB}
implies that $!^t(\epsilon') = \mathord !^t(\epsilon'')$ and $x'_j =
x''_j$. Since $\sigma_2 \cat \strace{\epsilon'} \generates{\Gamma}\strace{e} =
\trace{c.v_1\ldots v_k}$ and $\rho_2 \cat \strace{\epsilon''}
\generates{\Gamma} \strace{e'} = \trace{c.v_1\ldots v_{j-1}.v'_j.v_{j+1}\ldots
  v_k}$  and $j \in \mathord !^t(\epsilon'')$ (as $j \in \mathord
!^t(\epsilon')$), we have that $x'_j = v_j$ and $x''_j = v'_j$. Hence $v_j =
v'_j$. This is a contradiction, so $!(\epsilon') \subseteq \mathord
!(\epsilon)$.  \qed


\proof[Proof of Proposition \ref{proposition:adataindepimplication}]
Since $\shiftedhat{T}$ is a subset of $T$, we have that $\Gamma, \Gamma' \in
\mathit{Env}(\shiftedhat{T})$ implies $\Gamma, \Gamma' \in \mathit{Env}(T)$,
since every partial function from $\Var$ to $\shiftedhat{T}$ is also a partial
function from $\Var$ to~$T$. We now prove the result using an induction on
$n$, the number of times Translation Rule~4 of COSE
(p.~\pageref{rule:translationrule4}) had to be applied in order to obtain the
transition in~(\ref{equation:adataindepimplicationeq1}). 

\para{Base case.} Suppose that $n = 0$. We separately consider the cases for
$a$ being $\tau$ or visible.

\subpara{Case 1.} Suppose that $a = \tau$. 
Then, the translation rules of COSE (see
Section~\ref{section:translationrules}) imply that the transition
in~(\ref{equation:adataindepimplicationeq1}) can be a result of either
Translation Rule~1 (p.~\pageref{rule:translationrule1}) or Translation Rule~3
(p.~\pageref{rule:translationrule3}).

For Translation Rule~1,
it must be that $Proc(t) \trans[\epsilon][s] P(t)$ for some visible symbolic
event $\epsilon = \construct$ such that $\#\$^t(\epsilon) > 0$ and some
symbolic state $P(t)$. In addition, $Proc'(t) = Replace_{\$ \mapsto
  !}^t(c, Proc(t))$ and $\Gamma' = \Gamma \oplus \set{x_i \mapsto v_i |
  i \in \mathord \$^t(\epsilon)}$, where $v$ is a function in $\$^t(\epsilon)
\rightarrow \shiftedhat{T}$. Then, $v$ is also a function in $\$^t(\epsilon)
\rightarrow T$, so Translation Rule~1 implies that  
\begin{eqnarray*}
(Proc(t), \Gamma, T) & \trans[\tau] &
  (Replace_{\$ \mapsto !}^t(c, Proc(t)), 
    \Gamma \oplus \set{x_i \mapsto v_i | i \in \mathord \$^t(\epsilon)}, T) \\
& = & (Proc'(t), \Gamma', T).
\end{eqnarray*}

\noindent
If the transition in~(\ref{equation:adataindepimplicationeq1}) results
from Translation Rule~3, then $Proc(t)
\trans[\tau][s] Proc'(t)$ and $\Gamma = \Gamma'$. The same rule then
yields $(Proc(t), \Gamma, T) \trans[\tau] (Proc'(t), \Gamma', T)$.

\subpara{Case 2.} 

Suppose that $a = c.v_1 \ldots v_k$ is a visible event.
Then, the translation rules of COSE imply that the transition
in~(\ref{equation:adataindepimplicationeq1}) must be the result of Translation
Rule~2 (p.~\pageref{rule:translationrule2}). The rule implies that $Proc(t)
\trans[\epsilon][s] Proc'(t)$ for some visible symbolic event $\epsilon =
\construct$ such that $\#\$^t(\epsilon) = 0$. In addition, $\Gamma' = \Gamma
\oplus \set{{x_i \mapsto v_i} | i \in \mathord ?^t(\epsilon)}$, and for all $i$
in $?^t(\epsilon)$,\, $v_i$ is in $\shiftedhat{T}$. However, since
$\shiftedhat{T} \subseteq T$, we have that for all $i$ in $?^t(\epsilon)$,\,
$v_i$ is in $T$. Therefore, Translation Rule~2 implies that  
\[
(Proc(t), \Gamma, T) \;\trans[a]\; 
  (Proc'(t), \Gamma \oplus
       \set{x_i \mapsto v_i | i \in \mathord ?^t(\epsilon)}, T)
\; = \; (Proc'(t), \Gamma', T).
\]
This completes the base case.

\para{Inductive case.} 

Suppose the result holds for some $n = k$, where $k \geq 0$. Suppose that the
transition in~(\ref{equation:adataindepimplicationeq1}) requires $k+1$
applications of Transition Rule~4. Then it must be that $Proc(t)
\trans(3)[cond][s] P(t)$ and $(P(t),\Gamma,\shiftedhat{T}) \trans[a]
(Proc'(t),\Gamma', \shiftedhat{T})$. The latter transition requires $k$
applications of Transition Rule~4, so the inductive hypothesis implies that
$(P(t),\Gamma,T) \trans[a] (Proc'(t),\Gamma', T)$. Hence, by Translation
Rule~4, $(Proc(t),\Gamma,T) \trans[a] (Proc'(t),\Gamma', T)$, which completes
our proof. 
\qed


\proof[Proof of Corollary~\ref{corollary:adataindepimplication}]
Let $s$ be an event-sequence with $(Proc(t), \Gamma, \shiftedhat{T})
\mapstotrans[s] (Proc'(t), \Gamma', \shiftedhat{T})$ and $s \hide \set{\tau} =
tr$. Then the result follows from a simple induction on the length of $s$
using Proposition~\ref{proposition:adataindepimplication}. 
\qed

\section{Proofs for Section~\ref{section:typereduction}}
\label{sec:traces_proofs}

\subsection{Proofs for Section~\ref{section:tracesresults}}
\label{ssec:bigtracesproposition_proof}

\proof[Proof of Proposition \ref{bigtracesproposition}] We prove the result
using a structural induction on $Proc(t)$.  We give just the cases for prefix
and conditional choice.


\para{Prefix}
Suppose that $Proc(t) = \alpha \then P(t)$ where $\alpha =
c'\S'_1x'_1\mathrm{:}X'_1\ldots\S'_kx'_k\mathrm{:}X'_k$. We consider two
cases. 

\subpara{Subcase 1.} 

Suppose that $tr = \nil$. Then, 
$\sigma \in \set{\tau}^*$, and
$Proc(t) \starit{\trans}[\tau][s]\trans[\epsilon][s]$. Using Translation
Rule~3 (p.~\pageref{rule:translationrule3}) and 
Remark~\ref{remark:translationrules1and2} we get that  
\begin{equation}\label{six}
\align
\forall v' \in \set{1 \upto k} \rightarrow \Value | \\
\qquad 
  (\forall i \in \mathord \$^t(\epsilon) \union \mathord ?^t(\epsilon) \spot
    v'_i \in T) \land
 (\forall i \in \mathord !(\epsilon) \spot  v'_i = \Gamma_{init}(x_i)) \spot\\
\qquad \qquad \trace{c.v'_1\ldots v'_k} \in \tsem{Proc(t), \Gamma_{init}} 
\endalign
\end{equation}
Since $tr = \nil$, assumption (iii) of the proposition 
implies that
$\strace{\epsilon} \generates{\phi(\Gamma_{init})} \trace{e}$. Therefore,
by assumption~(iv), 
\[
\forall i \in \mathord !^t(\epsilon) \spot 
  v_i = (\phi(\Gamma_{init}))(x_i) \land v_i \in \set{0 \upto B-1}.
\]
Suppose that 
\[
\begin{align}
v' \in \set{1 \upto k} \rightarrow \Value \mbox{ is such that } \\
\qquad
(\forall i \in \mathord \$^t(\epsilon) \union \mathord ?^t(\epsilon) \spot
     v'_i \in T) \land 
(\forall i \in \mathord !(\epsilon) \spot v'_i = v_i)
\end{align}
\]
Then
\(
\forall i \in \mathord !^t(\epsilon) \spot 
  v'_i = (\phi(\Gamma_{init}))(x_i) \land v'_i \in \set{0\upto B-1}.
\)
However, the properties of $\phi$ imply that for all variables $var$ and all
values $val$, we have that  
\[
(\phi(\Gamma_{init}))(var) = val \land val \in \set{0 \upto B-1} 
  \implies \Gamma_{init}(var) = val,
\]
so
\begin{equation}\label{four}
\forall i \in \mathord !^t(\epsilon) \spot v'_i = \Gamma_{init}(x_i).
\end{equation}
In addition, from the definition of $\generates{}$,
\(
\forall i \in \mathord !^\nont(\epsilon) \spot v_i = \phi(\Gamma_{init})(x_i).
\)
But we know that $\forall i \in \mathord !^\nont(\epsilon) \spot v'_i = v_i$, so 
\[
\forall i \in \mathord !^\nont(\epsilon) \spot
   v'_i = (\phi(\Gamma_{init}))(x_i) = \Gamma_{init}(x_i)
\]
with the last equality following from the fact that for all $i$ in
$!^\nont(\epsilon)$, $x_i$ must be of a non-$t$ type. Hence and from
(\ref{four}), 
\[
	\forall i \in \mathord !(\epsilon) \spot v'_i = \Gamma_{init}(x_i).
\]
Therefore, (\ref{six}) implies that 
\( \trace{c.v'_1\ldots v'_k} \in \tsem{Proc(t), \Gamma_{init}} \).
We have shown:
\[
\align
\forall v' \in \set{1 \upto k} \rightarrow \Value | 
  (\forall i \in \mathord \$^t(\epsilon) \union \mathord ?^t(\epsilon) \spot
     v'_i \in T) \land 
  (\forall i \in \mathord !(\epsilon) \spot  v_i' = v_i) \spot\\
\qquad \trace{c.v'_1\ldots v'_k} \in \tsem{Proc(t), \Gamma_{init}},
\endalign
\]
which is what we wanted to show.

\subpara{Subcase 2.} 

Suppose that $tr = \trace{e'}\cat tr'$ for some visible event $a$ and some
trace $tr'$. Then $\phi(tr) = \trace{\phi(e')} \cat \phi(tr')$. So clearly
$\phi(tr)$ is non-empty and we know that
$\sigma \generates{\phi(\Gamma_{init})} \phi(tr)$. By the definition of
$\generates{}$, it must be that there is at least one visible symbolic event
within $\sigma$. Hence, $\sigma = \sigma_1\cat\strace{\epsilon'}\cat\sigma_2$
for some visible symbolic event $\epsilon'$ and some symbolic traces
$\sigma_1$ and $\sigma_2$ such that $\sigma_1$ is in $\set{\tau}^*$
($\sigma_1$ cannot contain any conditional symbolic events because $Proc(t)$
is a prefix). Then,
\begin{equation}
\label{equation:maintracespropeq1}
Proc(t) \starit{\trans}[\tau]\trans[\epsilon'] P'(t)
\qquad\mbox{and}\qquad
\sigma_2\cat\strace{\epsilon} \in \stsem{P'(t)},
\end{equation}
where $P'(t)$ is like $P(t)$, but with some substitutions of concrete values
for the non-$t$ type input variables of $\alpha$, as dictated by the SSOS
firing rules for prefix (Section~\ref{section:SSOSfiringrules}). 
We aim to apply the inductive hypothesis to $P'(t)$.

We can infer using Translation Rule~3 (p.~\pageref{rule:translationrule3}) and
Remark~\ref{remark:translationrules1and2} that
\begin{eqnarray*}
(Proc(t), \phi(\Gamma_{init})) & \starit{\trans}[\tau]\trans(3)[\phi(e')] &
  (P'(t), \phi(\Gamma_{init}) \oplus \mathit{Match}(\epsilon', \phi(e'))),
  \end{eqnarray*}
where, recall from Section~\ref{section:generates}, $\mathit{Match}(\epsilon',
\phi(e'))$ is a map from type $t$ input variables of $\epsilon'$ to the
corresponding concrete values of event $\phi(e')$. We have that configuration
$(P'(t), \phi(\Gamma_{init}) \oplus \mathit{Match}(\epsilon', \phi(e')))$ is
unique (thanks to Proposition~\ref{prop:environmentuniqueness}).  We know from
assumption~(ii) that \( \trace{\phi(e')} \cat \phi(tr') \cat\trace{e} \in
\tsem{Proc(t), \phi(\Gamma_{init})} \).  Hence
\begin{eqnarray}
\label{toshow1}
\phi(tr')\cat\trace{e} & \in &
  \tsem{P'(t), \phi(\Gamma_{init}) \oplus \mathit{Match}(\epsilon', \phi(e'))}.
\end{eqnarray}

Similarly, since $Proc(t) \starit{\trans}[\tau]\trans[\epsilon'] P'(t)$ and
$tr = \trace{e'} \cat tr' \in \tsem{Proc(t),\Gamma_{init}}$ (from
assumption~(i)), using Translation 
Rule~3 (p.~\pageref{rule:translationrule3}), 
Remark~\ref{remark:translationrules1and2} and
Proposition~\ref{prop:environmentuniqueness} we can infer that 
\begin{equation}
\label{equation:maintracespropeq2}
(Proc(t),\Gamma_{init}) \; \starit{\trans}[\tau]\trans[e'] \;
  (P'(t), \Gamma_{init} \oplus \mathit{Match}(\epsilon', e')) \; \Trans[tr'].
\end{equation}
Hence,
\begin{eqnarray}
\label{toshow2}
tr' & \in & \tsem{P'(t), \Gamma_{init} \oplus \mathit{Match}(\epsilon', e')}.
\end{eqnarray}

Finally, assumption~(iii) implies that
\[
\sigma \cat \strace{\epsilon} 
  \;=\; \sigma_1 \cat \strace{\epsilon'} \cat \sigma_2 \cat \strace{\epsilon} 
  \;\generates{\phi(\Gamma_{init})} \;
  \phi(tr) \cat \trace{e}\; = \;  \trace{\phi(e')}\cat\phi(tr')\cat \trace{e}.
\]
So, by the definition of $\generates{}$ (p.~\pageref{def:generates})
\begin{eqnarray}\label{toshow3}
\sigma_2\cat\strace{\epsilon} 
  & \generates{\phi(\Gamma_{init})\oplus \mathit{Match}(\epsilon',\phi(e'))} &
  \phi(tr')\cat \trace{e}.
\end{eqnarray}

We can now deduce the inductive hypothesis for $P'(t)$, with $tr'$ in place
of~$tr$,\, $\sigma_2$ in place of~$\sigma$, and
$\Gamma_{init}\oplus \mathit{Match}(\epsilon',e')$ in place of
$\Gamma_{init}$: (\ref{toshow2})~gives us condition~(i); (\ref{toshow1})~gives
us condition~(ii), observing that
\(
\phi(\Gamma_{init}) \oplus \mathit{Match}(\epsilon',\phi(e'))
= \phi(\Gamma_{init}\oplus \mathit{Match}(\epsilon',e'))
\); and (\ref{equation:maintracespropeq1}) and (\ref{toshow3}) give us
condition~(iii).  Hence
\[
\align
\forall v' \in \set{1 \upto k} \rightarrow \Value | 
  (\forall i \in \mathord \$^t(\epsilon) \union \mathord ?^t(\epsilon) \spot
     v'_i \in T) \land 
  (\forall i \in \mathord !(\epsilon) \spot  v_i = v'_i) \spot\\
\qquad tr'\cat\trace{c.v'_1\ldots v'_k} \in 
  \tsem{P'(t), \Gamma_{init} \oplus \mathit{Match}(\epsilon',e')}. 
\endalign
\]
This, combined with (\ref{equation:maintracespropeq2}), gives us that 
\[
\align
\forall v' \in \set{1 \upto k} \rightarrow \Value | 
  (\forall i \in \mathord \$^t(\epsilon) \union \mathord ?^t(\epsilon) \spot
     v'_i \in T) \land 
  (\forall i \in \mathord !(\epsilon) \spot  v_i = v'_i) \spot\\
\qquad \trace{e'}\cat tr'\cat\trace{c.v'_1\ldots v'_k} \in
   \tsem{Proc(t), \Gamma_{init}}.
\endalign
\]
However, $tr = \trace{e'}\cat tr'$, so the result holds.

\para{Conditional choice}
Suppose that $Proc(t) = \If cond \Then P(t) \Else Q(t)$. If $cond$ is not in
$Cond$, then $cond$ immediately evaluates to $\True$ or $\False$ and the
result is immediately implied by the inductive hypothesis for $P(t)$ or
$Q(t)$, respectively. So suppose $cond$ is in $Cond$. Then, by the SSOS firing
rules for conditional choice (see Section~\ref{section:SSOSfiringrules}) it
must be that
\[
\sigma = \strace{cond}\cat\rho
  \mbox{\quad or \quad}\sigma = \strace{\neg cond}\cat\rho
\]
for some symbolic trace $\rho$. We now perform a case analysis on the truth
value of the evaluation of $cond$ within the environments $\Gamma_{init}$ and
$\phi(\Gamma_{init})$.

\subpara{Case 1.} 
Suppose that $\eval{cond}{\phi(\Gamma_{init})} = \eval{cond}{\Gamma_{init}} =
True$.  Then it must be that $\rho \in \stsem{P(t)}$. From assumption~(iii) we
have that
\begin{eqnarray*}
\rho\cat \strace{\epsilon} & \generates{\phi(\Gamma_{init})} &
  \phi(tr)\cat \trace{e}.
\end{eqnarray*}
In addition, from assumptions~(i)~and~(ii) we have that
\[
tr \in \tsem{P(t), \Gamma_{init}}
\mbox{\quad and \quad} 
\phi(tr)\cat \trace{e} \in \tsem{P(t), \phi(\Gamma_{init})} .
\]
The result is now implied in this case by the inductive hypothesis for $P(t)$
and the fact that $(Proc(t), \Gamma_{init}) \Trans[\nil]
(P(t),\Gamma_{init})$.

\subpara{Case 2.} Suppose that $\eval{cond}{\phi(\Gamma_{init})}
= \eval{cond}{\Gamma_{init}} = False$. 
This case is like Case 1, above, with $Q(t)$ in place of $P(t)$.

\subpara{Case 3.} Suppose that $\eval{cond}{\phi(\Gamma_{init})} =
True \land \eval{cond}{\Gamma_{init}} = False$.  Then, by
assumption~(i)~and~(ii),
\[
tr \in \tsem{Q(t), \Gamma_{init}}
\mbox{\quad and \quad} 
\phi(tr)\cat \trace{e} \in \tsem{P(t), \phi(\Gamma_{init})}.
\]
Since $Proc(t)$ satisfies \textbf{RevPosConjEqT}, we have that
\( (Q(t),\phi(\Gamma_{init})) \trefinedby (P(t),\phi(\Gamma_{init})) \),
so
\[
	\phi(tr)\cat \trace{e} \in \tsem{Q(t), \phi(\Gamma_{init})}.
\]
Let $\rho'\cat \strace{\epsilon'} \in \stsem{Q(t)}$ be such that
$\rho'\cat\strace{\epsilon'} \generates{{\phi(\Gamma_{init})}} \phi(tr)\cat 
\trace{e}$. Then, 
by Lemma~\ref{lemmaD}, $!(\epsilon') \subseteq \mathord !(\epsilon)$. Hence,
$!^t(\epsilon') \subseteq \mathord !^t(\epsilon)$, so assumption~(iv) implies
that 
\[
	\forall i \in \mathord !^t(\epsilon') \spot v_i \in \set{0 \upto B-1}.
\]
Therefore, by the inductive hypothesis for $Q(t)$,
\begin{equation}\label{tracespropositioncondchoiceeq1}
\align
\forall v' \in \set{1 \upto k} \rightarrow \Value | \\
\qquad 
  (\forall i \in \mathord \$^t(\epsilon') \union \mathord ?^t(\epsilon') 
     \spot v'_i \in T) \land 
  (\forall i \in \mathord !(\epsilon') \spot v'_i = v_i) \spot\\
\qquad\qquad tr\cat \trace{c.v'_1\ldots v'_k} \in \tsem{Q(t),\Gamma_{init}}.
\endalign
\end{equation}
Suppose $v' \in \set{1 \upto k} \rightarrow \Value$ is such that
\[
(\forall i \in \mathord \$^t(\epsilon) \union \mathord ?^t(\epsilon)
   \spot v'_i \in T) \land 
(\forall i \in \mathord !(\epsilon) \spot v'_i = v_i).
\]
The fact that $!(\epsilon') \subseteq \mathord !(\epsilon)$ implies that
\begin{equation}\label{equation:tracespropositione1}
\forall i \in \mathord !(\epsilon') \spot v'_i = v_i.
\end{equation}
In addition, we know that both $\epsilon$ and $\epsilon'$ give rise to
$c.v_1 \ldots v_k$, so  
\begin{eqnarray*}
\$^t(\epsilon') \union \mathord ?^t(\epsilon') \union \mathord !^t(\epsilon') 
 & = &
  \$^t(\epsilon) \union \mathord ?^t(\epsilon) \union \mathord !^t(\epsilon),
\end{eqnarray*}
which means that
\begin{eqnarray*}
\$^t(\epsilon') \union \mathord ?^t(\epsilon') & \subseteq &
  \$^t(\epsilon) \union \mathord ?^t(\epsilon) \union \mathord !^t(\epsilon).
\end{eqnarray*}
Therefore, since $\forall i \in \mathord !^t(\epsilon) \spot v'_i \in T$ (as
$\forall i \in \mathord !^t(\epsilon) \spot v'_i  = v_i$ and, by
assumption~(iv), $\forall i \in \mathord !^t(\epsilon) \spot v_i \in T$) and
$\forall i \in \mathord \$^t(\epsilon) \union \mathord ?^t(\epsilon) \spot
v'_i \in T$ (by our assumption about $v'$),
\[
\forall i \in \mathord \$^t(\epsilon') \union \mathord ?^t(\epsilon') \spot
   v'_i \in T.
\]
Combining this with (\ref{equation:tracespropositione1}) and
(\ref{tracespropositioncondchoiceeq1}), we get that 
\[
\align
\forall v' \in \set{1 \upto k} \rightarrow \Value | 
  (\forall i \in \mathord \$^t(\epsilon) \union \mathord ?^t(\epsilon) \spot
      v'_i \in T) \land 
  (\forall i \in \mathord !(\epsilon) \spot v'_i = v_i) \spot\\
\qquad tr\cat \trace{c.v'_1\ldots v'_k} \in \tsem{Q(t),\Gamma_{init}}.
\endalign
\]
The result now follows, because $(Proc(t), \Gamma_{init}) \Trans[\nil]
(Q(t),\Gamma_{init})$.

\subpara{Case 4.} Suppose that $\eval{cond}{\phi(\Gamma_{init})} =
False \land \eval{cond}{\Gamma_{init}} = True$. 
This case is not possible since $cond$ is a conjunction of equality tests and
for no function $\phi$ we can ever have $x = y$ and $\phi(x) \not = \phi(y)$. 
\qed



We now prove Proposition~\ref{prop:noEqT-threshold}.  We will need the
following lemma which shows that in this case non-$t$ equivalent symbolic
traces are in fact non-$\tau$ equivalent.
\begin{lem}
Suppose $\sigma$, $\sigma'$ are symbolic traces that contain no conditional
symbolic events, $\sigma \nontequiv \sigma'$, neither $\sigma$ nor $\sigma'$
ends in~$\tau$, and
\[
P(t) \SymTrans{\sigma} Q(t) \qquad\mbox{and}\qquad 
P(t) \SymTrans{\sigma'} Q'(t).
\]
Then $\sigma \nontauequiv \sigma'$ and $Q(t) = Q'(t)$.
\end{lem}
\proof
We prove the result by induction on the number of visible symbolic events in
$\sigma$ and~$\sigma'$.  The base case of $\sigma = \sigma' = \trace{}$ is
trivial.  

Suppose $\sigma = \sigma_0 \cat \tau^a \cat \trace\epsilon$,\, $\sigma' =
\sigma_0' \cat \tau^b \cat \trace{\epsilon'}$, and $\sigma_0$, $\sigma_0'$ do
not end in~$\tau$.  Then $\sigma_0 \nontequiv \sigma_0'$,\, $\epsilon
\nontequiv \epsilon'$, and
\[
P(t) \SymTrans{\sigma_0} Q_0(t) 
  \mapstotrans(3)[\tau^a \cat \trace\epsilon][s]  Q(t) 
\qquad\mbox{and}\qquad 
P(t) \SymTrans{\sigma_0'} Q_0'(t) 
  \mapstotrans(3)[\tau^b \cat \trace{\epsilon'}][s] Q'(t)
\]
for some $Q_0(t)$ and $Q_0'(t)$.  Then by the inductive hypothesis, $\sigma_0
\nontauequiv \sigma_0'$ and $Q_0(t) = Q_0'(t)$.  Then by Lemma~\ref{lemmaB},
the $t$ parts of $\epsilon$ and $\epsilon'$ are equal, so $\epsilon =
\epsilon'$; hence $\sigma \nontauequiv \sigma'$.  Finally, by
Lemma~\ref{lemmaX}, $Q(t) = Q'(t)$.  \qed


\proof[Proof of Proposition \ref{prop:noEqT-threshold}]
Let $\sigma \cat
\trace{\epsilon}$ be a symbolic trace of~$\Spec(t)$.  If $\sigma' \cat
\trace{\epsilon'}$ is a symbolic trace of~$\Spec(t)$ such that $\sigma \cat
\trace{\epsilon} \nontequiv \sigma' \cat \trace{\epsilon'}$, then by the above
lemma, $\epsilon = \epsilon'$.  Hence $!^t(\sigma, \epsilon)(\Spec(t)) =
\mathord{!}^t(\epsilon)$.  So in Theorem~\ref{maintracesthm},
\begin{eqnarray*}
\Thresh_{\mathrm{T}} & = & 
  \max\set{ \# !^t(\epsilon)(\Spec(t)) | 
     \sigma \cat \trace\epsilon \in \stsem{P(t)} } \\
& \le & 
 \max\set{\# !^t(\alpha) | \mbox{$\alpha$ is a construct of $\Spec(t)$}}
\end{eqnarray*}
with equality in the normal case that every construct is reachable. 
\qed

\subsection{Proofs for Section \ref{section:failuresresults}}
\label{ssec:bigfailuresproposition_proof}

\proof[Proof of Proposition~\ref{bigfailuresproposition}]
We prove the result using a structural induction on $Proc(t)$.  We give just
the cases for prefix and conditional choice. 


\para{Prefix}
Suppose that $Proc(t) = \alpha \then Proc'(t)$ for some construct $\alpha
= \construct$ and some process syntax $Proc'(t)$. We now consider two cases.

\subpara{Subcase 1.} Suppose that $tr = \nil$.
The fact that $(\phi(tr), X) \in \fsem{Proc(t), \phi(\Gamma_{init})}$ implies
that there exists an environment $\Gamma$ with $\dom (\Gamma) = \$^t(\alpha)$
such that   
\begin{eqnarray*}
(Proc(t),\phi(\Gamma_{init})) & \Trans[\nil] &
  (P(t),\phi(\Gamma_{init})\oplus \Gamma) \refuses X,
\end{eqnarray*}
where $P(t)$ is like $Proc(t)$, but with some substitutions of concrete values
for the nondeterministic input variables of non-$t$ types of $\alpha$ and with
the effects of the application of $Replace_{\$ \mapsto !}$, as dictated by the
SSOS firing rules for prefix (see
Section~\ref{section:SSOSfiringrules}). Then, 
\begin{eqnarray*}
(Proc(t),\Gamma_{init}) & \Trans[\nil] & (P(t),\Gamma_{init}\oplus \Gamma)
\end{eqnarray*}
by resolving the nondeterministic selections of $\alpha$ (if any) in the same
way. We now show that $(P(t),\Gamma_{init}\oplus \Gamma) \refuses X$. Observe
that the only difference between the initial events of two configurations
$(S,\Gamma_1)$ and $(S,\Gamma_2)$ are the output values of type $t$  that come
from the environments $\Gamma_1$ and $\Gamma_2$. Therefore, 
\[
\align
\initials(P(t),\Gamma_{init}\oplus \Gamma) =\\
\qquad \set{c.v'_1\ldots v'_k  |
  \align
  (\forall i \in \mathord !^t(\alpha) \spot 
     v'_i = \left(\Gamma_{init}\oplus \Gamma\right)(x_i))\\
  \land \exists 
    \align
    c.v_1\ldots v_k \in \initials(P(t),\phi(\Gamma_{init}) \oplus \Gamma) \spot\\
    \qquad \forall i \in \set{1 \upto k} \setminus \mathord !^t(\alpha) \spot
       v'_i = v_i )}.
    \endalign
  \endalign
\endalign
\]
Let $v$ be such that $c.v_1 \ldots v_k$ is in
$\initials(P(t), \phi(\Gamma_{init}) \oplus \Gamma)$ and let $v'$ be such that
$c.v'_1 \ldots v'_k$ is in $\initials(P(t), \Gamma_{init} \oplus \Gamma)$ with
$\forall i \in \set{1 \upto k} \setminus \mathord !^t(\alpha) \spot v'_i =
v_i$. Also, let $i \in \mathord!^t(\alpha)$.  Hence $v_i =
(\phi(\Gamma_{init}) \oplus \Gamma)(x_i)$ and $v_i' =
(\Gamma_{init} \oplus \Gamma)(x_i)$. Hence, by assumption (iii) of the
proposition, $v'_i \in \set{0 \upto B-1}$. So, thanks to the properties of
$\phi$, $\phi(v'_i) = v'_i$.  Hence $(\phi(\Gamma_{init}) \oplus \Gamma)(x_i)
= v'_i$ since $x_i \nin \dom (\Gamma)$.  Therefore
\[
\forall i \in \mathord !^t(\alpha) \spot 
  v'_i \; = \; (\Gamma_{init}\oplus \Gamma)(x_i) 
  \; = \; (\phi(\Gamma_{init})\oplus \Gamma)(x_i) \; = \;  v_i.
\]
Hence,
\begin{eqnarray}
\label{initialsequality}
\initials(P(t),\Gamma_{init}\oplus \Gamma)
  & = & \initials(P(t),\phi(\Gamma_{init}) \oplus \Gamma).
\end{eqnarray}
Since $(P(t),\Gamma_{init}\oplus \Gamma)$ is stable,
$(P(t),\Gamma_{init} \oplus \Gamma) \refuses Y$ for all
$Y \subseteq \Sigma \setminus \initials(P(t),\linebreak[1]
{\Gamma_{init}\oplus \Gamma})$. 
However, from the fact that
$(P(t), \phi(\Gamma_{init}) \oplus \Gamma) \refuses X$ we can infer that
$X \subseteq \Sigma \setminus \initials(P(t),\phi(\Gamma_{init})\oplus \Gamma)$,
so by (\ref{initialsequality}) $(P(t),\Gamma_{init}\oplus \Gamma) \refuses
X$. This implies that $(\nil, X) \in \fsem{Proc(t), \Gamma_{init}}$, as
required.

\subpara{Subcase 2.} 

Suppose that $tr \not = \nil$.
Then $tr = \trace{e}\cat tr'$ for some visible event $e$ that matches~$\alpha$
and trace $tr'$. Let $\Gamma = \phi(\Gamma_{init}) \oplus
Match(\alpha,\phi(e))$ and $\Gamma' = \Gamma_{init} \oplus
Match(\alpha,e)$. From the assumptions of the proposition we can infer that 
\[
tr' \in \tsem{P(t), \Gamma'} \qquad\mbox{and}\qquad 
(\phi(tr'),X) \in \fsem{P(t),\Gamma},
\]
where $P(t)$ is like $Proc'(t)$, but with some substitutions of concrete
values for the non-$t$ type input variables of $\alpha$, as dictated by the
SSOS firing rules for prefix (see
Section~\ref{section:SSOSfiringrules}). Assumption~(iii), combined with the
fact that $(Proc(t), \Gamma_{init}) \Trans[\trace{e}] (P(t), \Gamma')$,
implies that if $P$ is a configuration such that $(P(t), \Gamma') \Trans[tr']
P$, then every output of type~$t$ of every event in $\initials(P)$ is in
$\set{0 \upto B-1}$. Observe that $\Gamma = \phi(\Gamma')$. So, by the
inductive hypothesis for~$P(t)$, 
\(	(tr', X) \in \fsem{P(t), \Gamma'} \),
which implies that
\(	(tr, X) \in \fsem{Proc(t), \Gamma_{init}} \).

\para{Conditional choice} 

Suppose that $Proc(t) = \If cond \Then P(t) \Else Q(t)$ for some process
syntaxes $P(t)$ and $Q(t)$. If $cond$ is not in $Cond$, then it immediately
evaluates to $\True$ or $\False$, in which case the result is implied by the
inductive hypothesis for $P(t)$ or $Q(t)$, respectively. For a condition
$cond$ in $Cond$ we perform a case analysis on the result of the evaluation of
$cond$ within environments $\Gamma_{init}$ and $\phi(\Gamma_{init})$.

\subpara{Case 1.} Suppose that $\eval{cond}{\phi(\Gamma_{init})}
= \eval{cond}{\Gamma_{init}} = True$. 
Then
\[
	(\phi(tr),X) \in \fsem{P(t), \phi(\Gamma_{init})}\mbox{\quad and \quad} tr \in \tsem{P(t), \Gamma_{init}}.
\]
In addition, assumption~(iii), combined with the fact that
$(Proc(t), \Gamma_{init}) \Trans[\nil] (P(t), \Gamma_{init})$, implies that if
$P$ is a configuration such that $(P(t), \Gamma_{init}) \Trans[tr] P$, then
every output of type~$t$ of every event in $\initials(P)$ is in $\set{0 \upto
B-1}$. Then, the inductive hypothesis for $P(t)$ implies that 
\(
	(tr, X) \in \fsem{P(t), \Gamma_{init}}.
\)
Therefore,
\(
	(tr, X) \in \fsem{Proc(t), \Gamma_{init}}.
\)

\subpara{Case 2.} Suppose that $\eval{cond}{\phi(\Gamma_{init})}
= \eval{cond}{\Gamma_{init}} = False$. 
This case is like Case 1, above, with $Q(t)$ in place of $P(t)$.

\subpara{Case 3.} Suppose that $\eval{cond}{\phi(\Gamma_{init})} =
True \land \eval{cond}{\Gamma_{init}} = False$. 
Then 
\[
	(\phi(tr),X) \in \fsem{P(t), \phi(\Gamma_{init})}\mbox{\quad and \quad} tr \in \tsem{Q(t), \Gamma_{init}}.
\]
However, $Proc(t)$ satisfies $\textbf{RevPosConjEqT}_{\mathrm{F}}$, so
$(Q(t), \phi(\Gamma_{init})) \frefinedby (P(t), \phi(\Gamma_{init})$. Hence, 
\[
	(\phi(tr),X) \in \fsem{Q(t), \phi(\Gamma_{init})}.
\]
In addition, assumption~(iii), combined with the fact that
$(Proc(t), \Gamma_{init}) \Trans[\nil] (Q(t), \Gamma_{init})$, implies that if
$P$ is a configuration such that $(Q(t), \Gamma_{init}) \Trans[tr] P$, then
every output of type~$t$ of every event in $\initials(P)$ is in $\set{0 \upto
B-1}$. The inductive hypothesis for $Q(t)$ implies now that 
\( 	(tr, X) \in \fsem{Q(t), \Gamma_{init}}, \)
which implies that
\(	(tr, X) \in \fsem{Proc(t), \Gamma_{init}}. \)

\subpara{Case 4.} 

Suppose that $\eval{cond}{\phi(\Gamma_{init})} = False \land
\eval{cond}{\Gamma_{init}} = True$.  This case is not possible, since $cond$
is a conjunction of equality tests and for no function $\phi$ we can ever have
$x = y$ and $\phi(x) \not = \phi(y)$.
\qed



\vspace{-43 pt}
\end{document}